\documentclass[12pt,a4paper]{report}
\usepackage[a4paper,left=20mm,right=20mm,top=1in]{geometry}
\usepackage[table]{xcolor}
\usepackage{cite}      
\usepackage{graphicx}  
\usepackage{psfrag}   
\usepackage{subfig}
\usepackage{dblfloatfix}
\usepackage{array}
\usepackage{placeins}
\usepackage{titlesec}
\usepackage{etaremune}
\usepackage{pbox}
\usepackage{multirow}
\usepackage[utf8]{inputenc}
\usepackage{amssymb,amsmath,enumerate}   
\usepackage{cleveref}
\usepackage{array}
\usepackage{bibentry}
\usepackage[pagewise]{lineno}
\usepackage{color}
\usepackage{etoolbox}
\newcolumntype{M}[1]{>{\centering\arraybackslash}m{#1}}
\definecolor{LightCyan}{rgb}{0.88,1,1}
\usepackage{setspace}
\usepackage{graphicx}

\usepackage{array}
\newcolumntype{M}[1]{>{\centering\arraybackslash}m{#1}}
\graphicspath{{Images/}}
\begin{document}
\begin{center}
    \centering
    {\normalsize \textbf{A Report on}\par}
    {\normalsize July 2018\par}
    \vspace{3cm}
 {\Huge \textbf{ New Reconfigurable L-Band Digital Aeronautical Communication System}\par}
    \vspace{5cm}
{\normalsize Submitted By\par}
    {\Large \textbf{Niharika Agrawal}\par}
     \vspace{1cm}
      {\normalsize Advisor: Dr. Sumit J. Darak \par}
     \vspace{2cm}
    {\normalsize Department of Electronics and Communication Engineering\\
Indraprastha Institute of Information Technology - Delhi\\
New Delhi-110020 (India)\par}
\end{center}
\doublespacing
\setcounter{page}{1}%
\begin{abstract}

The Air Traffic Management (ATM) system handles the communication between the aircrafts and between aircraft
and ground terminals in the controlled airspace. Existing
ATM system can support low data rate services and fewer
communication links due to limited bandwidth. To meet the
spectrum demand of ever-increasing air traffic and enable a
variety of services to support different phases of flight, international civil aviation organization (ICAO) proposed Future
Communication Infrastructure (FCI) system. It consists
of several communication data links between satellite stations,
aircraft and ground terminals. Such FCI
system is also used for other Communications, Navigation, and
Surveillance (CNS) applications. The work presented
in this report focuses on the air-to-ground and air-to-air communication links which are the most important data links of the FCI system.
	
Recently, $L-$band (960-1164 MHz) has been identified for FCI and corresponding system is also referred to as $L-$band (960-1164 MHz) Digital Aeronautical Communication System (LDACS). The LDACS is based on opportunistic spectrum access based inlay approach over multiple  1 MHz available frequency bands between incumbent distance measuring equipment (DME) signals in $L$-band. 
The existing OFDM based LDACS has fixed transmission bandwidth of 498 KHz due to the high out-of-band emission of
OFDM and hence, cannot adapt the transmission bandwidth
as per the desired service and deployment. This leads to less than 50\%
vacant spectrum utilization and limits the usefulness for other
CNS applications. 

Our first contribution deals with the proposal for new frame structure for LDACS which supports wide range of the tunable bandwidth ranging from 186 KHz to 732 KHz. It is a generalized version of existing protocol which supports only 498 KHz bandwidth. For the proposed frame structure, we design reconfigurable filtered OFDM (Ref-OFDM) based LDACS transceiver using a reconfigurable linear phase multi-band finite impulse response (FIR) filter. It can adapt the bandwidth on-the-fly without the need of changing the filter coefficients and can be easily extended to multi-band filter for simultaneous transmission in multiple bands. It offers higher spectral efficiency due to lower OOB via filtering, lower interference to incumbent users, support for scalable bandwidth via reconfigurable filter and coexistence of asynchronous users making the proposed work an attractive solution for next-generation LDACS. The use of filter however leads to increase in the complexity compared to OFDM and hence, the design of area and power efficient reconfigurable filter is the challenging task and focus of ongoing work. Future work will focus on the theoretical analysis of the Ref-OFDM and multi-user multi-band deployment which has not been considered yet in the literature.

From architecture perspective, we plan to design and implement end-to-end LDACS transceiver on system on chip (SoC). We propose to follow novel hardware-software co-design approach to divide LDACS architecture for software (ARM processor) and programmable hardware (FPGA/ASIC) implementation and analyze optimum configuration for a given area, delay, OOB and power constraints. The first phase of the work involving design and implementation of OFDM based LDACS transceiver has been completed for seven different configurations. The final aim is to integrate LDACS transceiver with RF front-end (AD9361) for validation using the real radio signals with channels specific to LDACS deployment environment. 
\end{abstract}

\chapter*{Publications}
\renewcommand{\theenumi}{J\arabic{enumi}}
\section*{Journals}
\begin{enumerate}
\item N. Agrawal, S.J. Darak and Faouzi bader \lq\lq New Reconfigurable L-Band Digital aeronautical Communication System,\rq\rq submitted in ~\emph{IEEE Transactions on Aerospace and Electronic Systems,} March 2017 (Revision).
\item N. Agrawal, S. Garg, S.J. Darak and Faouzi bader \lq\lq Hardware - Software co simulation of LDACS - DME coexistence for Air to Ground Communications, \rq\rq in ~\emph{IEEE Transactions on Very Large Scale Integration (VLSI) Systems}, 2018 (Under Submission).
\end{enumerate}

\renewcommand{\theenumi}{C\arabic{enumi}}
\section*{Conferences}
\begin{enumerate}
\item 
N. Agrawal, S. J. Darak and F. Bader, \lq\lq Reconfigurable Filtered OFDM Waveform for Next Generation Air-to-ground Communications,\rq\rq ~\emph{2017 IEEE/AIAA 36th Digital Avionics Systems Conference (DASC),} St. Petersburg, FL, 2017, pp. 1-7. (Second Best Paper Award)
\item
S. Garg, N. Agrawal, S. J. Darak and P. Sikka, \lq\lq Spectral Coexistence of Candidate Waveforms and DME in Air-to-ground Communications: Analysis via Hardware Software Co-design on Zynq SoC,\rq\rq ~\emph{2017 IEEE/AIAA 36th Digital Avionics Systems Conference (DASC),} St. Petersburg, FL, 2017, pp. 1-6.
\end{enumerate}

\tableofcontents
\listoffigures
\listoftables
\chapter{Introduction}
\section{Background}
\begin{figure}[h!]
\centering
  \includegraphics[scale=0.4]{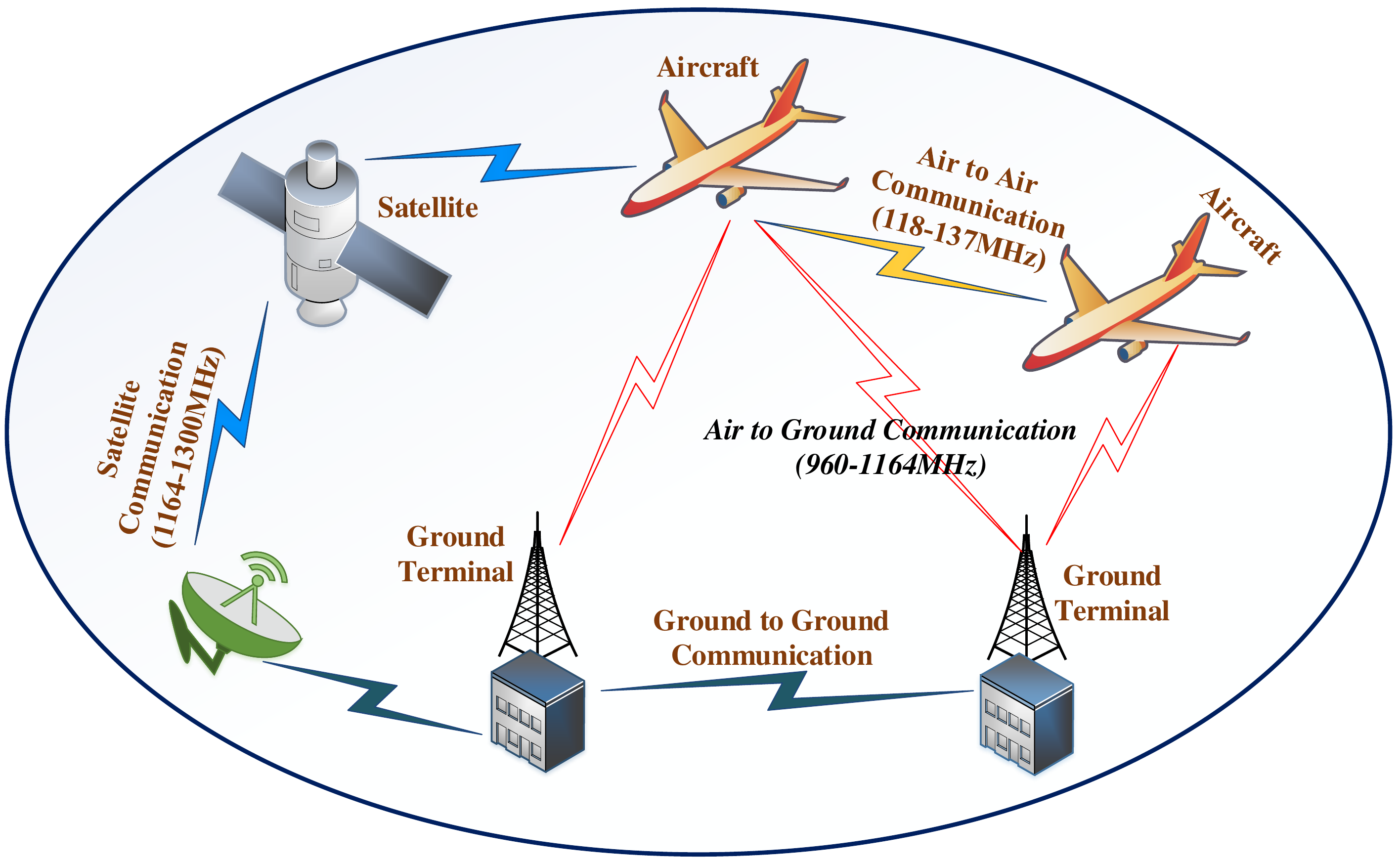}
  \caption{Various communication links in the future communication infrastructure (FCI) system.}
  \label{fci}
\end{figure}

The Air traffic management (ATM) system enables reliable communication between the aircrafts in airspace, ground terminals as well as aircrafts and ground terminals. Existing ATM systems are designed to support only low data rate services making them inefficient for upcoming delay sensitive high data rate services with ever increasing air-traffic. Research projects such as Next Generation Air Transportation System (NextGen) and Single European Sky ATM Research (SESAR) are dedicated for modernization of the existing ATM system. They aim to develop an efficient and reliable future communication infrastructure (FCI) which can offer various services ranging from data to multimedia \cite{1,2,3}. It consists of several communication data links between satellite stations, aircraft and ground terminals as shown in Fig.~\ref{fci}. Such FCI system is also used for other communications, navigation, and surveillance (CNS) applications. This work mainly focuses on the air-to-ground communication (A2GC) which is one of the important data links of the FCI system.\\

\begin{figure}[h!]
\centering
  \includegraphics[scale=0.5]{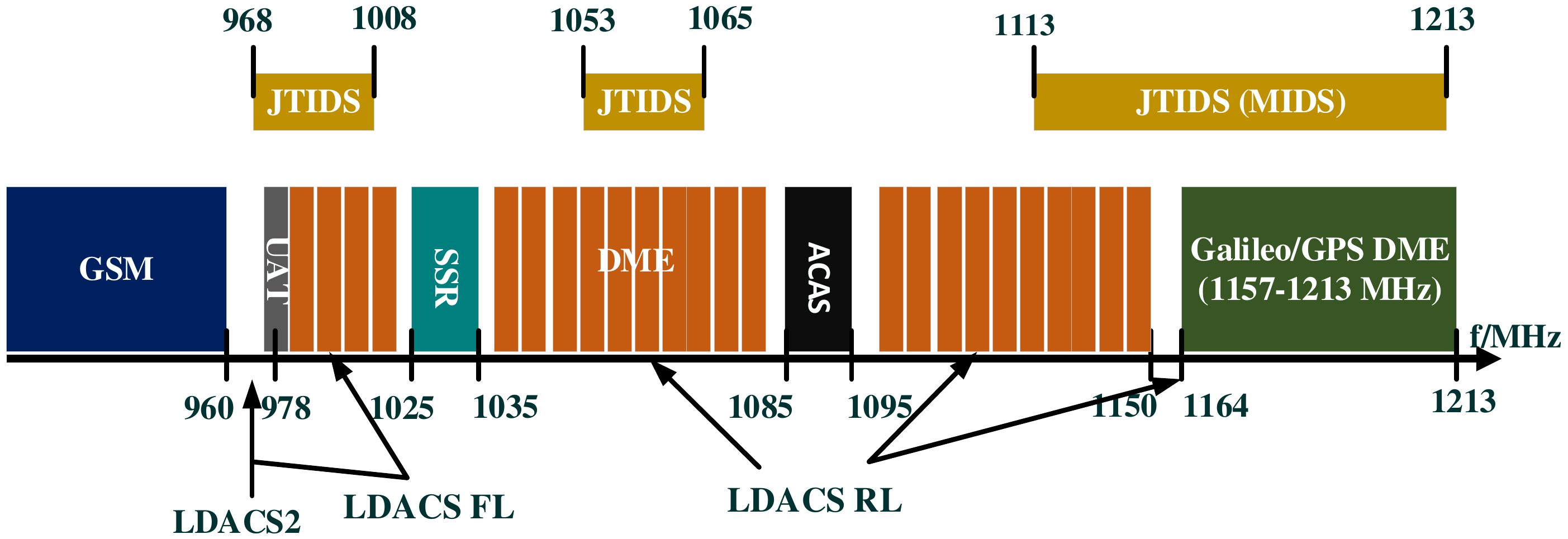}
  \caption{L-band spectrum occupancy and incumbent users.}
  \label{lband}
\end{figure}

\begin{figure}[h!]
\centering
  \includegraphics[scale=0.5]{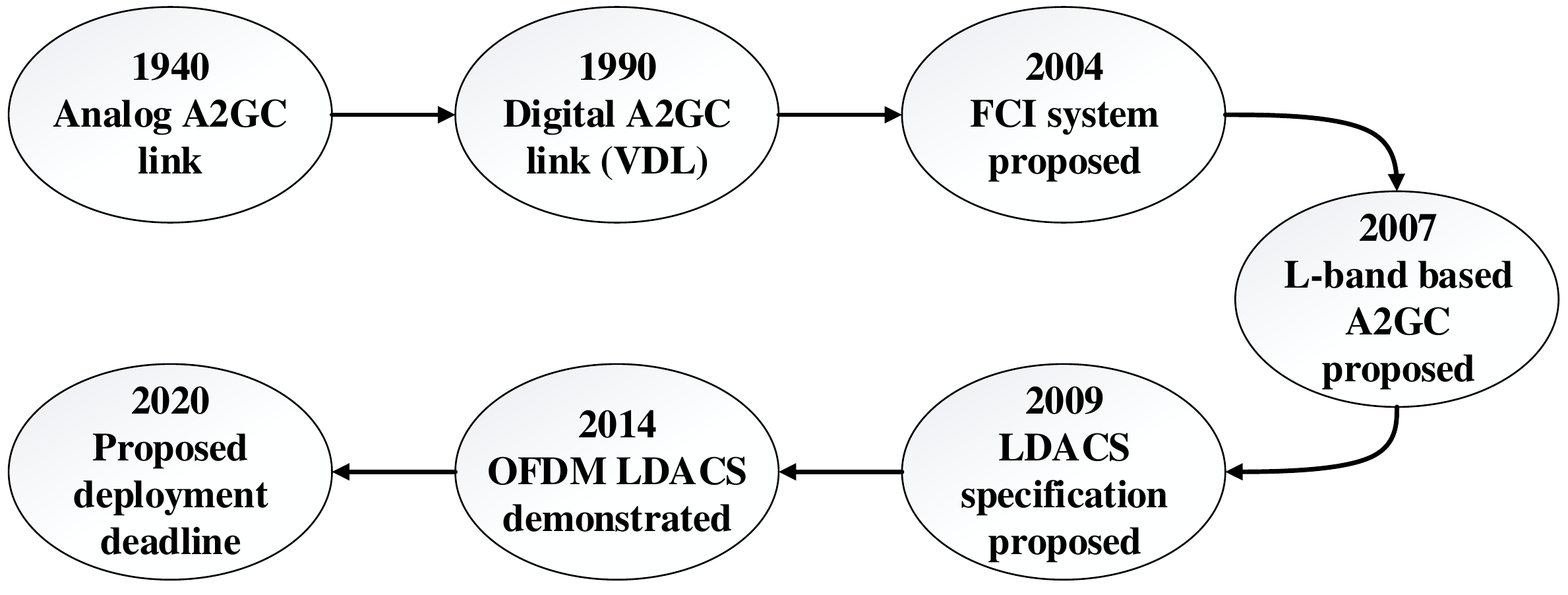}
  \caption{Brief history about the evolution of A2GC system.}
  \label{ev}
\end{figure}

The brief history of the evolution of the A2GC system is shown in Fig.~\ref{ev}. In 1940, the first A2GC link was deployed using the analog modulation based communication system. To improve the robustness and throughout, A2GC link was digitized in 1990's and deployed in the 19 MHz VHF band (118-137 MHz). Hence, it is referred to as VHF Data Link (VDL) \cite{2}. In the following years the air traffic volume has increased dramatically, which implied that the aeronautical communication systems operated in the VHF band suffered from severe congestion in some regions of the world. With an increase in the air-traffic and the need to support futuristic delay sensitive multimedia services which demand wider bandwidths, L-band (960-1164MHz) digital aeronautical communication system (LDACS) has been recently proposed. Though L-band is being used by other legacy systems such as distance measuring equipment (DME), joint tactical information distribution system, radars etc., spectrum measurement studies show that major portion of the L-band is underutilized as shown in Fig.~\ref{lband}. This lead to an inlay approach based LDACS where transceivers can exploit frequency bands between adjacent legacy signals. \\

In 2009, LDACS specifications were finalized and the first prototype was demonstrated in 2014 \cite{2,3,4,5}. Till today, there is an active research on the design of robust and low complex LDACS transceivers. \\

Existing LDACS consists of two sub-systems: 1) LDACS1: Broadband multicarrier system based on
orthogonal frequency division multiplexing (OFDM). It is similar to IEEE 802.16 standard and employs inlay approach between incumbent distance measuring equipment (DME) signals, and 2) LDACS2: Narrowband single carrier system based on time division duplex approach. It is similar to the global system for mobile communication (GSM) and uses Gaussian minimum shift keying modulation scheme \cite{6}. For the next-generation A2GC system, LDACS1 seems to be a better choice due to the capability to support high-speed delay-sensitive multimedia services and compatibility with the cellular communication standards. Hence, the work presented in this report will focus on LDACS1 and we will refer to it as LDACS hereafter. \\
 
\section{Motivation}
The existing OFDM based LDACS protocol \cite{SBr,phl,sp} has lower computational complexity, good peak-to-average-power ratio satisfactory performance in various channel conditions and compatibility with cellular systems \cite{5,6,7}. However, the disadvantages such as additional control signalling for time and frequency synchronization, spectrum inefficiency due to cyclic prefix and high out-of-band emission limits the transmission bandwidth of LDACS to approximately 498 KHz for 1 MHz vacant spectrum as shown in Fig.~ \ref{mt}. This leads to less than 50\% vacant spectrum utilization and limits the usefulness for other CNS applications. Furthermore, in multi-user network scenario where multiple aircrafts and ground terminals share L-band, OFDM is not an efficient choice due to these disadvantages. \\

\begin{figure}[h!]
\centering
  \includegraphics[scale=0.5]{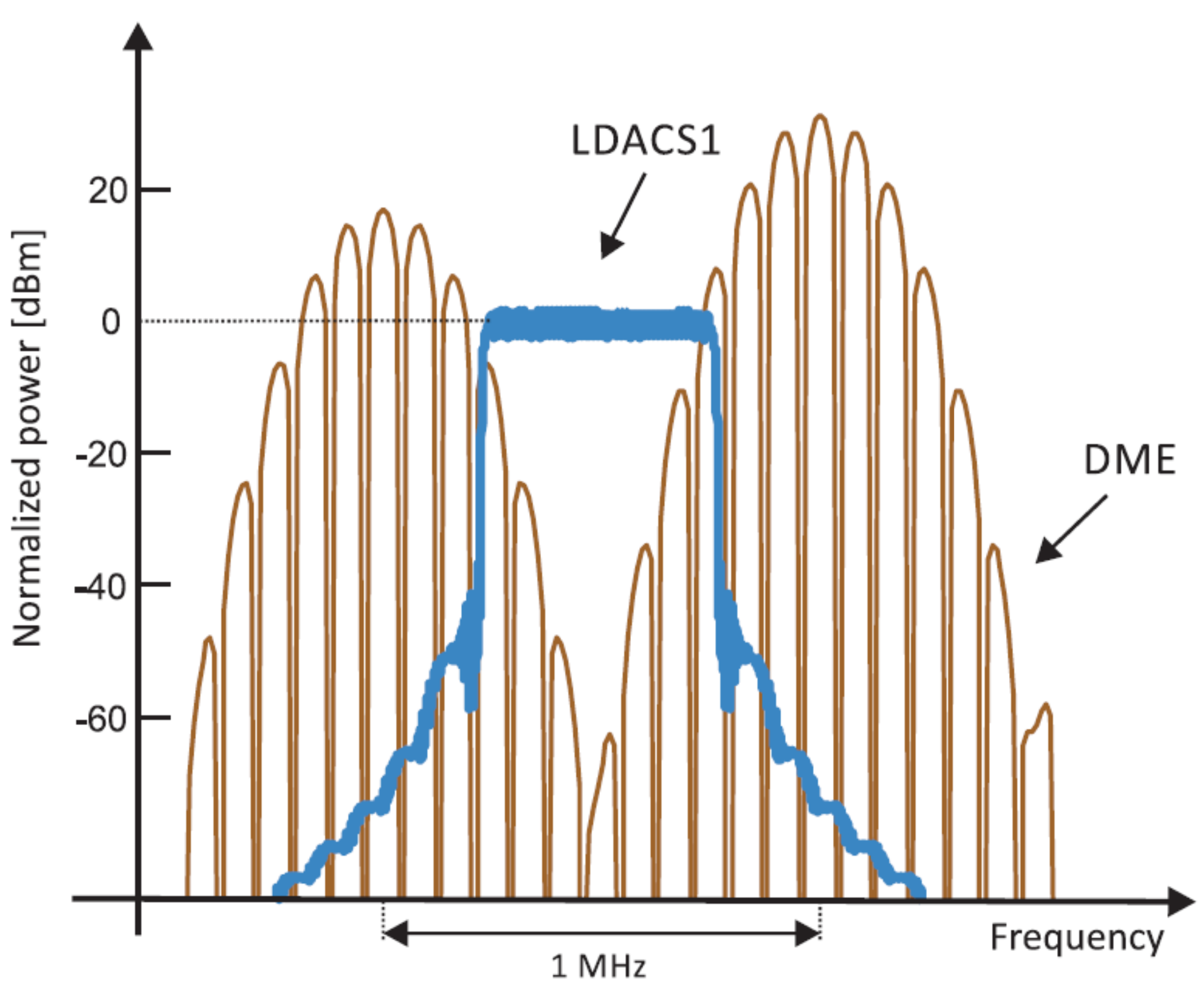}
  \caption{Brief history about the evolution of A2GC system.}
  \label{mt}
\end{figure}

Currently, to overcome the drawbacks of existing OFDM based LDACS, \cite{Hj} presents FBMC based LDACS transceiver. It offers superior performance over the existing one but from the architecture perspective, the complexity of FBMC is very high as it uses subcarrier filtering. In addition to that the receiver design of FBMC is challenging due to complex synchronization and channel equalization techniques. Furthermore, FBMC architecture is significantly different from that of OFDM which makes compatibility issues between existing and upcoming standards. \\

To solve the issues associated with FBMC based LDACS there is a need of the revised LDACS protocol along with a new waveform which can adapt wider 	transmission bandwidth as well as multiple narrowband transmissions to meet the spectrum requirements. The requirements of the new waveform are:
\renewcommand{\theenumi}{\arabic{enumi}}

\begin{enumerate}
    \item It should allow transceivers to adapt the transmission bandwidth over a wide range without compromising on the interference constraints of the incumbent L-band users.
    \item The implementation complexity of transmitter and receiver should be as low as possible.
    \item It should be compatible with existing OFDM based LDACS transceivers. Ideally, transceivers should be capable of dynamically switching between existing and new waveform on-the-fly.\\
\end{enumerate}

To adapt any of the transmission bandwidth we need to design a new frame structure of the revised LDACS protocol that supports multiple transmission bandwidths. The requirements of the frame structure are :
\begin{enumerate}
    \item It should be compatible with the existing LDACS frame structure for 498 KHz bandwidth so that there is no need to redesign other transceiver parameters.
    \item The sub-carrier spacing should be same as before as it depends on the LDACS deployment environment and hence, can not be changed.
    \item The number of symbols per frame should be fixed and independent of the transmission bandwidth.
    \item It is preferable to use identical synchronization and pilot symbol patterns as in existing LDACS protocol.\\
\end{enumerate}

\section{Research Objectives and Contribution} 
To meet an ever increasing demand of the spectrum for communication between aircrafts and ground terminals, OFDM based L-band (960-1164MHz) Digital Aeronautical Communication System (LDACS) was proposed. The drawbacks linked with the existing OFDM based LDACS transceivers leads to the need of new flexible and efficient LDACS protocol. To fulfil the above mentioned requirements of this revised LDACS protocol, we have proposed a new L-band digital aeronautical communication system (LDACS) using reconfigurable filtered OFDM (Ref-OFDM). The proposed Ref-OFDM waveform has the similar architecture as existing OFDM based LDACS fulfils the above mentioned requirements of the improved protocol.\\

The proposed protocol enables transceivers to dynamically adapt the transmission bandwidth over a wide range to meet the desired quality of service and high out-of-band attenuation leads to significant improvement in the vacant spectrum utilization. It is designed by replacing the conventional filter in F-OFDM with the reconfigurable linear phase finite impulse response filter. This filter is bandwidth reconfigurable which uses one filter to get all the desired bandwidths. This helps us to design a low complexity transmitter and receiver architecture. \\

Along with that, there is need to design and implement it on a platform like ZSoC which gives flexibility of choosing the part to be implemented on processor (PS) or FPGA (PL) for efficient and flexible prototyping of LDACS transceiver in real time. In this report the OFDM based LDACS transceiver architecture is designed for implementation on ZSoC using hardware software co-design workflow of MATLAB and Simulink. Various configurations of the architecture are realized by dividing it into two sections, one for PL and other for PS. Such co-design approach gives the flexibility to choose which part of the system to implement on PL and which on PS to meet the given area, delay and power constraints. \\

The research objectives are the following:

\begin{enumerate}
\item To propose a revised Ref-OFDM based LDACS protocol with the new frame structure:
\begin{itemize}
    \item It can accommodate different service needs with affordable computational complexity by using the same filter reconfigured to meet various transmission bandwidth constraints.
    \item It can adapt to different transmission bandwidths. It is a generalized version of existing protocol which supports only 498 KHz bandwidth.
    \item It can adapt the bandwidth on-the-fly without the need of changing the filter coefficients and can be easily extended to multi-band filter for simultaneous transmission in multiple bands.
    \item The computational complexity of Ref-OFDM is slightly higher than the OFDM but lower than other waveforms makes it an attractive solution for next generation LDACS.
\end{itemize}

\item To design and implement the various configurations of OFDM, F-OFDM and WOLA-OFDM based transceiver efficiently on Xilinx Zynq SoC along with the RF front end transmission:
\begin{itemize}
    \item Various configurations of the architecture are realized by dividing it into two sections, one for PL and other for PS. 
    \item AD9361 integration with all these configuration models validates the LDACS-DME coexistence in the real time scenario
    \item F-OFDM \cite{FO} offers better spectral containment but at the risk of exhausting the resources and higher power consumption but the utilization of the FPGA slices and LUTs for the PL is less than 27\% and 35\% respectively leaving enough resources for higher LDACS layers.\\
\end{itemize}
\end{enumerate}

The contributions of this report, which are published or under review as mentioned in ‘Publication’ section, are elaborated below:
\begin{itemize}
    \item  A low complexity reconfigurable fixed coefficient filter is proposed in [C1] to support variable bandwidth (186-732 KHz) baseband bandpass responses. The proposed filter is designed using a combination of coefficient decimation method (CDM) and Modified CDM. In addition to that A new revised frame structure is proposed to support various services that demand distinct bandwidths as the existing LDACS can support at the most 498 KHz bandwidth due to the high out-of-band emission of OFDM and does not allow multiple users to share the frequency band especially when transmitting over a narrow bandwidth. we also extended the above reconfigurable filter for the case where user simultaneously transmits in the multiple bands in [J1]. Numerically, Ref-OFDM offers around 40 dB better out-of-band emission than OFDM which in turn leads to a significant increase in the transmission bandwidth for a given BER and interference constraints. Additionally, We theoretically analyze the bit-error-rate (BER) performance of the Ref-OFDM in presence of the DME interference and LDACS wireless channels in [J1]. 
    
    \item In [C2], the OFDM based LDACS transceiver is implemented on Xilinx ZSoC consisting of programmable logic (PL) as Kintex FPGA and processing system (PS) as ARM Cortex A9. We demonstrate the flexibility offered by hardware software codesign approach to decide which part of the transceiver to implement on PL and which on PS to meet the given area, delay and power constraints. Seven configuration variants (V1-V7) of the architecture are realized by dividing it into two sections, one for PL and other for PS. In each subsequent variant one block move to the PL. We have also extended the OFDM for windowed and filtered versions of OFDM. In addition to that, we have applied  pipelining  by  introducing  delays  to  reduce  the propagation  time  and  critical  path  delay. As expected, F-OFDM offers better side lobe attenuation at the cost of high resource utilization but in all the configurations it does not exhaust the resources available on FPGA. \\
\end{itemize}

\section{Organization}
The rest of the report is organized as follows. Chapter 2 presents the research challenges in the LDACS-DME coexistence scenario. We review some of the most important work on LDACS-DME coexistence systems and candidate waveforms. The literature review on the design and implementation done on ZSoc is done in the latter part of chapter 2. In chapter 3, the LDACS-DME coexistence scenario along with the proposed LDACS protocol and Ref-OFDM is described. The reconfigurable filter design is also explained. The hardware implementation of transceivers on ZSoC is described in chapter 4.  Finally, Chapter 5 concludes the work presented in this report and possible future directions are mentioned. 
	
\chapter{Literature Review}
In this chapter, we will review the various works which analyze the performance of existing LDACS and contribute to improving its feasibility, robustness, and complexity. In addition to that the detailed literature review of the candidate waveforms followed by the ZSoC implementation is presented. \\

\section{LDACS-DME coexistence}
The LDACS system can be deployed in the L-band using two approaches: overlay and inlay approach. In overlay approach, it is deployed in the vacant spectrum where no other legacy system is present. This approach is easy and chosen for GSM like LDACS2 in 960-975 MHz vacant band. On the other hand, overlay approach is not preferable for OFDM based LDACS due to limited vacant spectrum in the L-band. Hence, LDACS is deployed using an inlay approach exploiting the multiple 1 MHz frequency bands between DME signals. \\

The specification of OFDM based LDACS physical layer is presented in \cite{SBr,phl,sp}. In \cite{SBr}, LDACS system performance has been analyzed for both inlay and overlay approach. Apart from the transmitter, the design of the L-DACS1 receiver is introduced using methods for mitigating interference from other L-band systems. The authors have applied additional methods such as pilot boosting and pilot erasure, leading to a system performance close to that achievable with perfect channel knowledge. The results in \cite{SBr} shows the compatibility of LDACS with the DME signal in the critical inlay approach.\\

In \cite{Dirk}, a new model has been proposed to evaluate and compare the performance of various FCI links as per the Required Communication Performance (RCP) metric introduced by the ICAO. Based on the analysis, they have also identified the desired characteristics for any data link in order to meet RCP requirements. The theoretical results presented in \cite{Ue} also confirm the feasibility of the inlay approach in $L$-band. The authors also analyzed the effect of the DME power and pulse rate on the BER of inlay approach based LDACS. Yhe simulation and experimental results are not presented in the paper. Also, the analysis is limited to the DME interference specific to European aerospace. The result obtained during compatibility measurements of LDACS carried out at labs of the German Air navigation Service Provider, DFS Deutsche Flugsicherung GmbH are presented in \cite{lc}. The authors have taken many possible interference scenarios for the inlay deployment and done experiments on their measurement setup. The experimental results presented show that LDACS co-existence with DME in the most strict inlay deployment option is feasible and also claimed that with the additional DUTs the compatibility should be proved also for other interference scenarios. \\

In \cite{UM,Si}, interference analysis is done via characterizing various incumbent users in $L$-band in terms of their spectral characteristics, transmit power and the duty cycle. The simulation results in \cite{UM} show slight degradation in the BER performance of LDACS for higher DME power. To mitigate this, two algorithms have been proposed for detecting the DME interference in \cite{Si,EK}. In \cite{Si}, the compensation of the impact of pulse blanking is proposed by reconstructing and subtracting ICI. The required shape of the subcarrier spectra is derived from the pulse blanking window. Supervised learning-based DME multipath mitigation technique, its performance, and sensitivity analysis is presented in \cite{EK}. Authors have also proposed an iterative receiver design for estimation of the transmitted data symbols and the channel coefficients of each subcarrier. The simulation results shows the impact of pulse blanking is reduced to an SNR loss resulting from erasing a certain fraction of the OFDM signal. \\

The works in \cite{ms,NF} validate the performance of LDACS in the presence of incumbent $L$-band users on the hardware testbed at intermediate frequency level. These tests validate the fixed point implementation of LDACS. Experiments in real radio environments are being carried out to analyze the effect of non-linearities in the power amplifier and analog-front-end on the performance of LDACS.\\

In \cite{Sr}, a novel sensing method for detecting the active LDACS transmissions via multiplier-less correlation-based method has been proposed. Results show that the proposed approach offers improved performance especially at low signal-to-noise ratio (SNR) and consumes much lower power than other architectures.\\

On the receiver side, reconfigurable low complexity filter and filter bank architectures for channelization and spectrum sensing applications have been proposed in \cite{Ambede,SD}. Such architectures are based on frequency response masking approach and they allow the LDACS receiver to receive and/or sense single as well as multiple frequency bands simultaneously. \\

Most of the existing works deal with the improving the performance of the OFDM based LDACS system. OFDM has high out-of-band emission which leads to less than 50\% vacant spectrum utilization. To design FCI system for various CNS application, it must offer large transmission capacity, low latency and high elasticity, along with the capability to support a wide variety of services. For this OFDM alone may not be sufficient and hence, transceivers which can support multiple waveforms need to be developed. A brief review of the candidate waveforms for an efficient transceivers is presented in the next section.\\

\section{Candidate Waveforms}
At present, Orthogonal Frequency Division Multiplexing (OFDM) is a widely adopted solution because of its robustness against multipath channels and Fast Fourier Transform (FFT) based  easy implementation. But the increased air traffic gives the challenges which OFDM can not address in future. Here, we will give a brief review about the candidate waveforms which can replace OFDM to further improve the transceiver system according to today's requirements. Various work has been done on the improvisation of these waveforms in terms of complexity, PSD, BER etc. \\

The replacement of OFDM based LDACS transceivers using F-OFDM and FBMC respectively are presented in\cite{Lu,Hj}. In \cite{Lu}, the broadband aeronautical communication with the wareform designed based on L-DACS1 and FOFDM is proposed and evaluated. The filters are elaborately designed to handle the requirements by taking into account the properties of the aeronautical communication. NC-OFDM is also adopted in each subband to integrate the non-continuous spectrum resources. The proposed system offers good spectrum efficiency with the same BLER performance. In \cite{Hj}. the FBMC waveform has been proposed as an alternative to OFDM in LDACS. The FBMC offers higher vacant spectrum utilization than OFDM due to sub-carrier filtering approach but the work in \cite{Hj} assumes fixed transmission bandwidth. 

In \cite{gf,gf1,uf}, Generalized Frequency Division Multiplexing (GFDM) and Universal Filtered Multicarrier (UFMC) has been proposed to overcome the drawbacks of OFDM respectively. A flexible multicarrier modulation scheme, named GFDM, has been proposed in \cite{gf,gf1} for the air interface of 5G networks. GFDM is based on the modulation of independent blocks, where each block consist of a number of subcarriers and subsymbols. The subcarriers are filtered with a prototype filter that is circularly shifted in time and frequency domain. \cite{uf} presents the UFMC where a group of subcarrier is filtered to reduce the OOB emission. UFMC does not require a CP and it is possible to design the filters to obtain a total block length equivalent to the CP-OFDM. However, because there is no CP, UFMC is more sensitive to small time misalignment than CP-OFDM and F-OFDM.\\

The waveforms presented in \cite{Lu, Hj, gf, uf}, are efficient in their own terms but are not suitable for LDACS transceivers. The F-OFDM and FBMC based LDACS in \cite{Lu,Hj} has the fixed transmission bandwidth and can not adapt the dynamic transmission bandwidth on the fly. Moreover, for each bandwidth both will require a different filter which increases the complexity very high. In addition to that the architectures of FBMC \cite{Hj} and GFDM \cite{gf} are significantly different from that of OFDM, the single transceiver cannot support both waveforms unless they are stacked in parallel. From the future perspective, it cannot be easily extended to multiple antenna configurations which is now a default configuration offering high data rates and better performance in deep fading. Time misalignment of UFMC discussed in \cite{uf} makes it inappropriate for A2GC. \\

To validate the performance in real time, the transceiver needs to be implemented on hardware. The various works done using the new age ZSoC platform to implement the communication transceivers is presented in the next section.\\

\section{Hardware Implementation}
The increased demand of spectrum has led to introduction of various new standards and protocols. For these systems, new reconfigurable transceivers are needed with the quality of spectrum sensing
to sense the vacant spectrum between adjacent DME signals. Such transceivers can be implemented on Zynq system on chip (ZSoC). It is a heterogeneous system which provides decision making capabilities and flexibility of choosing the part to be implemented on processor (PS) or FPGA (PL) for efficient and flexible prototyping of LDACS transceiver in real time.\\

In this section, we will review the numerous works done using Xilinx ZSoc platform that combines FPGA (PL) and ARM Cortex A9 (PS) processor. We will also review the work done to analyze the performance of existing LDACS. \\

In \cite{Jothan},authors have implemented a cognitive radio framework called Iris on Zynq SoC. Iris uses XML description to form a complete radio by linking the components together and run them on PS and PL. Video transmission using OFDM has been performed to evaluate the setup. Methods of profiling the software and accelerating critical functions are examined in the PL. A Zynq capable version of GNU Radio is presented in \cite{ryan}. They have also demonstrated the feasibility and usability of FPGA based SDR integrated with GNU Radio in GReas. In \cite{Baris}, Digital pre-distortion (DPD) required by 3G/4G base stations is implemented on Xilinx’s Zynq All Programmable SoC. DPD is an advanced digital signal-processing technique that mitigates the effects of power amplifier (PA) nonlinearity in wireless transmitters. The flexible design flow of ZSoC facilitates the generation of effective DPD solutions for modern wideband and multi-antenna transmitters. Using this design flow, a complete DPD feedback path on the Zynq SoC achieves up to 7x speed-up from hardware acceleration. \\

The authors in \cite{rold} describe and analyze the Zynq7000 SoC from the perspective of an evolvable hardware. In FPGA-based evolvable hardware the evolutionary algorithm (EA) generates candidate configurations that are used to configure chosen reconfigurable blocks of the FPGA. Their experiments confirmed the superiority  of the platform for evolvable hardware design in context of area overhead, execution time, reconfiguration time and throughput.  A Cognitive Radio Accelerated with Software and Hardware (CRASH) is introduce in \cite{john}. CRASH is a versatile heterogeneous computing framework for the Xilinx Zynq SoC. They have implemented spectrum sensing and the spectrum decision in three configurations : 1. both algorithms in the FPGA, 2. both in software only, and the last is spectrum sensing on the FPGA and spectrum decision on the CPU. Their experiments show that CRASH can successfully segment two cognitive radio algorithms, spectrum sensing and the spectrum decision, between the Zynq’s FPGA fabric and ARM processors. \\

Recently, few works dealing with the design and implementation of OFDM-based LDACS transceiver has been proposed in \cite{Thinh,TH1,SS,Ambede,SD}. A hardware architecture of a novel synchronization method for LDACS in \cite{Thinh}. This method is designed to achieve synchronization accuracy and to be much robust to large CFO than the SoA method. Their implementation results show that the proposed synchronizer has a reduced hardware usage and very low dynamic power consumption. In \cite{SS}, the design of LDACS transceiver via partial reconfiguration approach of the FPGA has been proposed. It offers significant improvement in power consumption of the transceiver without compromising on the performance. Authors in \cite{Ambede,SD} introduces reconfigurable low complexity filter and filter bank architectures for  channelization and spectrum sensing applications on the receiver side. These architectures are based on frequency response masking approach and allow the LDACS receiver to receive and/or sense single as well as multiple frequency bands simultaneously. \\

For understanding of the research done previously, the literature review done above is summarized in the next section. \\

\section{Summary}
The OFDM based LDACS coexists with the legacy DME signals by deploying it within the 1 MHz spectrum gap between two adjacent DME channels. The OFDM has several drawbacks such as additional control signalling for synchronization, high out-of-band emission and use of cyclic prefix leads to less than 50\% vacant spectrum utilization. Various works to improve the performance of the OFDM based LDACS system are reviewed in the first section of the chapter. According to the literature review, still there is a need of transceivers which can support a wide variety of services along with the large transmission bandwidth.\\

A brief review about the candidate waveforms is given in the second section of the chapter. In which we have reviewed the candidate waveforms such as F-OFDM \cite{Lu}, FBMC \cite{Hj}, GFDM \cite{gf} and UFMC \cite{uf} which can replace OFDM to further improve the transceiver system according to today's requirements. The implementation complexity of FBMC, GFDM and UFMC is very high. Additionally, the architectural difference of these waveforms with the existing OFDM does not provide the backward compatibility. While the F-OFDM has the backward compatibility but multiple filters for different transmission bandwidths increases the computational complexity of the system. For the real time hardware implementation, the various works done using the new age ZSoC platform to implement the communication transceivers is presented in the later part of the chapter. \\

ZSoC is a heterogeneous system which provides decision making capabilities and flexibility to choose the part to be implemented on processor (PS) or FPGA (PL) for efficient and flexible prototyping of LDACS transceiver in real time. Few works dealing with the design and hardware implementation of OFDM-based LDACS transceiver are presented in \cite{Thinh,SS,Ambede,SD}.\\

In the next chapter we will discuss about the proposed work in this report. \\

\chapter{Proposed work : Revised LDACS protocol for LDACS-DME coexistence}
Various work done for improving the LDACS transceivers for an efficient and reliable LDACS-DME coexistence are reviewed in the last chapter. The existing LDACS protocol has fixed transmission band-width of 498 KHz due to the high out-of-band emission of OFDM and hence, cannot adapt the transmission bandwidth as per the desired service. This leads to less than 50\% vacant spectrum utilization and limits the usefulness for other CNS applications. We have proposed a new revised protocol and reconfigurable waveform enabling LDACS transceivers to dynamically adapt the transmission bandwidth over a wide range to meet the desired quality of service. Also, high out-of-band attenuation allows wider transmission bandwidth as well as multiple narrowband transmissions leading to significant improvement in the vacant spectrum utilization. \\

The proposed LDACS protocol frame structure can adapt to different bandwidths (186-732 KHz). It is a generalized version of existing protocol which supports only 498 KHz bandwidth. In addition to that, the proposed reconfigurable filtered OFDM (Ref-OFDM) uses a reconfigurable linear phase multi-band finite impulse response (FIR) digital filter. It can adapt the bandwidth on-the-fly without the need of changing the filter coefficients and can be easily extended to multi-band filter for simultaneous transmission in multiple bands. Here, we have discussed the proposed Ref-OFDM based LDACS transceiver along with the revised protocol in detail.\\

\section{LDACS Deployment Environment}
In this section, we discuss LDACS deployment environment which is useful for better understanding of the proposed work.\\

\subsection{L-Band for A2GC}
The spectrum occupancy of L-band is shown in Fig. \ref{lband}. Various legacy or incumbent users in L-band are DME signals (960-1215 MHz), radar-based multi-functional information distribution system (MIDS), universal access transceiver (UAT) systems (978 MHz), secondary surveillance radar (SSR) (1030 MHz) and airborne collision avoidance system (ACAS) (1090 MHz) \cite{4}. The LDACS system can be deployed in the L band using two approaches: overlay and inlay approach. In overlay approach, it is deployed in the vacant spectrum where no other legacy system is present. This approach is easy and chosen for GSM like LDACS2 in 960-975 MHz vacant band. On the other hand, overlay approach is not suitable for OFDM based LDACS due to limited vacant spectrum in the L-band. Hence, an inlay approach is envisioned exploiting the multiple 1 MHz frequency bands between DME signals. \\

\begin{figure}[!h]
 	\vspace{-0.3cm}
 	\centering
 	\subfloat[]{\includegraphics[scale=0.4]{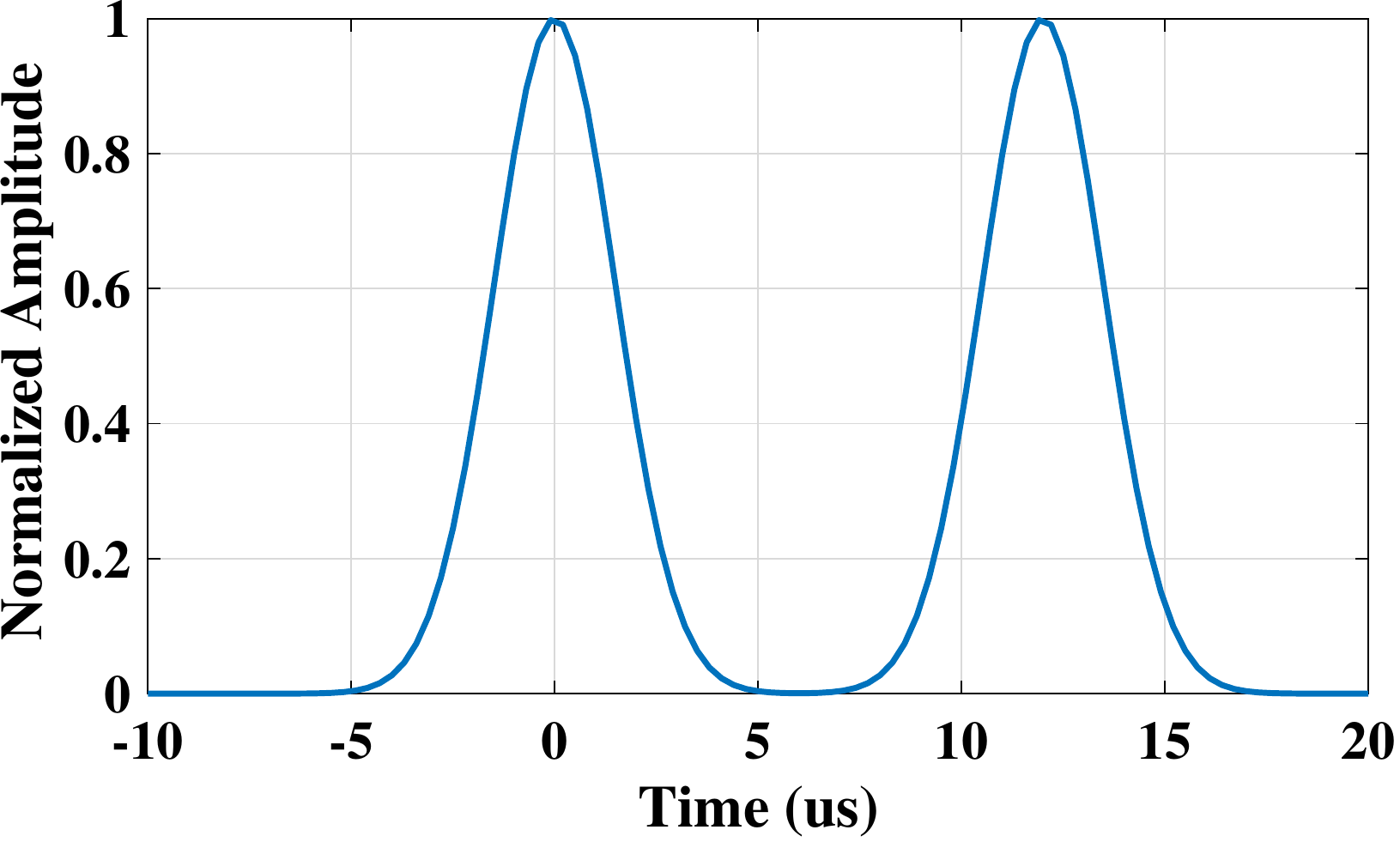}%
 		\label{v}}
 		\hspace{0.8cm}
 	\subfloat[]{\includegraphics[scale=0.4]{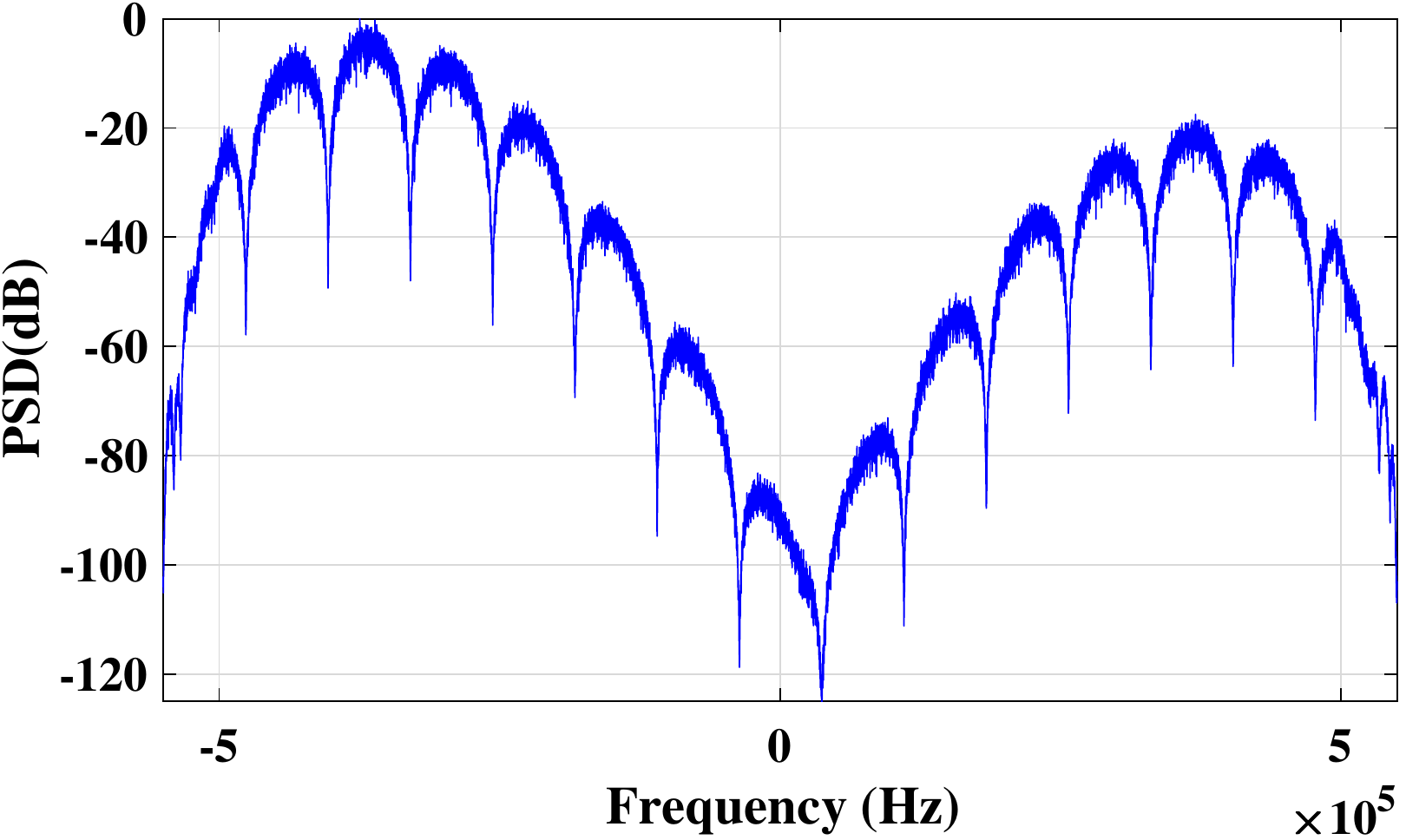}%
 		\label{vr4}}
 	\vspace{-0.2cm}
 	\caption{Legacy L-Band DME signal (a) Time domain response (b) Power spectral density.}
 	\label{dme}
 \end{figure}

The DME is a transponder-based navigation system used to measure the slant range distance \cite{Ue}. It is composed of Gaussian shaped pulse pairs. The time and frequency domain representations of the DME signal, shown in Fig.~\ref{dme} (a) and (b) respectively, are represented as \cite{Ue,Hj}.
 \begin{equation}
 S (t) = e^{\frac{-\alpha t^{2}}{2}} + e^{\frac{-\alpha (t-\Delta t)^{2}}{2}} 
 \label{1}
 \end{equation}
 \begin{equation}
 S (f) = A \sqrt{{\frac{8\pi}{\alpha}}} e^{\frac{-2\pi^{2}f^{2}}{\alpha}} e^{j\pi f\Delta t}cos(\pi f \Delta t)
\label{2}
 \end{equation}

where, $\alpha$ is pulse width of $4.5 * 10^{-11}$ $s^{-2}$, $\Delta t$ represents the spacing of the pulses (=12 $\mu s$) and A is constant depending on the power of DME signal.
  \\
Based on the existing spectrum allocation and occupancy measurements conducted worldwide, the identified spectrum for LDACS deployment are: 1) Forward link: 985.5-1008.5 MHz and 2) Reverse link: 1048.5-1071.5 MHz. In the future, the frequency band from 1104 MHz-1150 MHz might also be available for LDACS. Due to stringent interference specifications of DME signals, the design of reconfigurable transceiver for improving the spectrum utilization of existing LDACS is a challenging task and aim of the proposed work.\\

\subsection{Wireless channels}
For LDACS environment, three channels are modeled: Airport (APT), Terminal Maneuvering Area (TMA), En-routing (ENR) as shown in the Fig. ~\ref{chan}. They are modeled as wide sense stationary uncorrelated scattering channels and characterized using three properties: fading, delay paths, and Doppler frequency \cite{Eu}.\\

\begin{figure}[h!]
	\centering
	\includegraphics[scale=0.5]{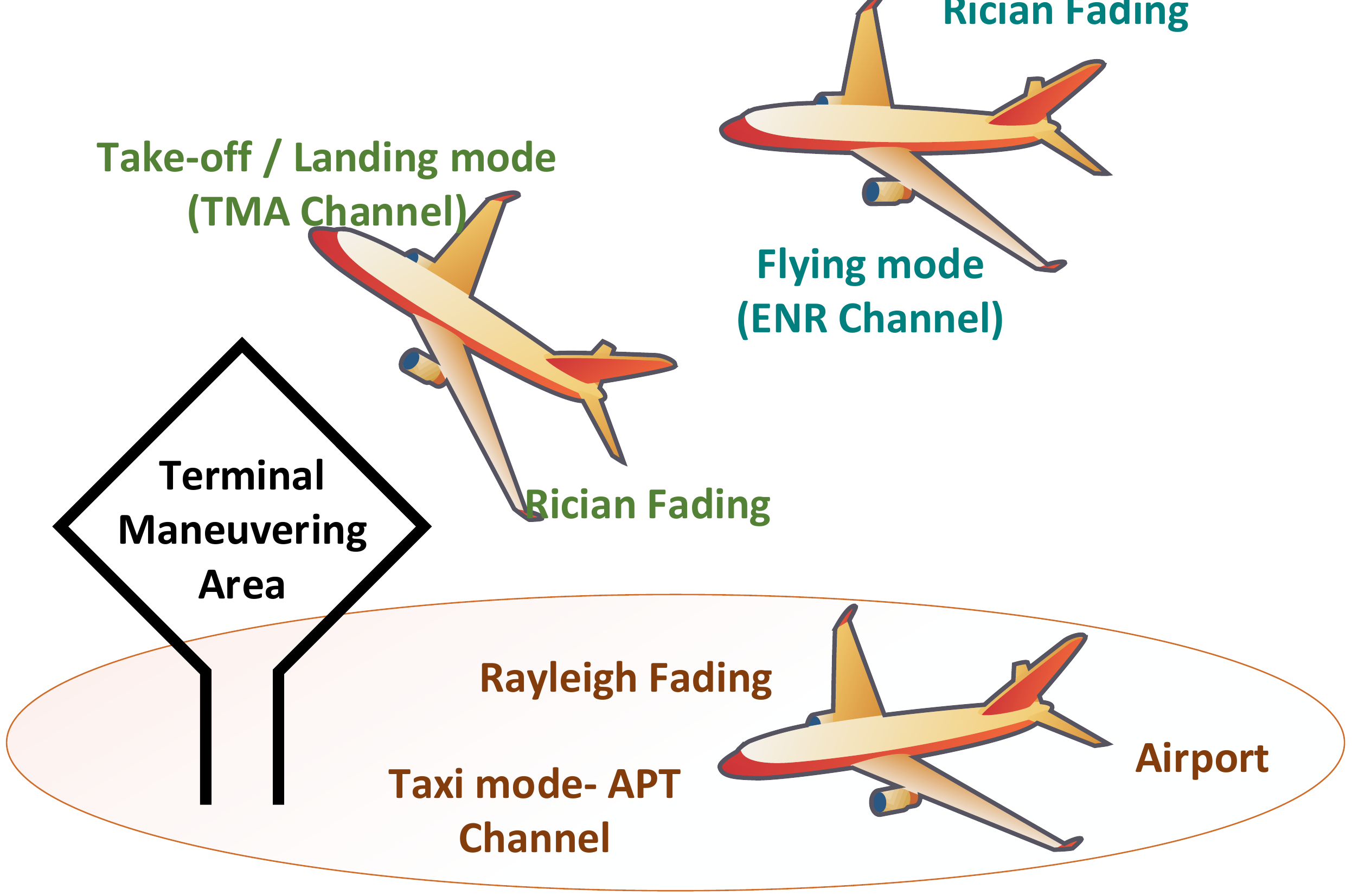}
	\caption{Channel models for LDACS environment}
	\label{mt}
\end{figure}

The A2GC link encounters APT channel during the arrival and departure phases of the aircraft i.e. when the aircraft is on the ground and moving slowly. In this case, the line-of-sight (LOS) path is assumed to be blocked resulting in Rayleigh fading with -100 dB factor \cite{Eu,Bv}. Due to low Doppler frequencies, the Doppler spectrum is Jakes distributed.\\

The TMA channel is modeled for the communication during the landing/take-off of the aircraft. Since the aircraft is at a low height, strongly scattered paths are observed in addition to LOS path. It experiences Rician fading with 10 dB Rician K factor which is the ratio of LOS component power to the power of scattered paths. For the worst case scenario, when the scatters from buildings are uniformly distributed, the Doppler spectrum follows Jakes distribution \cite{Eu,Bv}.\\

The ENR Channel is modeled for the communication during the flying mode. This channel typically consists of a strong path as well as other reflected and delayed paths. Therefore, the fading for this model is considered as Rician fading with the Rician k factor of 15 dB \cite{Eu,Bv,Bv1}. It is higher than the TMA channel due to strong LOS path. The Doppler power spectrum of the reflected path follows a Gaussian distribution.\\

The channel parameters are given in Table ~\ref{ch} \cite{Eu,Bv,Bv1,Ah}. Note that the Doppler frequency is obtained as $F_D=F_c \frac{v}{c}$. Here, the $F_c$ is the carrier frequency and is at most 1215 MHz, $v$ is the velocity of the aircraft is $m/s$ (1 Knots True Airspeed (KTAS)= 0.5144 $m/s$) and $c=3*10^8 m/s$.

\begin{table}[!h]
	\centering
	\caption{Channel Parameters}
	\label{ch}
	\resizebox{\linewidth}{!}{
		\begin{tabular}{ |p{2cm} | p{1.5cm}| p{2.5cm} |p{2cm} |p{2cm} |p{3cm} |}
			\hline
			\textbf{scenario}  & \textbf{Max Delay ($\mu$s)} & \textbf{Acceleration ($(m/s^2)$)} & \textbf{Harmonics} & \textbf{Velocity ((KTAS))} & \textbf{Doppler Frequency (Hz)} \\
			\hline 
			\rule{0pt}{10pt}
			APT &  3 & 5 & 8 & 200 & $(1215e6)\frac{200*.5144}{3e8}$ = 413 \\
			\hline
			\rule{0pt}{10pt}
			TMA &  20 & 50 & 8 & 300 & $(1215e6)\frac{300*.5144}{3e8}$ = 624 \\
			\hline
			\rule{0pt}{10pt}
			ENR  & 15 & 50 & 25 & 600 & $(1215e6)\frac{600*.5144}{3e8}$ = 1250\\
			
			\hline
		\end{tabular}
	}
\end{table}

\section{Revised LDACS Protocol: Frame Structure}
\begin{figure}[!b]
	\centering
	\includegraphics[scale=0.5]{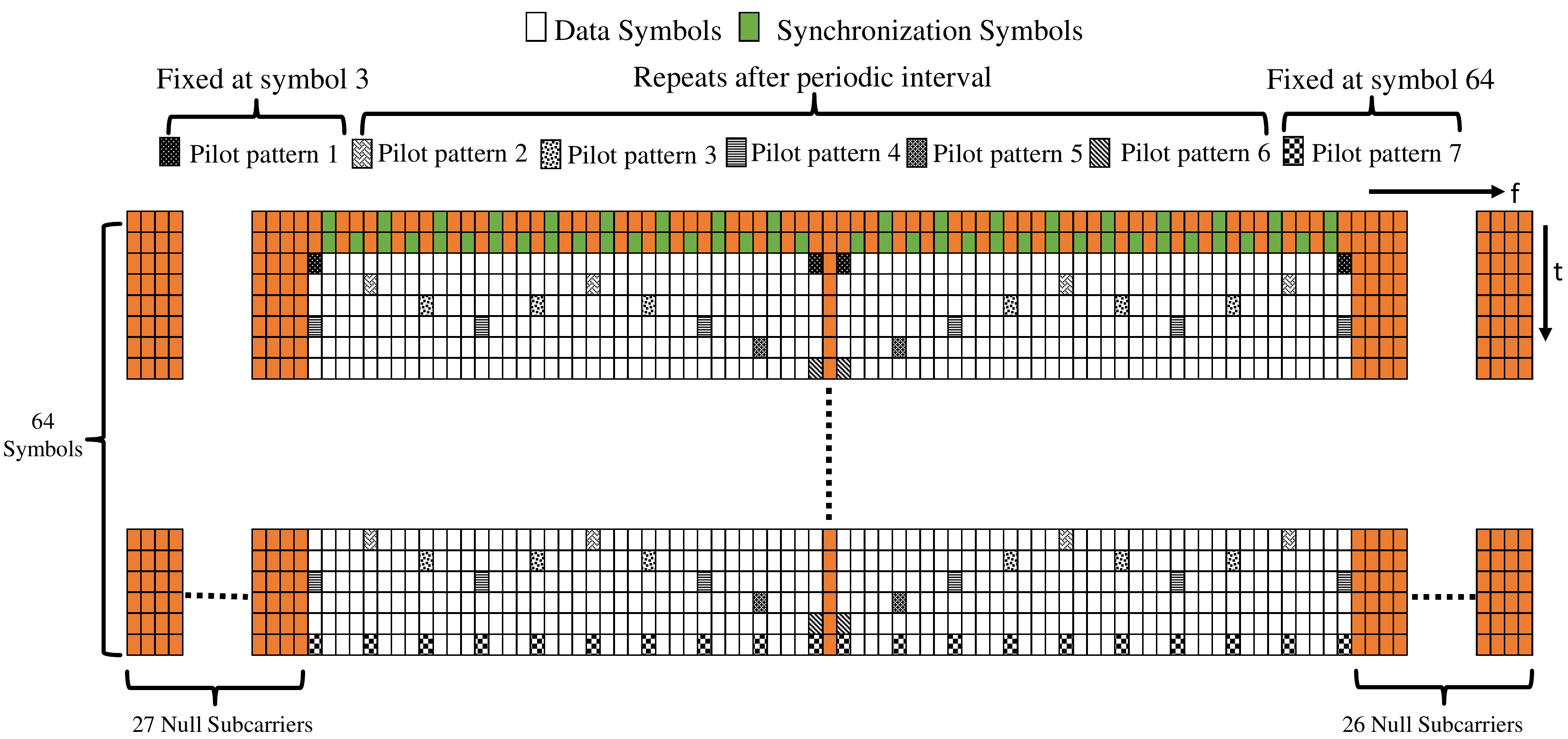}
	\caption{Proposed frame structure for the revised LDACS protocol depicting the data, pilot and synchronization symbol patterns and locations for 732 KHz transmission bandwidth.}
	\label{frame732}
\end{figure}

In this section, we present the frame structure of the revised LDACS protocol that supports multiple transmission bandwidths.Based on an experimental study of channel conditions between aircraft and ground terminals at different phases of the flight, the sub-carrier bandwidth is limited to 9.76 KHz and hence, the symbol duration is 120 $\mu$s \cite{5,SBr,phl,sp}. For these specifications and to support different bandwidths ranging from 100 KHz to 1 MHz, the FFT size in the proposed protocol is fixed and increased to 128 compared to 64 in the existing protocol. The proposed frame structure for the revised LDACS protocol depicting the data, pilot, and synchronization symbol patterns and their locations for 732 KHz transmission bandwidth are shown in Fig.~\ref{frame732}.\\

\begin{figure}[!t]
	\centering
	\subfloat[]{\includegraphics[scale=0.265]{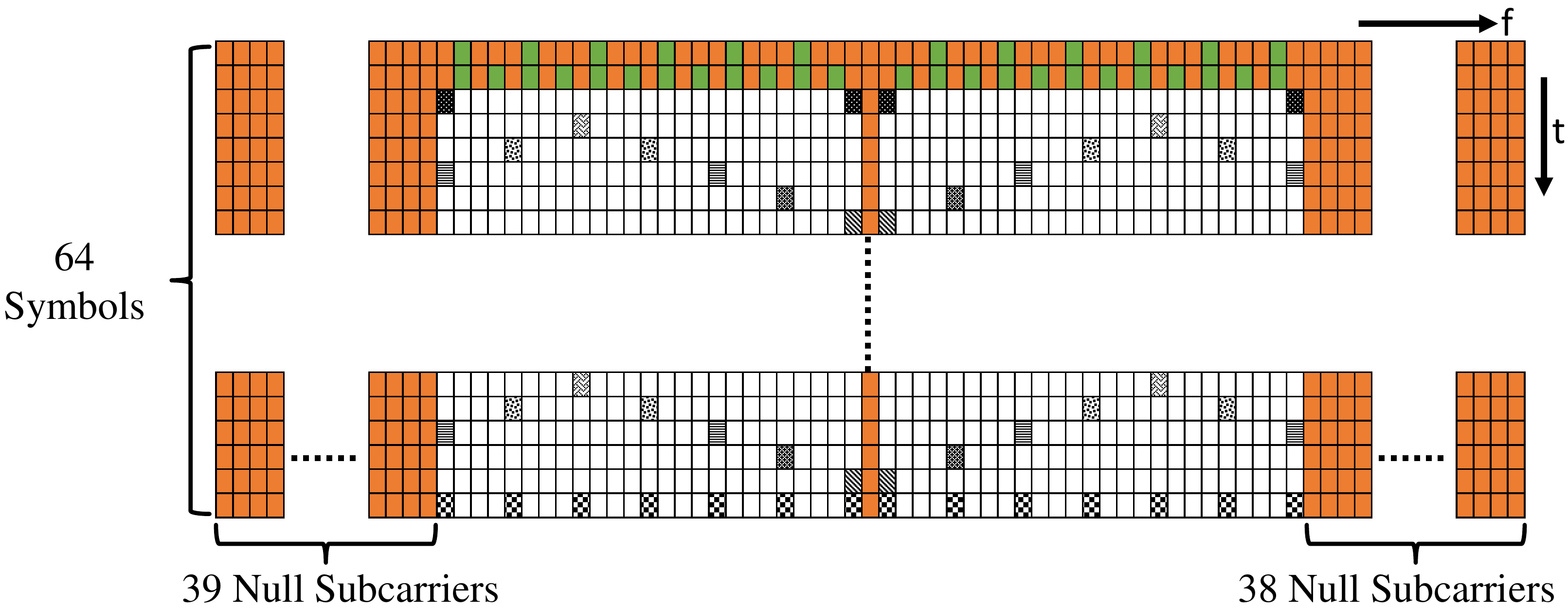}%
		\label{frame498}}
	\subfloat[]{\includegraphics[scale=0.265]{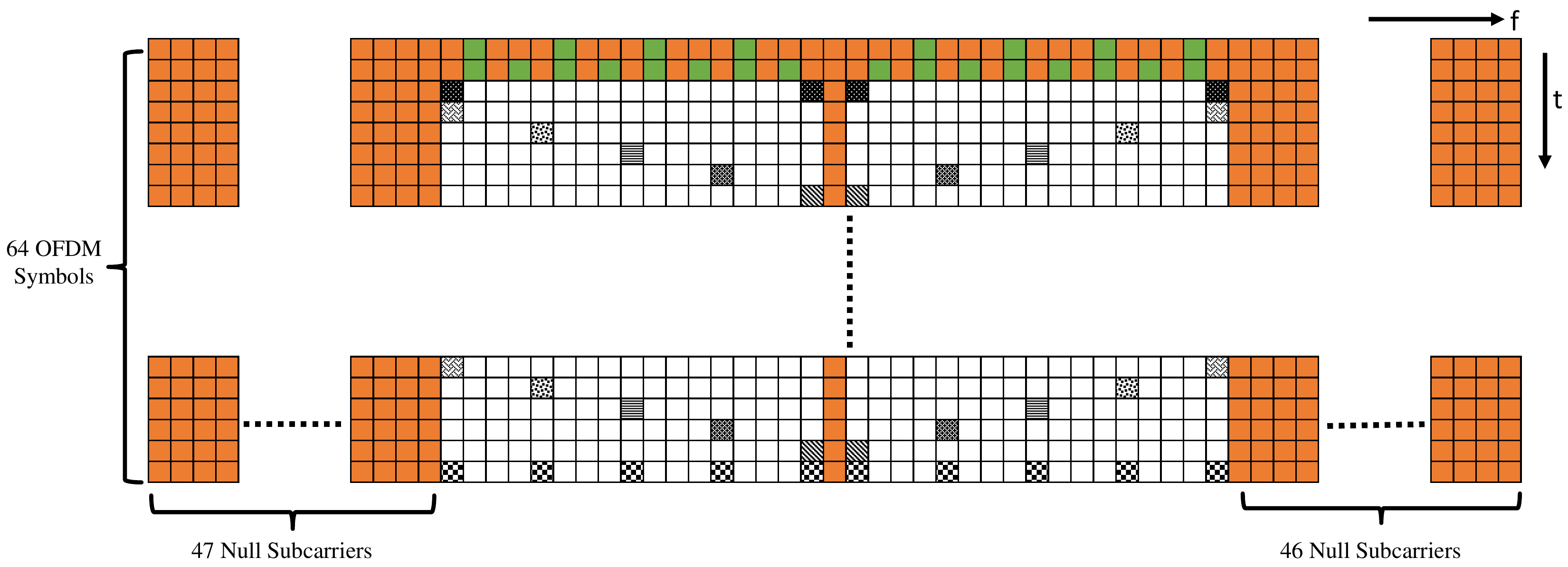}%
		\label{frame342}}
\hspace{0.2cm}
	\subfloat[]{\includegraphics[scale=0.265]{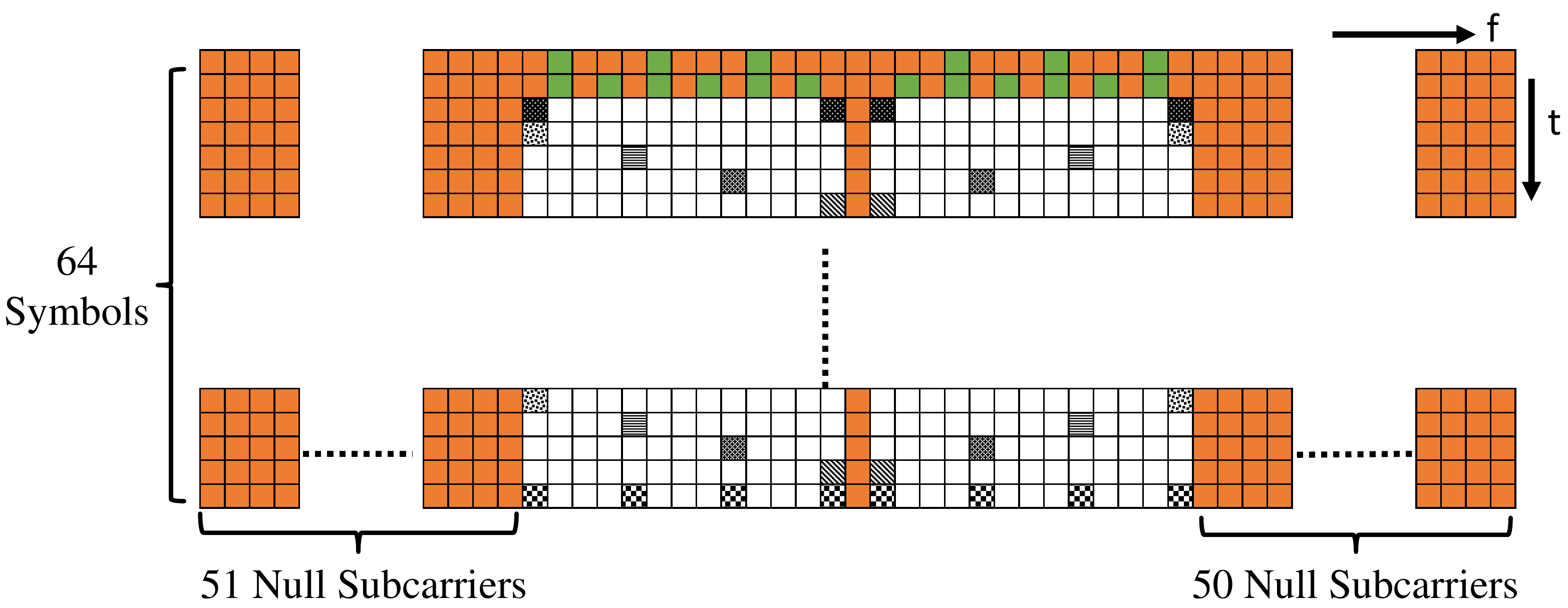}%
		\label{frame264}}
	\subfloat[]{\includegraphics[scale=0.28]{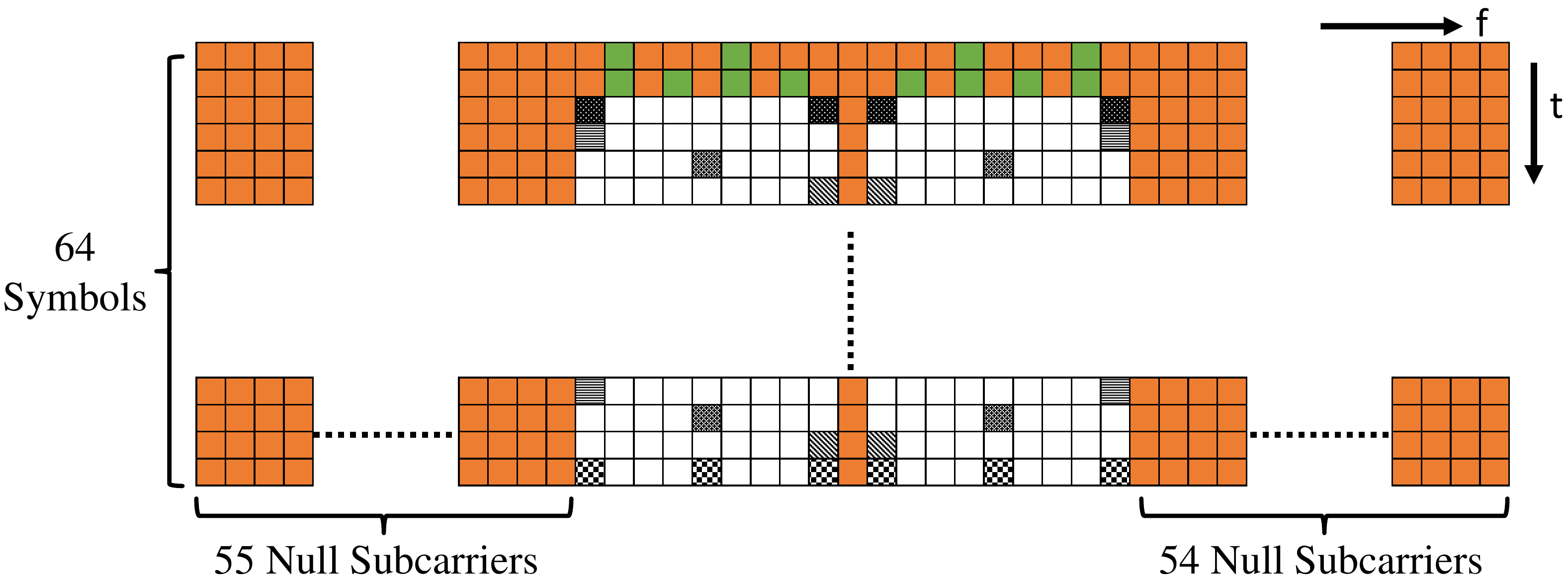}%
		\label{frame186}}
	
	\caption{Proposed frame structure for the revised LDACS protocol depicting the data, pilot and synchronization symbol patterns and locations for the transmission bandwidth of (a) 498 KHz, and (b) 342 KHz (c) 264 KHz (d) 186KHz.}
	\label{frame}
\end{figure}

The frame consists of 128 sub-carriers with the middle sub-carrier being the DC null sub-carrier. The first two symbols of each sub-carrier are reserved for synchronization. The frame consists of at the most 7 different pilot patterns which are critical for accurate channel estimation and equalization at the receiver. Similar to the existing protocol, we use fixed pilot patterns, pattern 1 ($P1$) and pattern 7 ($P7$), at the third and last symbols, respectively. Since the pilot symbols in $P7$ are separated by four sub-carriers on either side of the DC sub-carrier and the frequency resolution between adjacent sub-carrier is 9.76 KHz, the bandwidth can be incremented by 78 KHz only. Based on empirical observations, the bandwidth above 732 KHz is not feasible due to high interference to the DME signal. Also, the bandwidth below 186 KHz may not be suitable for the multi-carrier waveform. Thus, the proposed frame structure supports eight discrete bandwidths which are 732 KHz, 654 KHz, 576 KHz, 498 KHz, 420 KHz, 342 KHz, 264 KHz and 186 KHz. For these bandwidths, the number of symbols should be fixed and multiple of the number of repeating pilot patterns, $P2-P6$. For instance, for 732 KHz-342 KHz, all five patterns ($P2-P6$) are used while for the 264 KHz and 186 KHz bandwidth, the patterns used are $P2-P5$ and $P2-P4$, respectively. Hence, the number of symbols per frame are fixed to 64 out of which 2 are synchronization symbols and 2 are pilot patterns, $P1$, and $P7$. The number of null-sub-carriers on each side depends on the transmission bandwidth as shown in Fig.~\ref{frame}.\\

Note that the frame structure can be extended to other bandwidths by designing new pilot patterns and ensuring sufficient number of pilots for accurate channel estimation and equalization in LDACS deployment environment. Since the redesign of pilot patterns needs modifications of LDACS transceiver specifications, subsequent analysis and time-consuming experiments in the real radio environment, we leave it as part of future works.\\

\section{Ref-OFDM waveform based transceiver architecture}
\begin{figure}[h!]
	\centering
	\subfloat[]{\includegraphics[scale=0.3]{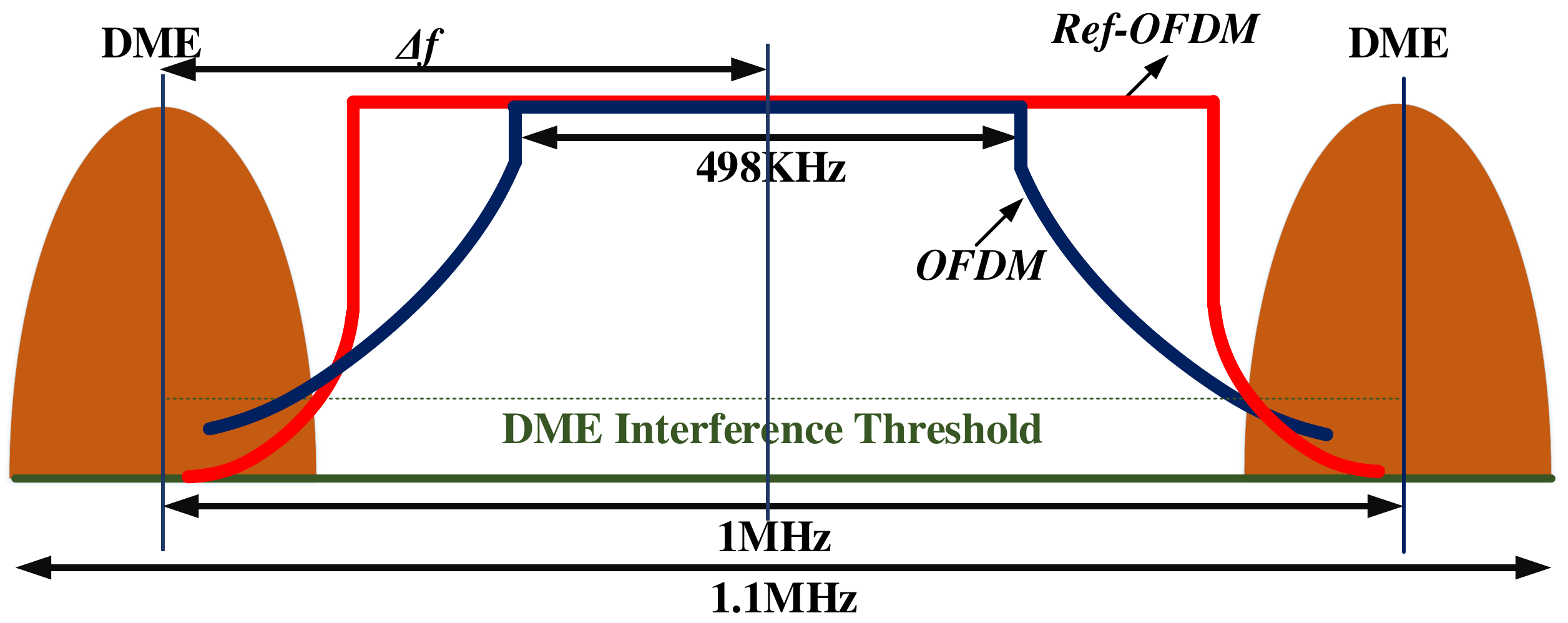}%
		\label{vr1}} \hspace{0.2cm}
	\subfloat[]{\includegraphics[scale=0.33]{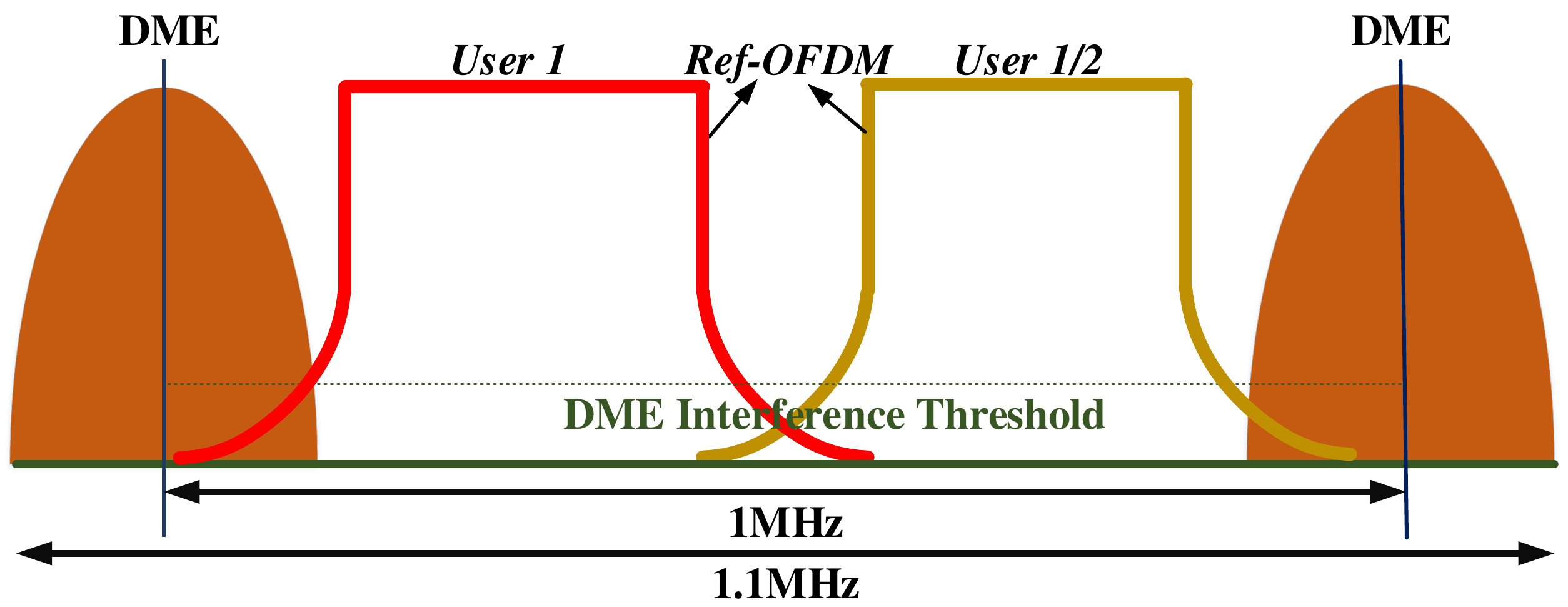}%
		\label{vr2}}
	\caption{(a) Single user, and (b) Multi-user LDACS deployment scenarios for a given DME interference threshold.}
	\label{scen}
\end{figure}

In this section, we present the design of the proposed Ref- OFDM waveform based transceiver for the revised LDACS protocol discussed in the previous section. The motivation behind the proposed transceiver is to support various services that demand distinct bandwidths. For example, consider the scenario in Fig.~\ref{scen} (a) where the existing LDACS can support at the most 498 KHz bandwidth due to the high out-of-band emission of OFDM. Similarly, it does not allow multiple users to share the frequency band especially when transmitting over a narrow bandwidth as shown in Fig.~\ref{scen} (b). The proposed waveform aims to overcome these drawbacks.

To begin with, we present detail design of the transmitter followed by the receiver. \\

\subsection{Ref-OFDM Transmitter}
\begin{figure}[!h]
	\captionsetup{justification=centering}
	\includegraphics[scale=0.6]{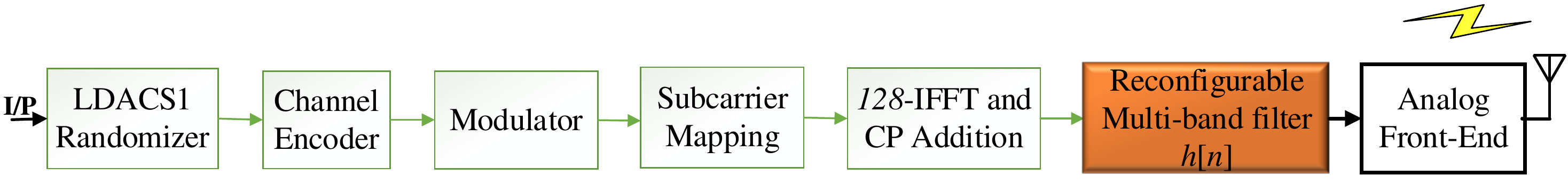}
	\centering
	\caption{Block diagram of the REF-OFDM based LDACS transmitter.}
	\label{tx}
\end{figure}

The block diagram of the Ref-OFDM transmitter is shown in Fig.~\ref{tx}. As per the LDACS specifications, it consists of randomizer block which randomizes the input data to be transmitted by XORing with the LDACS randomizer stream. The data is then encoded via Reed Solomen (RS) and Convolutional (CC) encoder with the coding rate as 0.9 and 0.5, respectively.  

A helical block interleaver is used to minimize the probability of burst errors. The output of the interleaver is modulated by an appropriate modulation scheme such as QPSK, 16 QAM, and 64 QAM followed by the symbol to frame mapping for a given transmission bandwidth. After conventional $N$ point Inverse fast Fourier Transform (IFFT) and Cyclic Prefix (CP) addition, the discrete time domain signal corresponding to the $k^{th}$ sub-carrier can be given as,

\begin{equation}
x[n]= \frac{1}{K} \sum \limits_{k=0}^{K-1} X_k e^{\frac{j 2\pi kn}{K}}
\label{3}
\end{equation}
where, $K$ is the IFFT size, $n$ is the discrete time index and $X_k$ is frequency domain response of the transmitted signal at the $k^{th}$ subcarrier. It is given by
\begin{equation}
X_k= \sum \limits_{n=0}^{K-1} x[n] e^{\frac{-j 2\pi kn}{K}}
\label{4}
\end{equation}

The $x[n]$ is then filtered using the proposed reconfigurable digital filter $f[n]$. Hence, the proposed waveform is referred to as Ref-OFDM waveform. The output of the filter $x'[n]$ is appropriately up-sampled and transmitted over the channel via analog front-end and antenna. 
The transmitted signal $x'[n]$ is the convolution of $x[n]$ and the filter $f[n]$ and can be expressed as,
\begin{equation}
x'[n]= f[n] \circledast x[n]
\label{5}
\end{equation}

Next, we present the design details of the reconfigurable filter.
\subsection{Filter Design}
For the revised LDACS specifications discussed in the previous Section, we need a filter which can support eight different bandwidths. One way to design such filter is the Velcro approach where eight distinct filters are stacked in parallel \cite{vdfr}. Such approach incurs huge area and power complexity and still offers limited flexibility. In a programmable filter, filter coefficients corresponding to different frequency responses are stored in the memory and retrieved when required \cite{pf}. Though less complex than Velcro approach, reconfiguration time of the programmable filters is high and it can not take the advantage of the methods which significantly reduce the complexity of fixed coefficient filter by replacing the computationally intensive coefficient multiplication operation with the simple shift and add operations. The proposed reconfigurable fixed coefficient filter offers variable bandwidth baseband bandpass responses and is based on the coefficient decimation method (CDM) and its extensions \cite{ambede2015design1,rmahesh}. Next, we will discuss the filter design approach and specifications for single band and multiple band transmission. \\

\subsubsection{Single Band Transmission}
\begin{figure}[h!]
	\centering
	\includegraphics[scale=0.7]{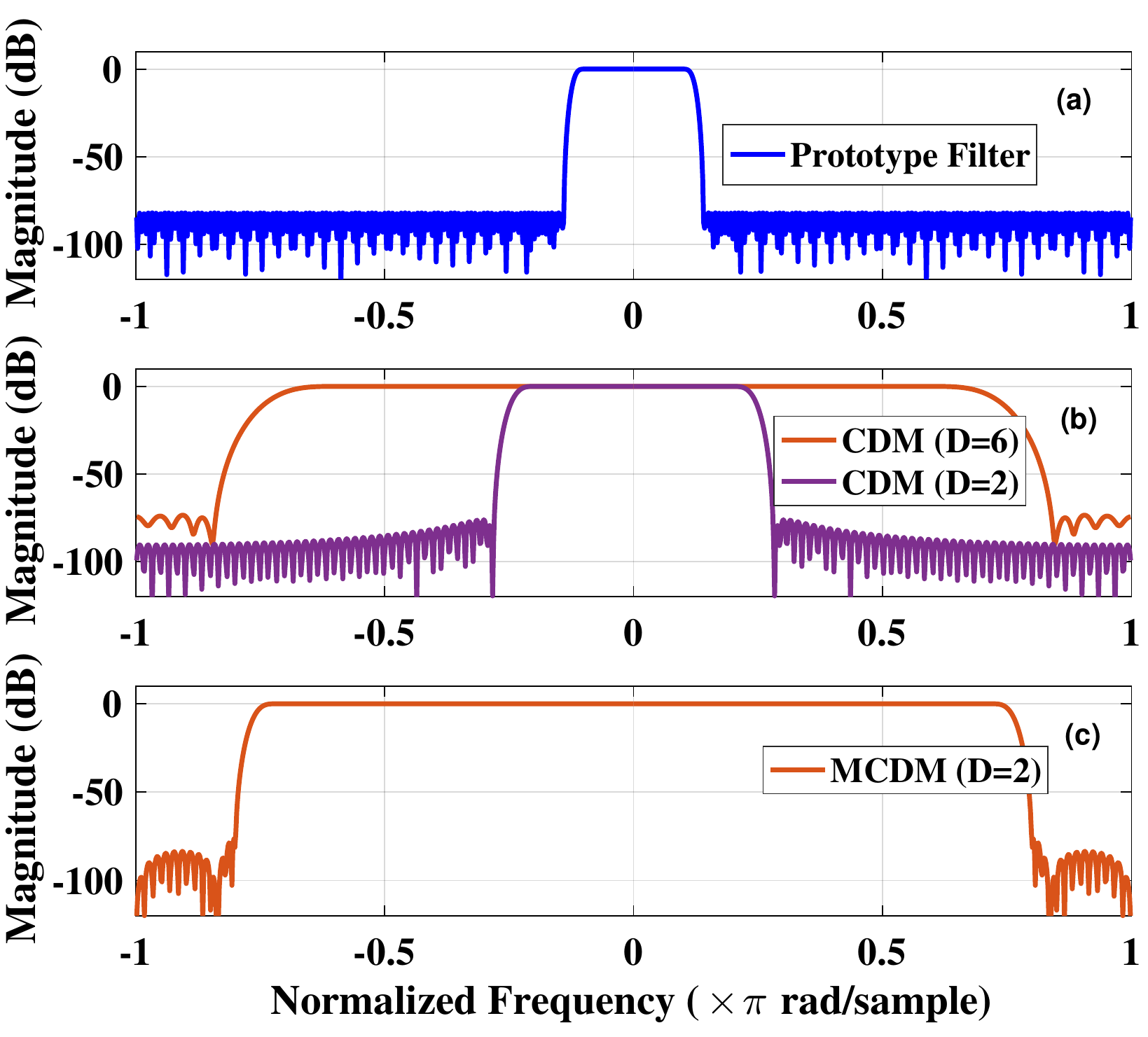}
	\vspace{-0.3cm}
	\caption{Reconfigurable filter design using CDM and MCDM. (a) Prototype baseband bandpass filter with $\omega_c=0.12 \pi$ (b) Baseband bandpass filter responses with the bandwidth $0.24 \pi$ and $0.72\pi$ obtained using the CDM approach with $D=2$ and $D=6$, respectively, (c) Baseband bandpass filter responses with the bandwidth $0.76\pi$ obtained using the MCDM with $D=2$.}
	\label{cm}
\end{figure}

Here, we present the reconfigurable filter design for single band transmission. To begin with, we discuss CDM using a suitable example. Consider the prototype baseband bandpass filter, $F(e^{j\omega_c})$, where $2\omega_c$ is the bandwidth of the filter. The filter coefficients are obtained using Parks-McClellan optimal FIR filter design method and the filter response for $\omega_c=0.12 \pi$ is shown in the Fig.~\ref{cm}(a). Note that all frequency specifications in this sub-section are normalized with respect to half the sampling frequency. The CDM can provide the frequency responses with the bandwidth integral multiple of the origil bandwidth, $2\omega_c$ using fixed-coefficient prototype filter. Let us consider this integer factor as $D \in \{1,2,3..\}$. In CDM with factor $D$, every $D^{th}$ coefficient of the prototype filter is kept unchanged and remaining coefficients are truncated to zero \cite{rmahesh}. This results in the multi-band frequency response, $F^{cdm}(e^{j\omega_c})$, which is given as 

\begin{equation}
F^{cdm}(e^{j\omega_c})=\frac{1}{D} \sum \limits_{i=0}^{D-1} F(e^{j(\omega_c-\frac{2\pi i}{D}}))
\label{6}
\end{equation}

Next, every $D^{th}$ coefficient of the filter is grouped together by removing the zero-valued coefficients to obtain the baseband bandpass response with the bandwidth $2D\omega_c$. For example, the frequency responses obtained using the prototype filter in Fig.~\ref{cm}(a) and the CDM with $D=2$ and $D=6$ are shown in Fig.~\ref{cm}(b). Note that, the filter coefficients are fixed and independent of $D$. However, the transition bandwidth and stop-band attenuation deteriorate by factor $D$. One way to overcome this deterioration is to over-design the prototype filter with higher order. An extension of CDM, referred to as modified CDM (MCDM) \cite{ambede2015design1}, offers decimated bandpass response with large bandwidth using a smaller value of $D$ than that required in the CDM. In MCDM, all coefficients except every $D^{th}$ coefficient of the prototype filter, $F(e^{j\omega_c})$ are discarded followed by sign reversal of every alternate retained coefficient \cite{ambede2015design1}. The filter response is then given by 

\begin{equation}
F^{mcdm}(e^{j\omega_c})=\frac{1}{D} \sum \limits_{i=0}^{D-1} F(e^{j(\omega_c -  \frac{\pi (2i+1)}{D})})
\label{7}
\end{equation}

For instance, MCDM with factor $D$ results in a baseband bandstop response and the corresponding bandpass response, with bandwidth $1-D\omega_c$, can be obtained by subtracting it from an appropriately delayed version of the input signal. For example, the MCDM with $D=2$, offers the bandpass response with the bandwidth of $0.76\pi$ as shown in Fig.~\ref{cm}(c). It has narrower transition bandwidth and better stopband attenuation than the bandpass response with bandwidth $0.72\pi$ obtained using the CDM in Fig.~\ref{cm}(b). 

The proposed filter is designed using a combination of CDM and MCDM. For easier understanding, we mention normalized bandwidths corresponding to actual transmission bandwidths in Table ~\ref{fi}. The maximum input frequency is 1250 MHz (=128 * 9.76 KHz) which corresponds to the sampling frequency of 2.5 MHz. For the desired values of bandwidths, we obtain the bandwidth of the prototype filter as $0.24\pi$ (i.e.,$\omega_c=0.12\pi$) and range of $D$ as $\{2,3,4,5,6,7\}$ via dynamic programming. For instance, the MCDM with $D=7$ and prototype filter with $\omega_c=0.12\pi$ give bandpass response with bandwidth $0.32\pi$ (i.e.,$\omega_{cd}=0.16\pi$). \\

\begin{table}[h!]
	\centering
	\caption{Reconfigurable Filter Design}
	\label{fi}
	
	\begin{tabular}{|m{3cm} | m{3cm}| m{3cm}| m{3cm}|}
		\hline
		\textbf{Bandwidth (KHz)} & \textbf{Desired cut-off frequency ($\omega_{cd}$)} & \textbf{Decimation Factor (D)} & \textbf{Filter Method}\\
		\hline
		186 & 0.16 & 7 & MCDM \\
		\hline
		264 & 0.22 & 2 & CDM \\
		\hline
		342 & 0.28 & 6 & MCDM \\
		\hline
		420 & 0.34 & 3 & CDM \\
		\hline
		498 & 0.40 & 5 & MCDM \\
		\hline
		576 & 0.46 & 4 & CDM \\
		\hline
		654 & 0.52 & 4 & MCDM \\
		\hline
		732 & 0.58 & 5 & CDM \\
		\hline
		
	\end{tabular}
\end{table}

\begin{figure}[h!]
	\centering
	\subfloat[]{\includegraphics[scale=0.45]{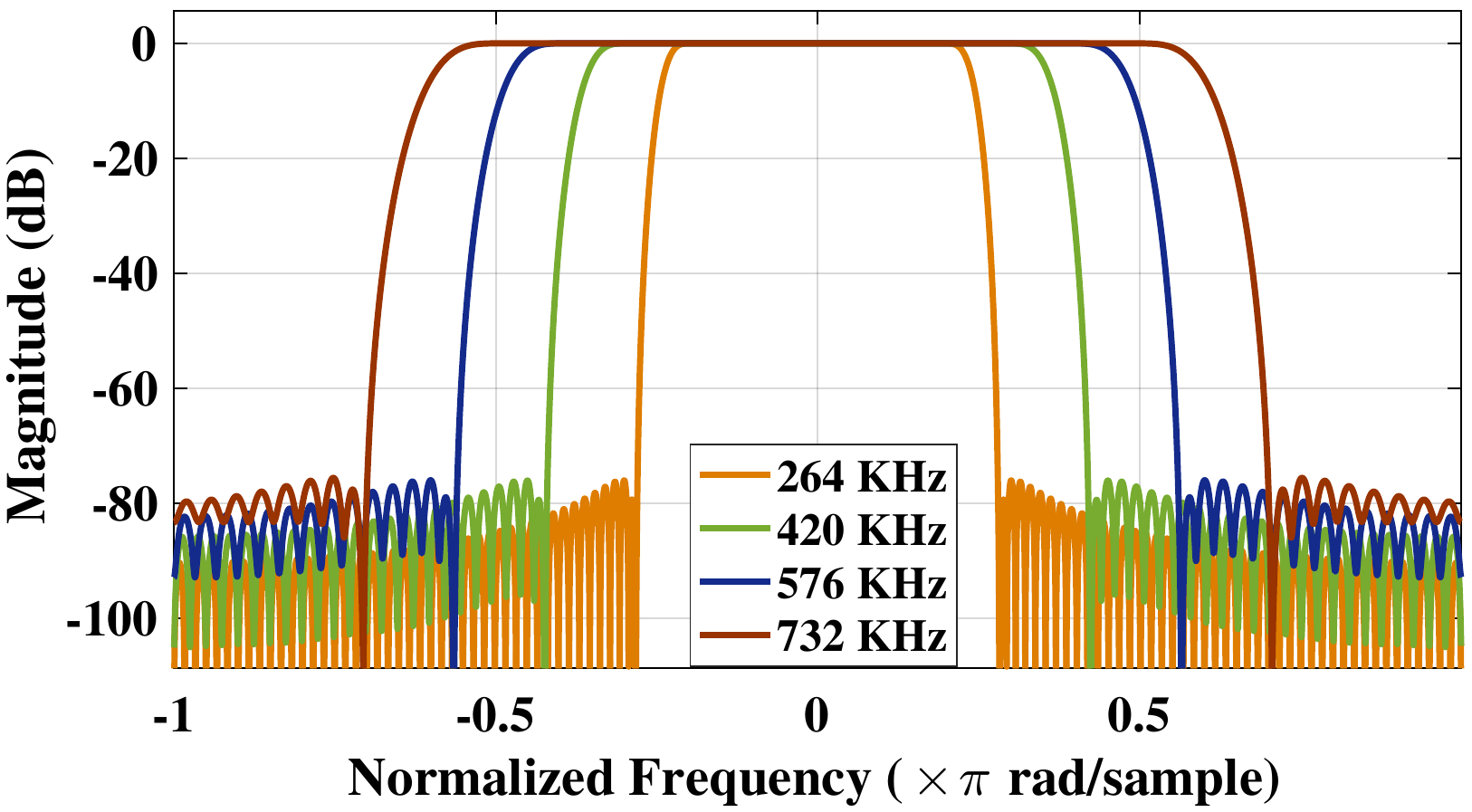}%
		\label{fil1}}
		\hspace{0.5cm}
	\subfloat[]{\includegraphics[scale=0.45]{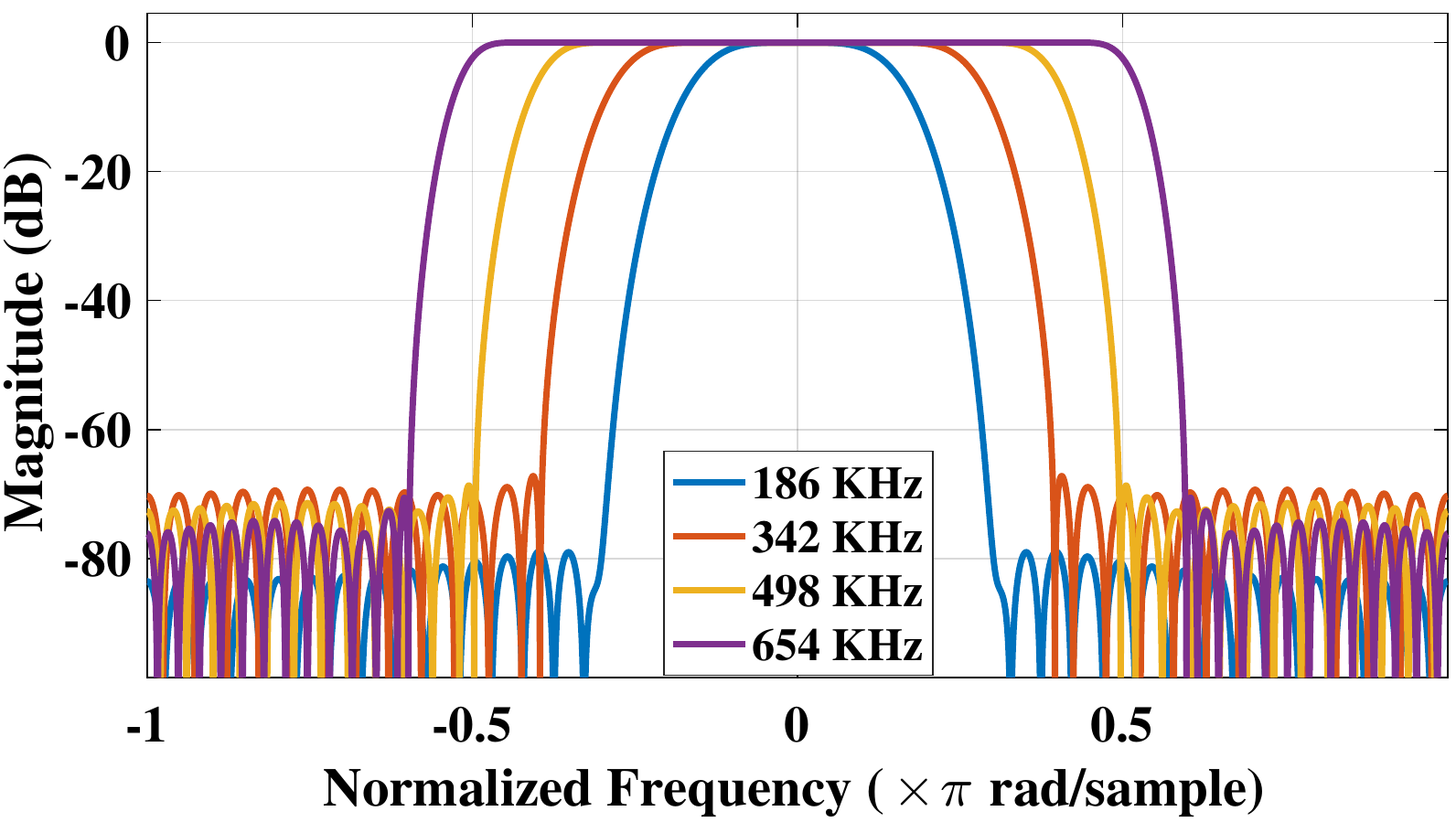}%
		\label{fil2}}
	\caption{Variable baseband bandpass frequency responses obtained using fixed-coefficient baseband bandpass prototype filter with $\omega_c = 0.12\pi$ and, (a) CDM , and (b) MCDM approach.}
	\label{fil}
\end{figure}

Since the CDM and MCDM result in deterioration of the filter response, the prototype filter needs to be over-designed such that the passband ripple, stopband attenuation and transition bandwidth of the prototype filter are $D_{max}(=7)$ times better than the respective desired values of these parameters. Based on these parameters, order and coefficients of the prototype filter are obtained. For example, for the desired stop-band attenuation, pass-band ripple and transition bandwidth of 70 dB, 0.1 dB and $0.1\pi$, respectively, the prototype filter order is 240 and bandwidth is $0.24\pi$, i.e. $\omega_c=0.12\pi$. Please refer to Table I for mapping between the desired bandwidth and corresponding $D$. The baseband bandpass responses for these bandwidths are shown in Fig.~\ref{fil}. 

\subsubsection{Simultaneous Transmission in Multiple Bands}
In this subsection, we extend the above reconfigurable filter for the case where user simultaneously transmits in the multiple bands as shown in Fig.~\ref{scen} (b). This is possible only when filter provides multi-band frequency response with no image on the other side of the DC frequency. Though CDM offers multi-band responses, the response is symmetrical with respect to the DC frequency for real prototype filter. In case of complex prototype filter (i.e. the prototype filter with complex-valued coefficients), the CDM cannot offer variable bandwidth responses for a given center frequency. To obtain asymmetrical frequency response with variable bandwidth and center frequency, we use conventional modulation based discrete Fourier Transform (DFT) filter bank (DFTFB) approach \cite{SG1}. \\

In DFTFB, the prototype filter is realized in the polyphase form and the resultant filter response is modulated using the DFT to obtain bandpass responses at the regular interval between -1 and 1 with no image on the other side of the DC frequency. For example, the DFTFB of order 4 needs 4-point DFT and offers four bandpass responses at the center frequencies of -1, -0.5, 0 and 0.5. Note that the bandwidth of all responses is same and equal to the bandwidth of the prototype filter. To obtain the control over the bandwidth, we replace the prototype filter of the DFTFB with the reconfigurable filter discussed in the previous sub-section. Thus, the bandwidth of all the sub-bands is same and can be tuned to one of the eight supported bandwidths on-the-fly. The control over the center frequency of the bandpass responses can be obtained by choosing the appropriate order of the DFT. For example, the DFTFB of order $K$ offers $K$ bandpass responses located uniformly between -1 and 1 at an interval of $2/K$ on the normalized frequency scale.  \\

\subsubsection{Filter Architecture}
\begin{figure}[h!]
	\includegraphics[scale=0.65]{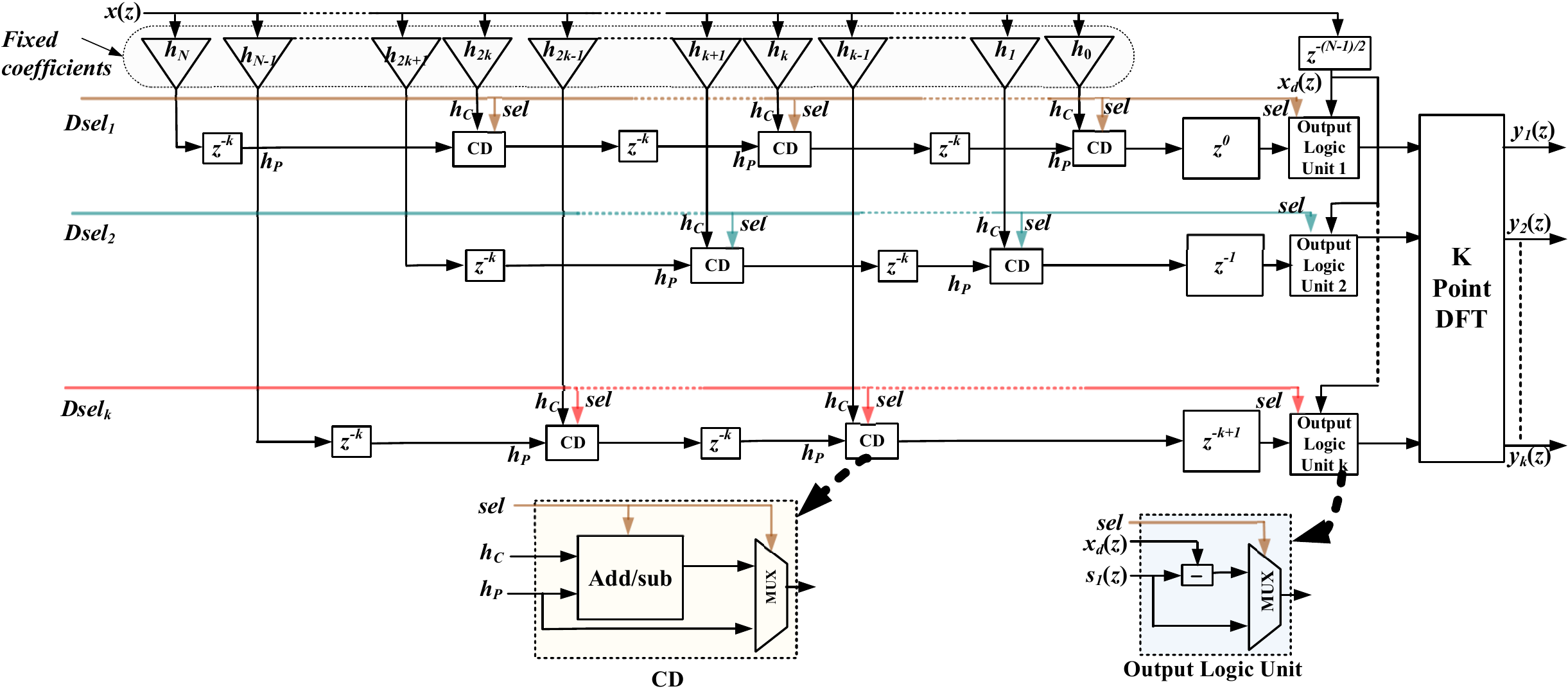}
	\centering
	\caption{Architecture of the proposed reconfigurable $K$-band filter.}
	\label{filter}
\end{figure}

The architecture of a $K$-band reconfigurable filter is shown in Fig.~\ref{filter}. It consists of $N^{th}$ order prototype filter with real and fixed valued coefficients as $\{h_0, h_1,..h_N\}$. It is implemented in the polyphase form with $K$ parallel branches. The sum of the output of all the polyphase branches provides the baseband bandpass response. To obtain the multi-band response, the output of the polyphase filter is given to the $K$-point DFT as shown in Fig.~\ref{filter}. To obtain the bandpass response with variable bandwidth, each adder in the conventional FIR filter is replaced with coefficient decimation (CD) block. The CD block either bypass the new coefficient, $h_C$ or perform addition operation in case of CDM. In case of MCDM, CD block needs to perform subtraction operation for alternate retained coefficients. The select signals are used to perform the desired operation on each of the coefficients. The output logic unit is used to obtain the bandpass response by subtracting the bandstop response obtained from the prototype filter and MCDM from the appropriately delayed version of the input signal. \\

\subsection{Ref-OFDM Receiver}
\begin{figure}[h!]
	\includegraphics[scale=0.6]{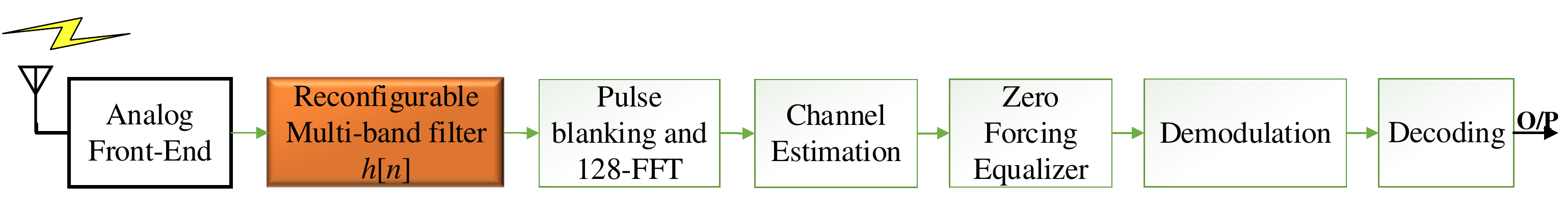}
	\centering
	\caption{Block diagram of REF-OFDM based LDACS receiver.}
	\label{rx}
\end{figure}

After passing the transmitted signal through the channel, an AWGN noise $(\Tilde{n_0}[n])$ and interference is added to the signal. As per the LDACS specifications, two types of interference are present, 1) DME interference, and (2) Gated Gaussian interference (GGI), which is a general interference present in all communication systems with ultra-wide bandwidth. The receiver performs all the functions similar to the transmitter in the reverse order as shown in the Fig.~\ref{rx}.

For the analysis, we consider the dominating DME interference. After addition of the interference the received signal $r[n]$ is given by the discrete convolution of the transmitted signal $x'[n]$ with the LDACS channel impulse response $h[n]$, corrupted by zero-mean additive white Gaussian noise $\eta[n]$ with variance $\sigma^2$ and DME interference represented as the discrete convolution of DME channel impulse response $h_d[n]$ with the DME signal $s[n]$. 

The channel for DME signal is given by $h_d[n]$ as it comes from the different node. 
\begin{equation}
r[n]= h[n] \circledast x'[n] + h_d[n] \circledast s[n]+ \Tilde{n_0}[n]
\label{8}
\end{equation}
where, Both the channels $h[n]$ and $h_d[n]$ assumes to have same statistics and can be represented as
\begin{equation}
h[n]=\sum_{l=1}^{L} h_l \delta[n-l]]
\label{9}
\end{equation}
and,
\begin{equation}
h_d[n]=\sum_{l=1}^{L} h_{d_l} \delta[n-l]]
\label{9}
\end{equation}
where, L is the total number of channel taps. $h_l$ and $h_{d_l}$ is the instantaneous value at the $l^{th}$ tap of the LDACS and DME channel impulse response respectively.

The receiver performs all the functions similar to the transmitter in the reverse order. In the beginning, the received digitized baseband signal is filtered using the same reconfigurable filter $f'[n]$ as transmitter. 

\begin{align}
\nonumber r[n]=& f'[n] \circledast h[n] \circledast f'[n] \circledast x[n] + f'[n] \circledast h_d[n] \circledast s[n]\\
&+ f'[n] \circledast \Tilde{n_0}[n]    
\label{10}
\end{align}

The filtered signal is then passed through the synchronization block to estimate time and frequency offsets. The coarse synchronization is based on correlation of synchronization symbols at beginning of forward link frame and fine synchronization is based on correlation of cyclic prefix of each OFDM symbol. Pulse Blanking technique used to remove the non-linearities and interference. After FFT, channel coefficients and transfer function are estimated using pilots followed by channel equalization via zero forcing approach. In the end, the symbols are demodulated and decoded to obtain the transmitted data. The received signal $R_k$ at $k^{th}$ subcarrier is as follows: 

\begin{equation}
R_k= F'_k H_k F'_k X_k + F'_k H_{d_k} S_k + F'_k \Tilde{N_0}{k}
\label{11}
\end{equation}
where, $F'_k$ is the frequency domain response corresponding to the filter $f'[n]$ at the $k^{th}$ subcarrier. Here, the length of the filter $f'[n]$, $\left \lceil  \frac{(L_{f}+1)}{D} \right \rceil$ will always be less than the total FFT points $K$. Hence, firstly the filter length is zero padded by the remaining length of $\left (K-\left \lceil  \frac{L_{f}+1)}{D} \right \rceil  \right )$ to make the filter length equal to the FFT length. Then,  We get $F'_k$ by taking K point FFT of the zero padded reconfigurable filter impulse response. The $F'_k$ can be expressed as,
\begin{equation}
F'_k= T^H. \left [f[n], \textbf{0}_{1 \times \left (K-\left \lceil  \frac{(L_{f}+1)}{D} \right \rceil  \right )}  \right ]
\label{12}
\end{equation}

Here, T represents the K point FFT matrix. 

Let $f[n]$ be the original set of coefficients. If we replace all the coefficients other than every $D^th$ coefficient by zeros,

\begin{equation}
f'[n] = f[n].b[n]
\label{13}
\end{equation}
where,
\begin{equation}
b[n] = \left\{\begin{matrix}
1 & \forall~n = mD;~ m=0,1,2... \\
0 & \mathrm{otherwise} 
\end{matrix}\right.
\label{14}
\end{equation}
The function \textbf{$b[n]$} is periodic with period $M$, and hence the Fourier series expansion is given by
\begin{equation}
b[n] = \frac{1}{D} \sum_{i=0}^{D-1} B(i) e^{\frac{j2 \pi in}{D}}
\label{15}
\end{equation}
where $B(i)$ are complex-valued Fourier series coefficients defined by, 
\begin{equation}
B(i)= \sum_{n=0}^{D-1} b[n] e^{\frac{-j2 \pi in}{D}}
\label{16}
\end{equation}
By substituting the equation \eqref{14} into \eqref{16} we will get,
\begin{equation}
B(i) = \left\{\begin{matrix}
1 & \forall~ k \\
0 & \mathrm{otherwise} 
\end{matrix}\right.
\label{new}
\end{equation}
Hence, from equation \eqref{new} and \eqref{15}, $b[n]$ can be expressed as,
\begin{equation}
b[n] = \frac{1}{D} \sum_{i=0}^{D-1} e^{\frac{j2 \pi in}{D}}
\label{new1}
\end{equation}
By putting this equation \eqref{new1} into equation \eqref{13}, we can easily compute $f'[n]$.

Next, we will analyze the BER performance of the Ref-OFDM based LDACS system given by the equation \eqref{11}.

\subsection{BER Analysis}

The Signal to interference plus noise ratio is the ratio of signal power and the sum of interference and noise power. For the received signal presented in \eqref{12}, the SINR for $k^{th}$ subcarrier can be given as equation \eqref{13}. 

\begin{equation}
SINR(k)= \frac{{F_{k}}'^{4} \left | H_{k}\right|^{2}  P}{{F_{k}}'^{2} P_{\tilde{N_0}} + {F_{k}}'^{2} \left | H_{d_k}\right |^{2}  P_{DME}}
\label{17}
\end{equation}
where, $F'_k$ is given by the equation \eqref{12}. Here, we are considering rayleigh fading channel $H_k$ for the analysis. The pdf of $\left |H_k  \right |^2$ follows the exponential distribution which can be given as,
\begin{equation}
p_\lambda (\lambda )=\frac{1}{\bar{\lambda} }  e^{\frac{-\lambda}{\bar{\lambda} }}
\label{18}
\end{equation}
where, $\bar{\lambda}$ is the mean of the variable $\lambda$ and is equal to variance of $H_k$. The DME channel $\left |H_{d_k}  \right |^2$ also follows the same exponential distribution as it is assumed to have the same distribution as the LDACS channel.

The term $P_{DME}$ indicates the DME interference power and can be expressed as,
\begin{equation}
P_{DME}= \int_{f_1}^{f_2} |S(f)|^2 df    
\label{19}
\end{equation}

Substituting \eqref{2} to \eqref{19} and putting $cos(\theta)=\frac{e^{j\theta}+e^{-j\theta}}{2}$ , 

\begin{align}
\nonumber P_{DME}= &A^2 \left (\frac{8\pi}{\alpha}  \right ) \int_{f_1}^{f_2} \left |e^{\frac{-2\pi^2f^2}{\alpha}}  \right | ^2 \left |e^{j2\pi f \Delta t}  \right |^2 \\
& \times \left |\frac{e^{j\pi f \Delta t}+e^{-j\pi f \Delta t}}{2}  \right |^2 df
\label{20}
\end{align}
By the Euler formula, $|e^{j \theta}|^2=1$ and assuming, $C_1=\frac{4 \pi^ 2}{\alpha}$ and $C_2=j2 \pi  \Delta t$ 

\begin{align}
\nonumber P_{DME} =  &A^2 \left (\frac{8\pi}{\alpha}  \right ) \left [2 \int_{f1}^{f2} e^{-C_1 f^2} df + \int_{f1}^{f2} e^{-C_1 f^2+C_2f} df \right. \\
&\left.+\int_{f1}^{f2} e^{-C_1 f^2-C_2 f} df \right ]
\label{21}
\end{align}
By further solving the equation we get,
\begin{align}
\nonumber P_{DME} =  &A^2 \left (\frac{8\pi}{\alpha}  \right ) \sqrt{\frac{\pi}{C_1}}\left [\left \{(erf(\sqrt{C_1} f_2)-erf(\sqrt{C_1} f_1))  \right \} \right .\\ 
\nonumber &+ \frac{1}{2} \left \{e^{\frac{C_2^2}{4C_1}}(erf(\frac{2C_1f_2-C_2}{2\sqrt{C_1}})-erf(\frac{2C_1f_1-C_2}{2\sqrt{C_1}})  \right \} \\
&\left . + \frac{1}{2} \left \{e^{\frac{C_2^2}{4C_1}}(erf(\frac{2C_1f_2+C_2}{2\sqrt{C_1}})-erf (\frac{2C_1f_1+C_2}{2\sqrt{C_1}})  \right \}  \right ]
\label{22}
\end{align}

Using \eqref{17}, BER of the $k^{th}$ received symbol for M-QAM can be expressed as,
\begin{align}
\nonumber P_{e_{MQAM}}^{\lambda_1,\lambda_2}(k) \cong &\frac{4}{log_2 M} \left (1-\frac{1}{\sqrt{M}}  \right ) \sum_{i=1}^{\sqrt{M}/2} Q(2i-1) \\
& \times \sqrt{\frac{3 log_2 M {F_{k}}'^{4} \left |H_{k}  \right |^{2} P}{\left (M-1  \right ) \left ({F_{k}}'^{2} P_{\tilde{N_0}} + {F_{k}}'^{2} \left |H_{D_k}  \right |^{2} P_{DME} \right )}}
\label{23}
\end{align}
where erfc is an error function \cite{JP}. Therefore, the Bit error rate (BER) averaged across the fading channel can be expressed as,
\begin{align}
\nonumber P_{e_{MQAM}}(k) = &E[P_{e_{MQAM}}^{\lambda_1,\lambda_2}(k)] \cong\int_{0}^{\infty}\int_{0}^{\infty} P_{e_{MQAM}}^{\lambda_1,\lambda_2}(k)~ \\
& \times p_{\lambda}(\lambda) d\lambda ~p_{\lambda_{d}}(\lambda_d) d\lambda_{d}
\label{24}
\end{align}

The average BER across all the subcarriers can be calculated as,
\begin{equation}
P_{e_{MQAM}} = \frac{1}{K} \sum_{k=0}^{K-1} P_{e_{MQAM}}(k)  
\label{25}
\end{equation}

The analytical expression in equation \eqref{24} can be evaluated by numerical methods. It provides the complete analysis of the proposed Ref-OFDM based LDACS-DME coexistence in terms of BER over the multi-path Rayleigh fading channel. Next, we present the simulation results to analyze the performance of various LDACS.

\section{Simulation Results}
In this section, we present extensive simulation results to compare the performance of the proposed LDACS protocol with the existing protocol and FBMC based LDACS in \cite{Hj} in realistic LDACS environment. Note that we consider the revised OFDM based LDACS which employs time domain windowing at the transmitter to improve the out-of-band attenuation. In addition, we also consider the LDACS protocol using GFDM waveform which has not been studied in the literature yet. The results include the comparison of these variants with respect to their out-of-band emission using the power spectral density (PSD) plots for various bandwidths, interference at the adjacent DME signal for these bandwidths, BER in presence of DME and GGI interference, and implementation complexity. The simulation parameters are chosen as per the LDACS specifications and are given in Table ~\ref{spec}. 
\begin{table}[!h]
	\centering
	\caption{Simulation Parameters}
	\label{spec}
	\renewcommand{\arraystretch}{1}
	\resizebox{\linewidth}{!}{
	\begin{tabular}{ |m{6cm} | m{6cm}|}
		\hline
		\textbf{Parameters}  & \textbf{Value} \\
		\hline
		Total Bandwidth & 1.25MHz\\
		Transmitted Bandwidth & Any of the supported bandwidths\\
		Length of FFT  & 128\\
		Used sub-carriers  & 18-74\\
		Sub-carrier spacing& 9.76KHz\\
		Total Symbol duration  & 120$\mu$s\\
		Modulation & QPSK \\
		Channel & ENR, APT, TMA\\
		CC encoder rate & 0.5 \\
		RS encoder rate & 0.9 \\
		\hline
	\end{tabular}
	}
\end{table}	\\

\subsection{Single Band Transmission}
\begin{figure}[h!]
	\centering
	\subfloat[]{\includegraphics[scale=0.5]{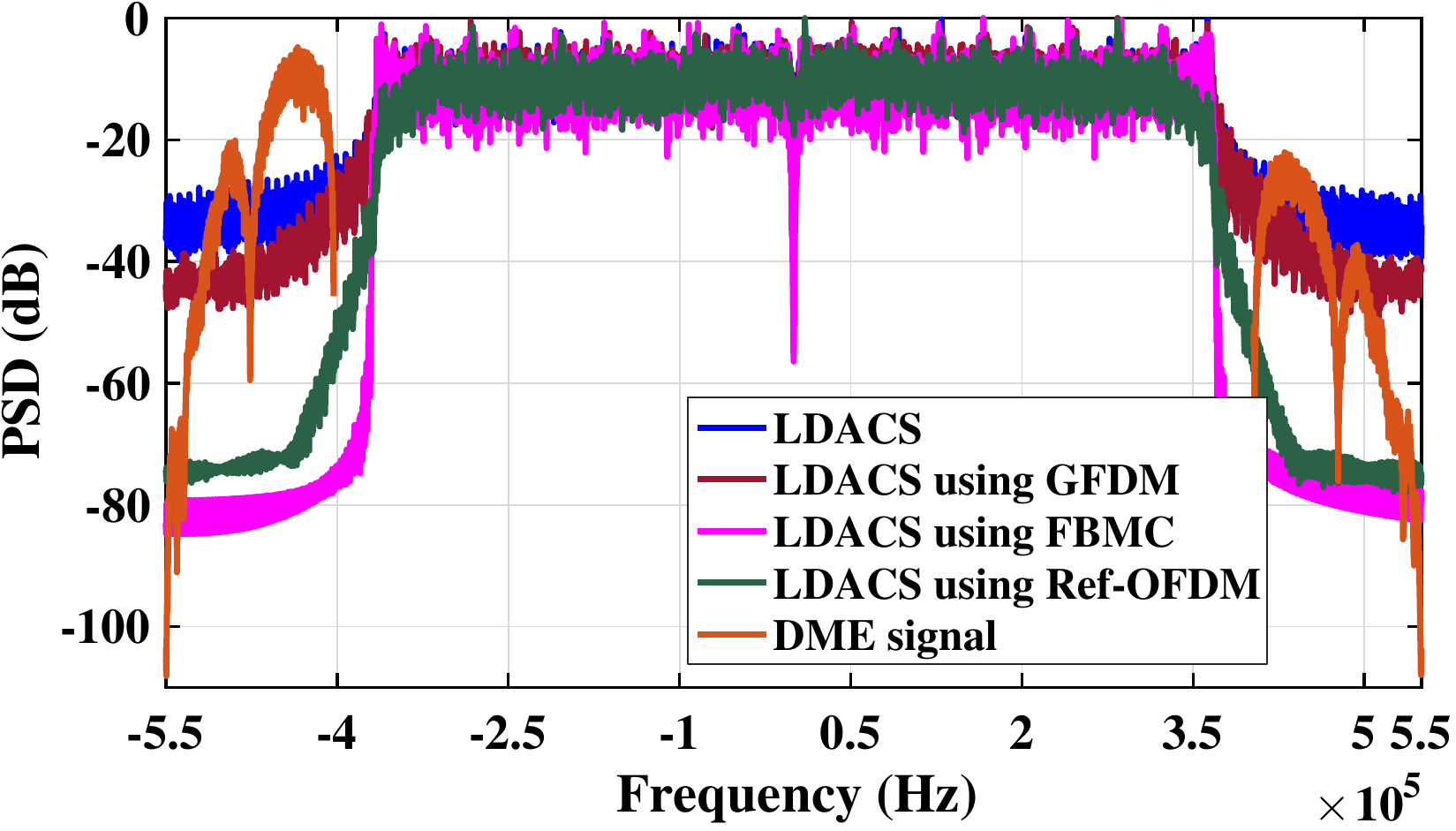}%
		\label{int2}}
        \hspace{0.1cm}
	\subfloat[]{\includegraphics[scale=0.5]{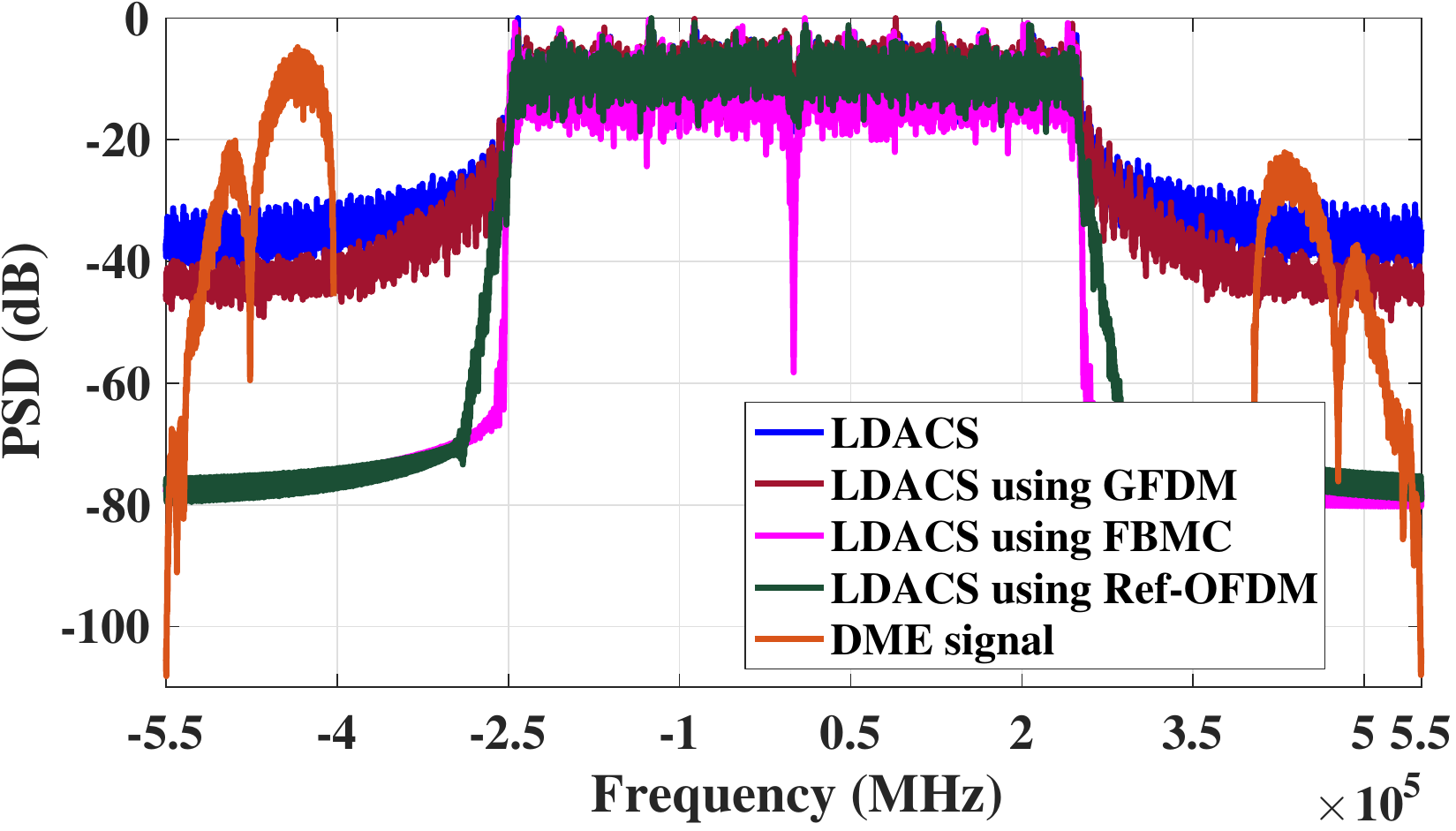}%
		\label{psd4}}
	\caption{The PSD comparison of various waveforms for ENR channel and two different transmission bandwidths, (a) 498KHz, and (b) 732KHz.}
	\label{psd}
\end{figure}

To begin with, we consider single user transmitting in the frequency band between adjacent DME signals. For illustration, we consider two bandwidths, 1) 732 KHz which incurs maximum interference to the DME, and 2) 498 KHz which is maximum bandwidth allowed in existing OFDM based LDACS. The corresponding PSD plots for ENR channel are shown in  Fig.~ \ref{psd}. For clarity of the plots, we only show the main lobes of the DME signals. It can be observed that the interference at the DME signal is quite high in case of OFDM and GFDM based LDACS (For actual values, please refer to discussion related to Table ~\ref{inter} at the end of this subsection). The proposed REF-OFDM and FBMC based LDACS can achieve the transmission bandwidths of up to 732 KHz due to high out-of-band attenuation leading to 50\% improvement in spectrum utilization. The PSD plots corresponding to other two channels are not shown to avoid repetitive plots. However, we have considered them for BER analysis. \\

Next, we compare the BER of these waveforms for three different channel conditions in presence of GGI interference only. We do not include BER plots for FBMC as they are overlapping with OFDM BER plots. We again consider 498 KHz and 732 KHz bandwidth and corresponding plots are shown in Fig. \ref{ber}. It can be observed that OFDM and Ref-OFDM waveforms have almost similar BER and perform better than GFDM. As expected, the performance is better in case of ENR channel due to strong LOS path. The PSD and BER plots show that the Ref-OFDM has better out-of-band attenuation than OFDM and GFDM without compromising on the BER performance. \\

\begin{figure}[h!]
	\centering
	\subfloat[]{\includegraphics[scale=0.45]{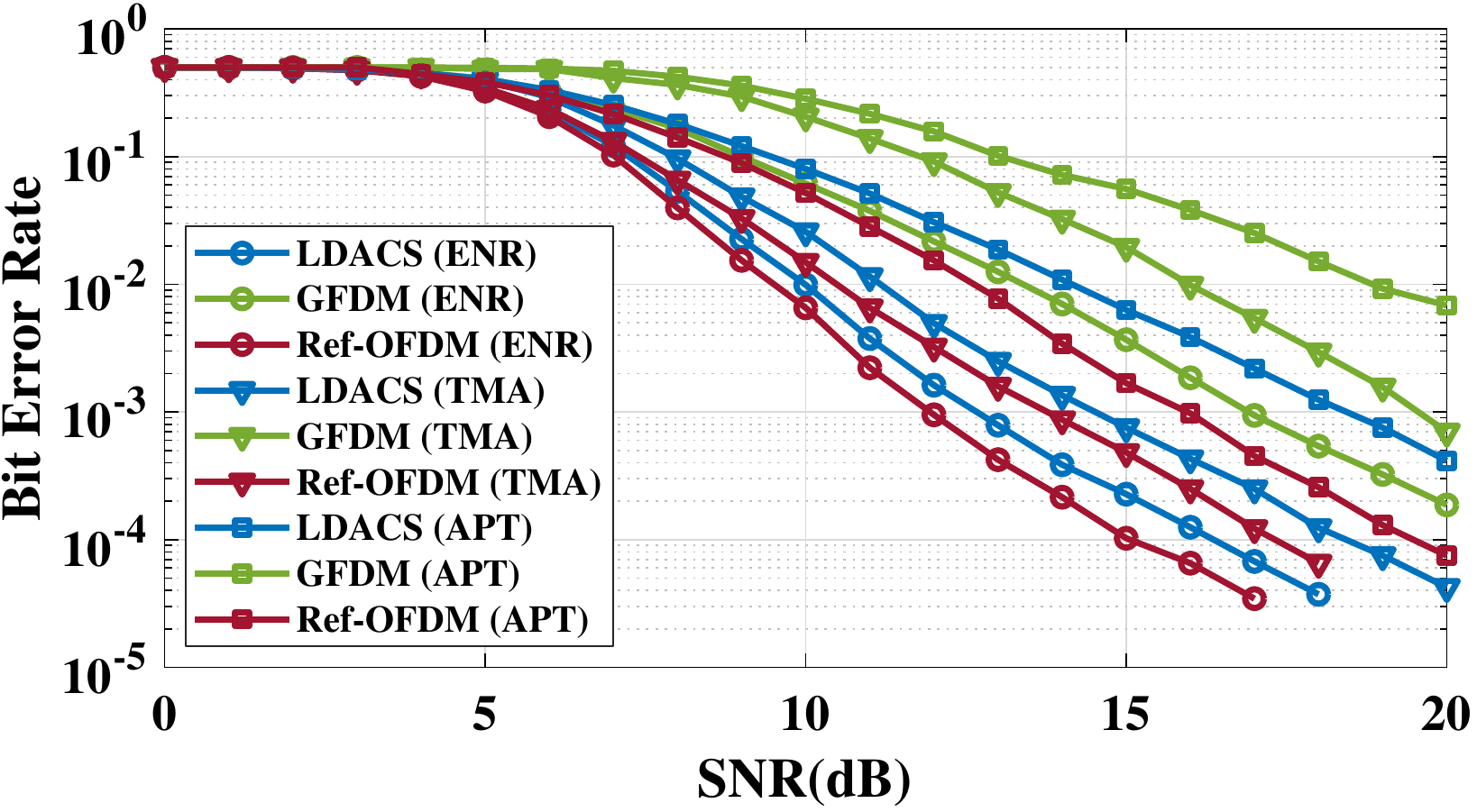}%
		\label{int2}}
		\hspace{0.5cm}
	\subfloat[]{\includegraphics[scale=0.45]{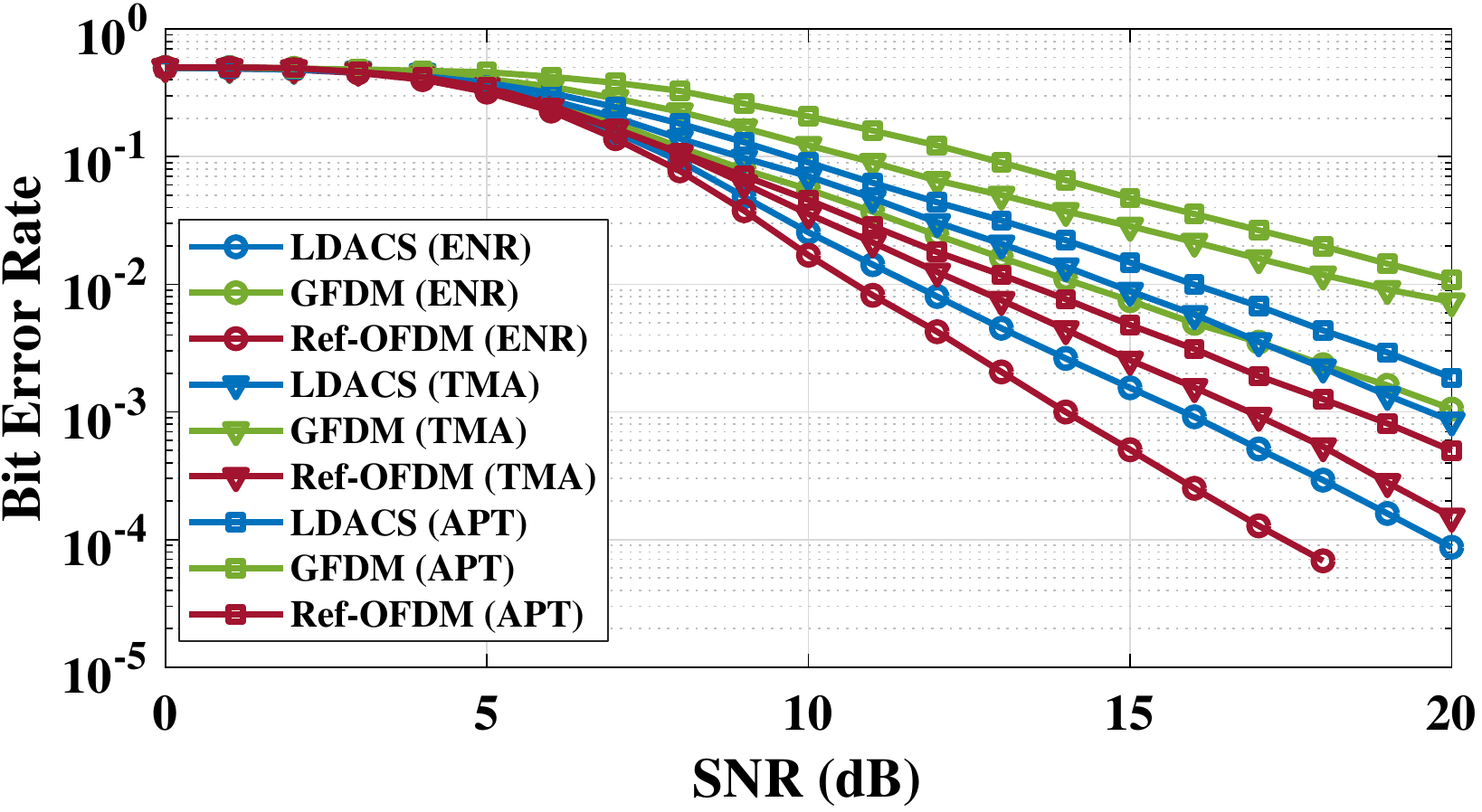}%
		\label{int3}}
		\caption{The BER comparison of various waveforms for two different transmission bandwidths, (a) 498KHz, and (b) 732KHz and three different channels.}
	\label{ber}
\end{figure}

\begin{figure}[!h]
	\centering
	\includegraphics[scale=0.5]{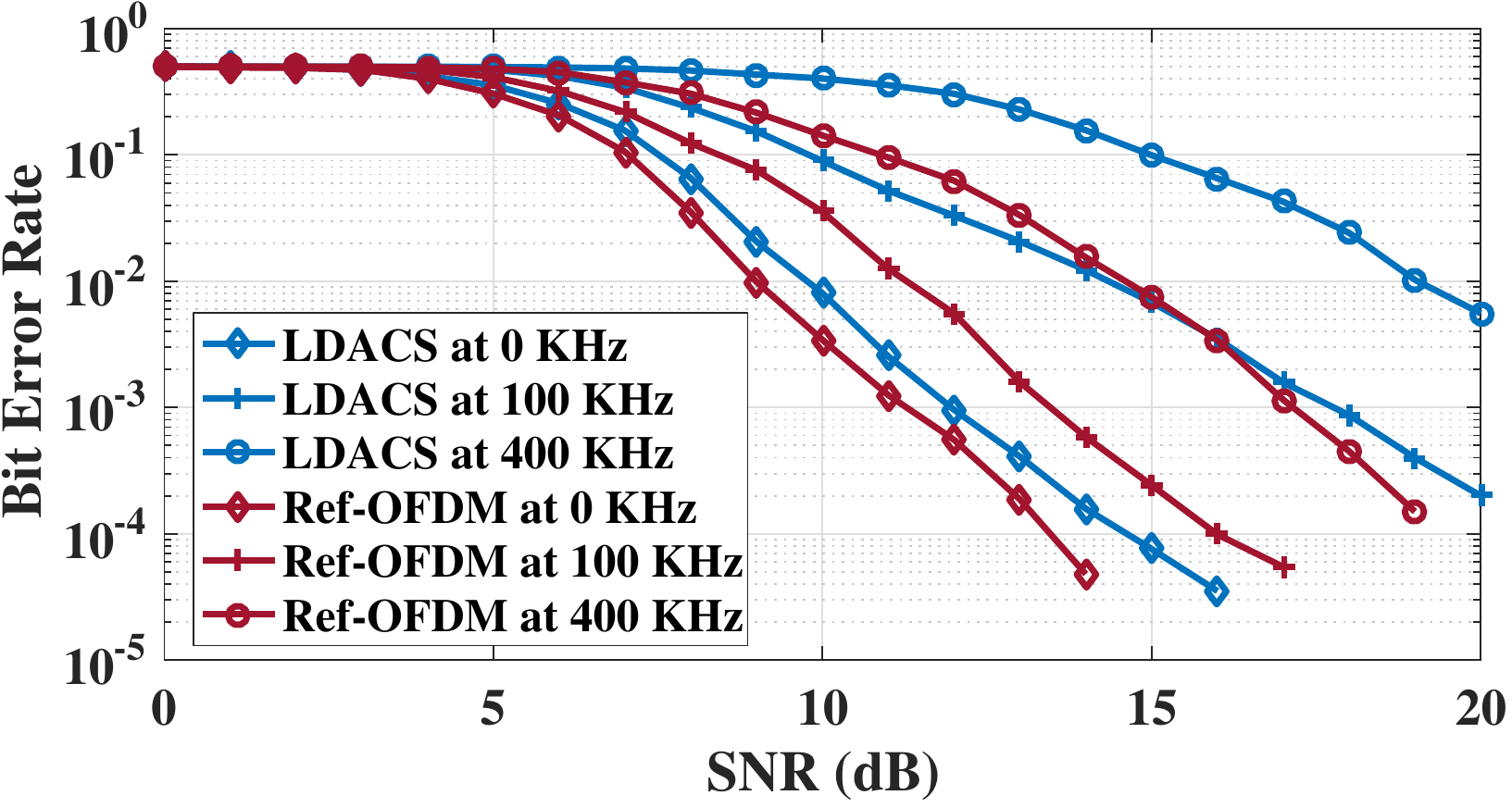}
	\caption{The BER comparison of Ref-OFDM and OFDM based LDACS in presence of DME interference for ENR channel with 342 KHz transmission bandwidth and three different center frequencies.}
	\label{dmeb}
\end{figure}

Next, we compare the BER of these waveforms in presence of DME interference for ENR channel. Here, we consider the transmission bandwidth of 342 KHz and three center frequency with a deviation of 0, 100 and 400 KHz from the baseband. As shown in Fig. \ref{dmeb}, the BER of the proposed Ref-OFDM is significantly better than existing LDACS for all the center frequencies considered here. We also observed that the difference between the BER performance of the Ref-OFDM and OFDM increases with the increase in the transmission bandwidth. The BER of GFDM is worse than that of OFDM while the BER of the FBMC is nearly identical to that of the Ref-OFDM. Similar behavior has also been observed for other channels. These results are not included here for clarity of the plots.\\

Next, we study the interference at the DME signals due to LDACS transmission with the transmission bandwidth of 732 KHz and 498 KHz. The interference ($I$) is calculated by the sum of PSD ($\phi(f)$) between two frequencies ($f_1$ and $f_2$) and can be represented by: 
\begin{equation*}
I=\int_{f_1}^{f_2} \phi(f) df
\end{equation*}

Here, we consider LDACS signal at different center frequencies located at an interval of 50 KHz with the DME signal located at the fixed center frequency. The corresponding interference values are shown in Table \ref{inter} where  refers to not applicable since corresponding center frequencies are not allowed for the requested bandwidth due to overlap with the main lobe of the DME signals. It can be observed that the proposed Ref-OFDM and FBMC based LDACS offer the lowest interference to the incumbent DME signals. In most of the cases, the interference is lower than 40 dB which is the desired threshold as per the LDACS requirement and approximately 35 dB better than existing LDACS. These results not only validates the superiority of the proposed waveform but also indicates the feasibility of multi-band multi-user deployment in case of Ref-OFDM and FBMC based LDACS.
\begin{table*}[!h]
	\centering
	\caption{Interference at DME in DB Due to Various Waveforms for Transmission Bandwidths of 498 KHz and 732 KHz}
	\label{inter}
	\renewcommand{\arraystretch}{1.2}
	\resizebox{\linewidth}{!}{
		\begin{tabular}{|m{3.5cm} | m{3cm}||m{3cm} | m{3cm}||m{3cm} | m{3cm}||m{3cm} |}
			\hline
			\multirow{2}{*}{\textbf{Bandwidth (BW)}} & \multirow{2}{*}{\textbf{Waveform}} & \multicolumn{5}{c|}{\textbf{Transmission Center frequency w.r.t the DME center frequency (r = 50 KHz)}}\\
			\cline{3-7}
			&&$\frac{BW}{2} + r$ & $\frac{BW}{2} + 2r$ & $\frac{BW}{2} + 3r$ & $\frac{BW}{2} + 4r $& $\frac{BW}{2} + 5r$ \\
			\hline
			
			\multirow{4}{*}{\textbf{498 KHz}} & OFDM & 7.1745 &  2.7038 & 0.9269  &   -0.1658   & -0.8701\\
			\cline{2-7}
			& GFDM & 5.9571  & -3.1187 & -7.3229  &   -8.7974 & -9.3662 \\
			\cline{2-7}
			& Ref-OFDM & 0.3596 & -38.7566  & -40.7823  & -42.1427 & -43.1139\\
			\cline{2-7}
			& FBMC & -2.0064 & -37.8831 & -39.8368 & -40.9379 & -41.6017 \\
			\hline
			\multirow{4}{*}{\textbf{732 KHz}} & OFDM & 6.4676  &  1.8911  & - & - & - \\
			\cline{2-7}
			& GFDM & 5.2885  & -5.6320  & -  & -  & -\\
			\cline{2-7}
			& Ref-OFDM & 
			-27.1217 & -41.6051 & -  & - & -\\
			\cline{2-7}
			& FBMC & -31.5356 & -45.7056 & - & - & -\\
			\hline
		\end{tabular}
	}
\end{table*}

\subsection{Multi-band Transmission}
\begin{figure}[h!]
	\centering
	\subfloat[]{\includegraphics[scale=0.45]{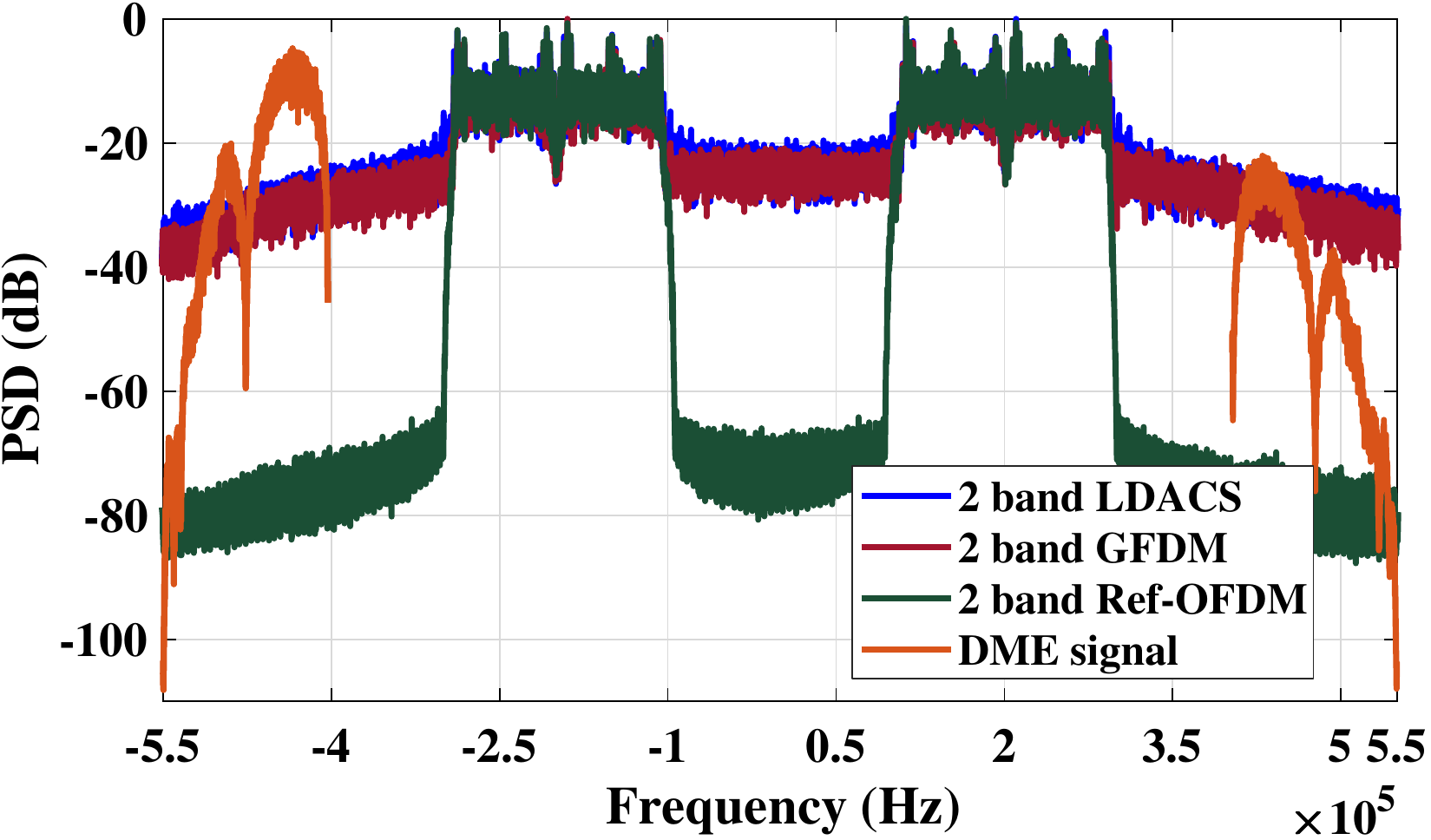}%
		\label{dm}}
		\hspace{0.5cm}
	\subfloat[]{\includegraphics[scale=0.5]{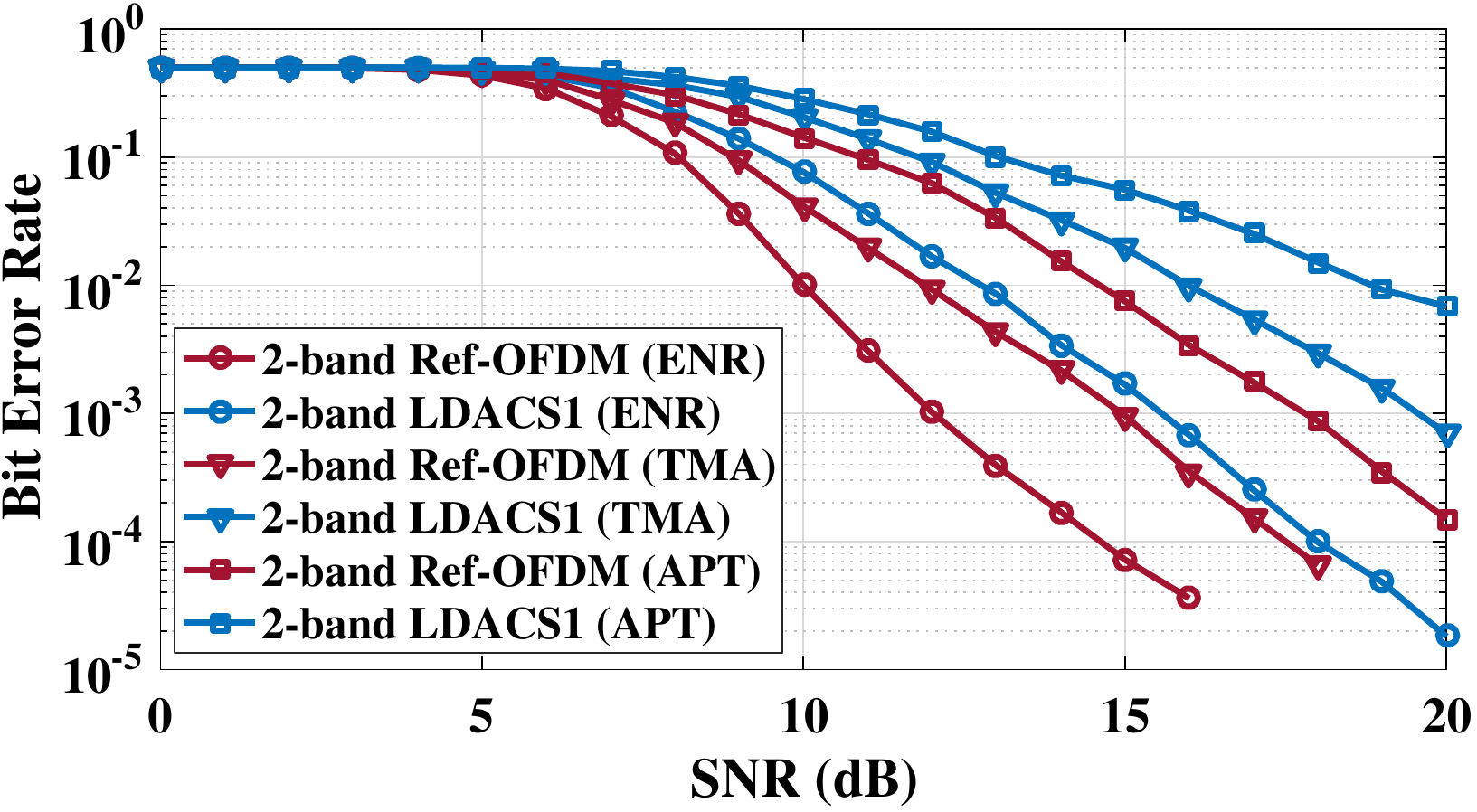}%
		\label{bm}}
		\caption{The performance comparison for 2-band transmission with bandwidth of 186 KHz (a) PSD for ENR channel (b) BER comparison}
	\label{pb}
\end{figure}

Here, we consider one user transmitting in two non-contiguous bands of 186 KHz bandwidth with baseband center frequencies of -200 KHz and 200 KHz. It can be observed from the PSD plots in Fig.~\ref{dm} that the performance of OFDM and GFDM based LDACS have degraded further compared to single band transmission. For instance, the interference at DME signal for existing OFDM based LDACS is -4.7 dB compared to -5.5 dB in case single-band transmission for a given center frequency. As expected, it is much higher than -41.5 dB interference at DME due to the proposed Ref-OFDM based LDACS.\\

For this case, we compare the BER performance of existing and proposed LDACS for three different channel conditions. As shown in Fig.~\ref{bm}, the performance of the proposed LDACS is significantly better than existing LDACS. Poor BER performance for narrow transmission bandwidth of 186 KHz confirms the non-feasibility of existing LDACS for multi-band deployment. \\

\subsection{Multi-user Transmission}
\begin{figure}[h!]
	\centering
	\subfloat[]{\includegraphics[scale=0.45]{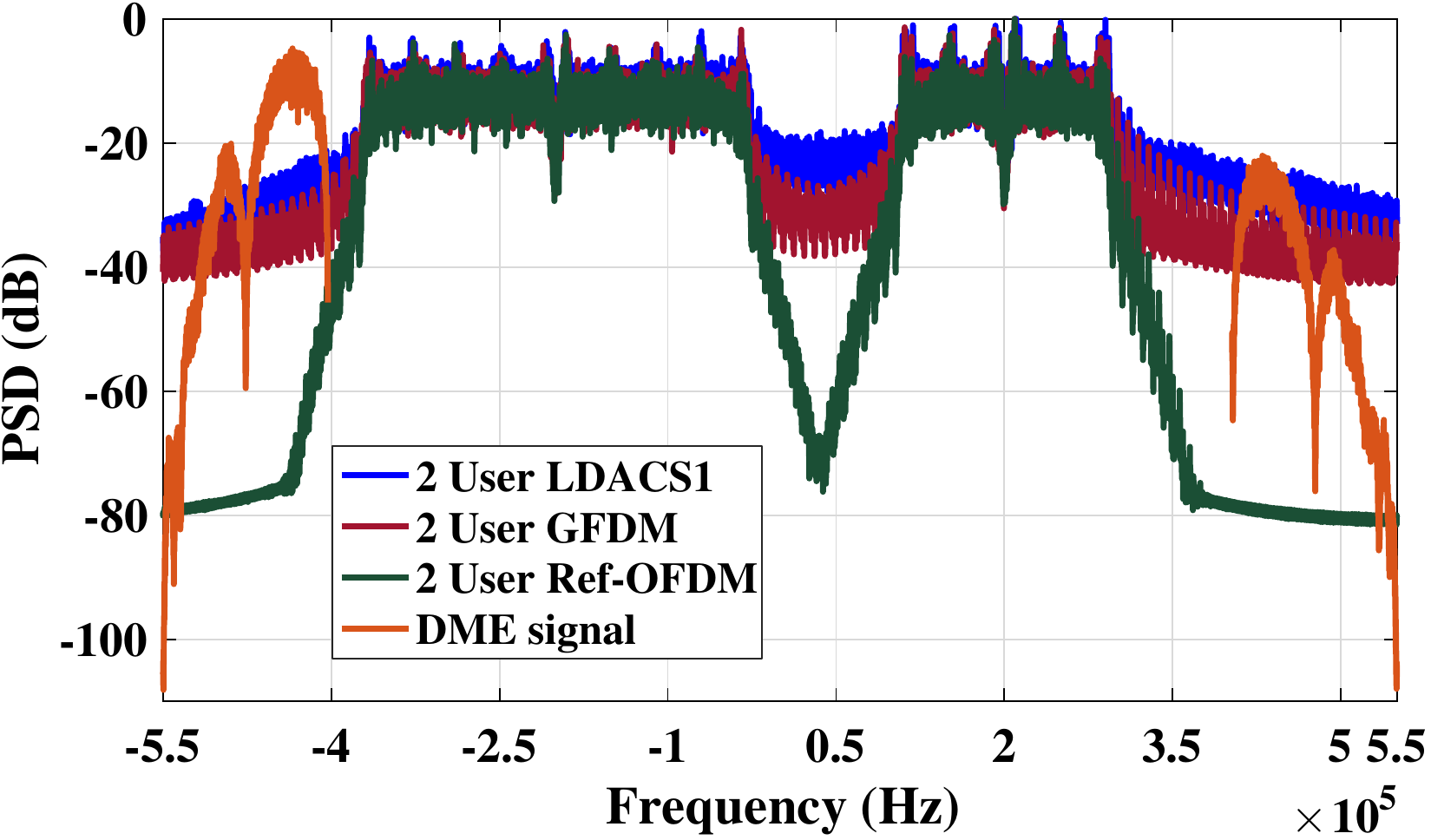}%
		\label{dmu}}
		\hspace{0.5cm}
	\subfloat[]{\includegraphics[scale=0.47]{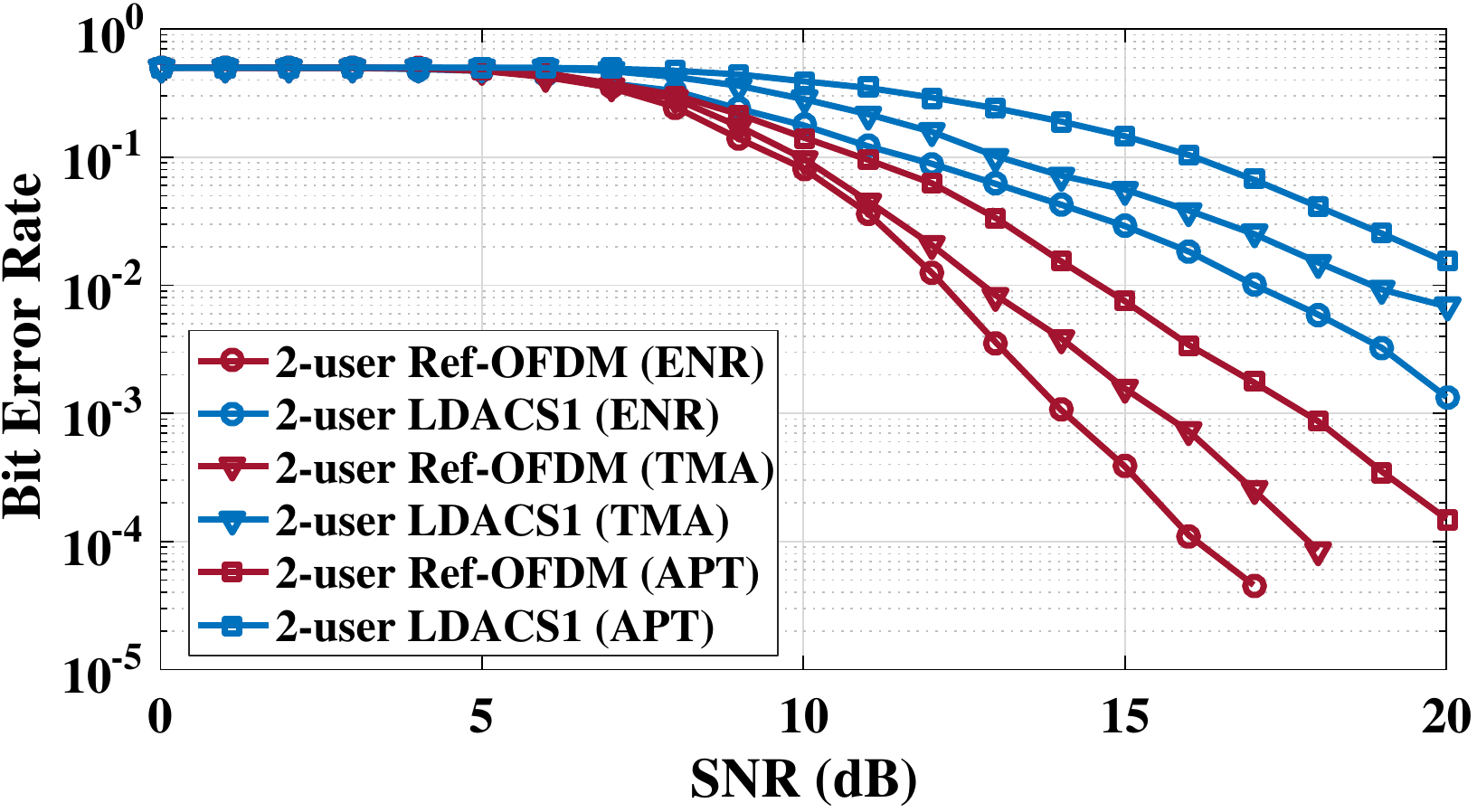}%
		\label{bmu}}
		\caption{The performance comparison for 2-user transmission with bandwidth of 186 KHz and 342 KHz (a) PSD for ENR channel (b) BER comparison}
	\label{ber}
\end{figure}

Next, we consider the scenario where two users share the frequency band where the transmission bandwidth of one user is 342 KHz and other user is 186 KHz. The center frequencies are same as that of 2-band transmission considered before. The corresponding PSD plots shown in Fig.~\ref{dmu} indicates very high interference to the DME signal from existing LDACS. For instance, the DME interference due to existing LDACS is 4 dB compared to -40 dB due to proposed LDACS.\\

For the above scenario, Fig.~\ref{bmu} shows the BER performance of 2-user case for all three channel scenarios. It can be observed that the proposed LDACS is significantly better than existing LDACS. These results also confirm the feasibility of multi-user transmission using proposed LDACS. \\

\subsection{Complexity Comparison}
\begin{figure}[h!]
\centering
\includegraphics[scale=0.55]{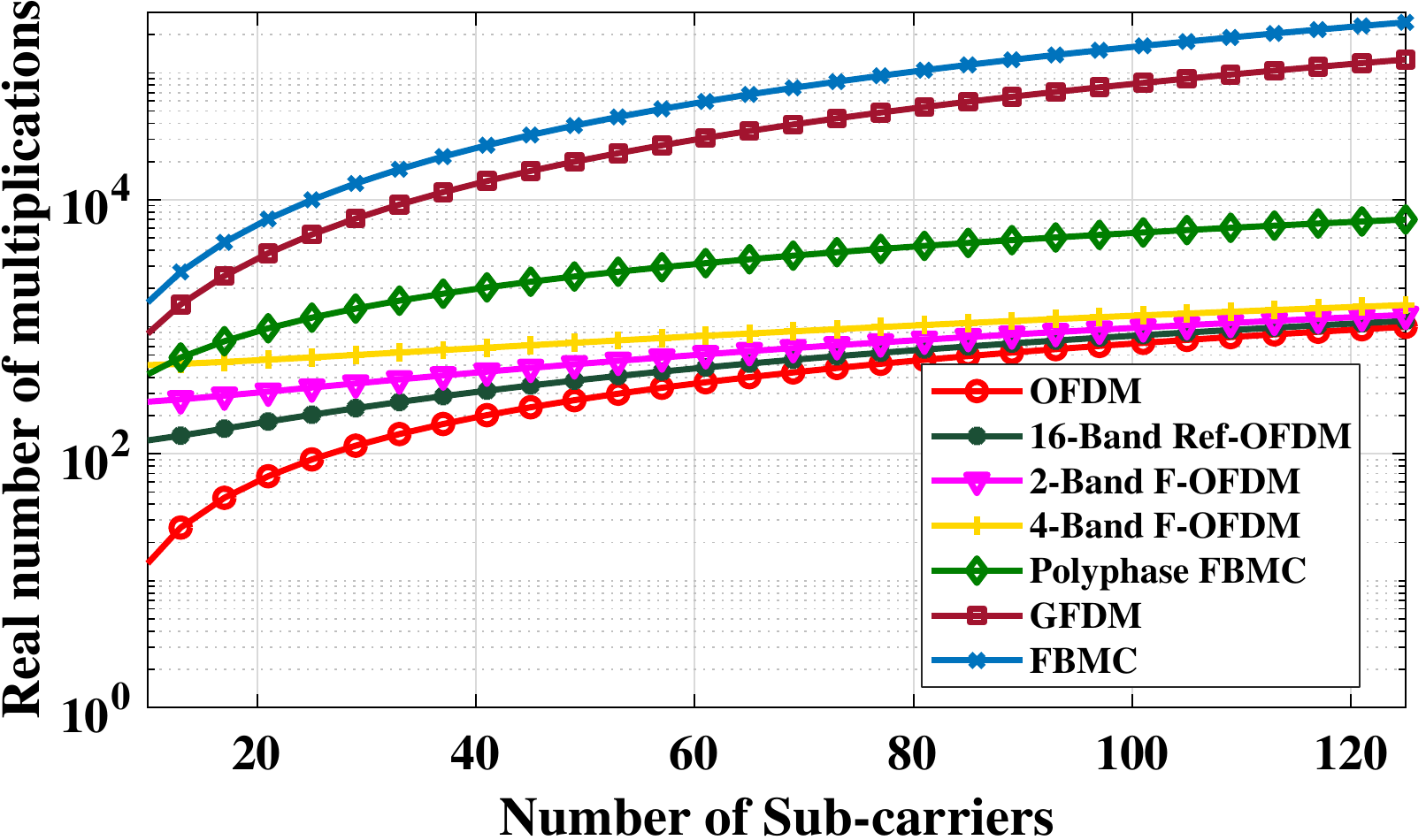}
\caption{computational complexity comparison of various waveforms}
\label{com}
\end{figure}

In this subsection, the complexity comparison of various waveforms in terms of the number of real multiplications for different numbers of sub-carriers is done. Here, we consider $K$-band transmissions in non-continuous bands where $K \in \{2,4\}$. For such transmissions, Ref-OFDM uses a single reconfigurable filter capable of offering 16-band response. We also consider OFDM with conventional filter design and referred to as filtered-OFDM (F-OFDM). As shown in Fig.~\ref{com}, as $K$ increases, the complexity of F-OFDM increases while that of Ref-OFDM remains the same. The complexity of GFDM and FBMC with polyphase filter implementation for single band transmission is much higher than 16-band Ref-OFDM waveform. Also, the complexity of Ref-OFDM waveform is close to that of OFDM for 128 sub-carrier case making the proposed protocol and waveform a good alternative to next generation LDACS. \\

For an easy understanding, the proposed Ref-OFDM based LDACS transceiver discussed above is summarized in the next section. \\
 
\section{Summary}
In this chapter, the need of new waveform for LDACS-DME coexistence scenario is explained briefly. For better understanding of LDACS inlay deployment, the coexistence environment is discussed first along with the brief description of considered real time wireless channels for A2GC. For efficient coexistence, a revised LDACS protocol with a new frame structure is proposed. It is compatible to the existing one. To dynamically adapt the various transmission bandwidths (186-732 KHz) a Ref-OFDM waveform is proposed for the LDACS transceivers. The reconfigurable filter is designed using combination of CDM and MCDM approach. A single prototype filter can serve all the transmission bandwidths by choosing an appropriate decimation factor. \\ 

Simulation results show significant improvement over the BER and at least 32 dB lower interference to incumbent $L$-band users than existing LDACS.  Furthermore, proposed work allows multi-band and multi-user transmission with adaptable bandwidth due to the proposed reconfigurable filter. Such transmission is not feasible in existing LDACS due to significant interference to incumbent $L$-band users. In addition, the computational complexity of Ref-OFDM is lower than other waveforms except OFDM making the proposed work an attractive solution for next-generation LDACS.\\

To validate the performance of the transceivers in real time, there is a need of the prototyping of LDACS transceivers on hardware. The next chapter discusses the design and implementation details of OFDM, WOLA-OFDM and F-OFDM on the Zynq System on Chip.

\chapter{Prototyping of transceiver models on ZSoC}
The increased demand has led to introduction of various new standards and protocols for LDACS transceivers as discussed in chapter 3. For such transceivers, heterogeneous Zynq System on Chip (ZSoC) platform, consisting of a processing system (PS) and a reconfigurable programmable logic (PL) on a single chip, is the preferred platform over two distinct chip based platforms \cite{zynq}.\\

In this chapter, we present the design and efficient implementation of the basic OFDM based transceivers for implementation on ZSoC using hardware software co-design workflow of MATLAB and Simulink along with the performance comparison of various waveforms. Such co-design approach gives the flexibility to choose which part of the system to implement on PL and which on PS to meet the given area, delay and power constraints. It also allows users to modify both the software and hardware according to their requirement. Based on the part implemented on PS and PL the design has seven configurations. The work presented here is the extension of the thesis \cite{sashath}.\\

Currently, the various configuration variants of basic OFDM transceiver is designed and implemented on ZSoC. These variants are realized by dividing the architecture into two sections, one for PL and other for PS. The work is also extended for WOLA-OFDM and F-OFDM transceivers. In this chapter first we will discuss the required Hardware-Software setup followed by the transceiver architecture. The Implementation using Hardware - Software codesign approach on ZSoC is discussed in the latter part of the chapter. Finally the experimental results are presented.\\

\section{Hardware - Software Setup}
In this section, the design details of ZSoC ZC706 hardware along with the software requirements to implement the transceiver design are given.\\

\subsection{Hardware design details}
The hardware required to validate the transceiver models, consists of Xilinx Zynq System on Chip ZC706 evaluation board along. A JTAG cable and an ethernet cable is used to make the connection between the host computer and the evaluation board.  \\

\subsubsection{Zynq System on Chip Architecture}
\begin{figure}[h!]
\centering
	\includegraphics[scale=0.6]{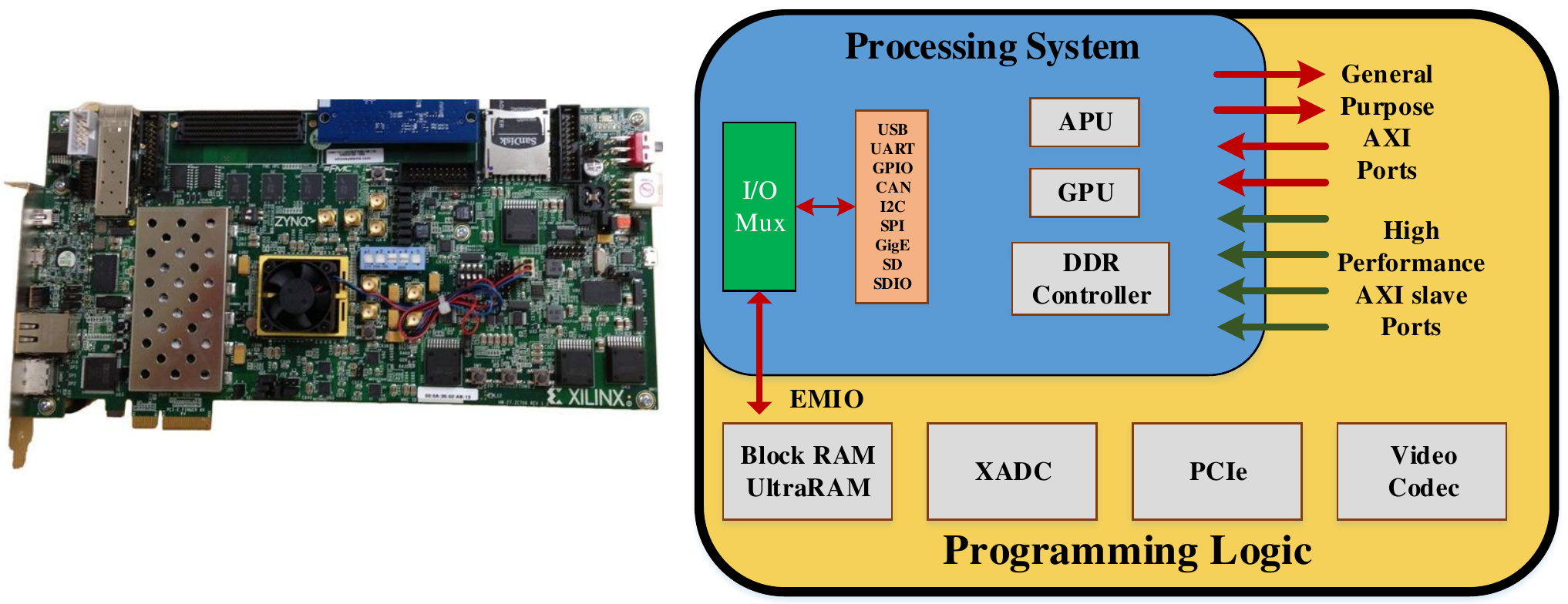}
	\caption{Spshot of Xilinx ZC706 evaluation board along with its important architectural features \cite{zynq}.}
	\label{Zyn}
\end{figure}

Xilinx ZSoC is a single chip platform which comes with higher degrees of flexibility, scalability and reconfigurability. The architecture of this is presented in Fig.~\ref{Zyn}. It provides the flexibility to design low end and high end applications on a single platform along with the flexibility of programming the processor system (PS) and programming logic (PL) separately according to the exact needs of corresponding application \cite{zynq_book}. ZSoc integrates the processor and FPGA with input output peripherals therefore, leading to lesser on board components. Due to this, it achieves better performance and leads to low power consumption compared to two chip platforms. ZC706 evaluation board is used to do all the analysis in our work. It consists of a dual core cortex A9 Advanced RISC Machines (ARM) as the software component (PS) and a Xilinx 28nm Kintex 7-series FPGA as the hardware component (PL) \cite{zynq}. Both PS and PL communicate with each other via Advanced eXtensible Interface (AXI) protocol. The specifications of the Zynq board are given in the Table.~ \ref{spec}.

\begin{table}
\centering
  \caption{Specifications of Zynq Board}
    \label{spec}
  \begin{tabular}{|c|c|}
        \hline
     \textbf{Device}& ZC706 \\
     \hline
     \textbf{FPGA}& Kintex-7 \\
     \textbf{Registers}& 4,37,200 \\
     \textbf{LUTs}& 2,18,600 \\
     \textbf{DSP slices}& 900 \\
     \textbf{BRAM blocks}& 545\\
     \textbf{Processor}& ARM Cortex 9 \\
     \hline
\end{tabular}
\end{table}

Programming System in ZSoC consists of the input/output peripherals, Application processor unit (APU), memory interfaces and interconnect. The APU has the dual core ARM Cortex-A9. the PS has a dual ported 256 KB on-chip RAM. The on-chip memory is accessible by both the CPU and the PL. Using on chip memory allows low latency access of data from the CPU, thereby increasing the speed of operation. Along with that it has 1GB of dynamic memory. dynamic memory controller allows 16 bit and 32 bit wide access to this 1 GB dynamic memory. It also allows PS and PL to share this memory. It has four 64-bit AXI slave ports, out of which two ports are dedicated to the PL, one to PS and one is shared by all the other AXI masters. In ZSoC, PS always boots first and thus making the architecture fully autonomous to PL. Along with memory, PS also has also has a total 130 IO port out of which 54 ports are used for multiplexed IO (MIO) which are shared by static/ flash memory interfaces and peripherals. The remaining 76 ports are dedicated for double data rate (DDR).\\

Programming Logic is similar to conventional FPGA which consists of flip flops, adders, look up tables, configurable logic blocks (CLBs) etc.
There are 8 LUTs, 16 flip flops and two 4-bit cascadeable adders per CLB. Along with CLBs, PL also consists of digital signal processing (DSP) blocks, 36 Kb Block RAM, PCI interface etc. ZC706 supports a wide range of voltage from 1.2V to 3.3V. It also has on-chip temperature and power supply sensors. The measurements are stored in dedicated registers and can be accessed using JTAG connection.\\

\begin{figure}[h!]
	\centering
	\includegraphics[scale=0.5]{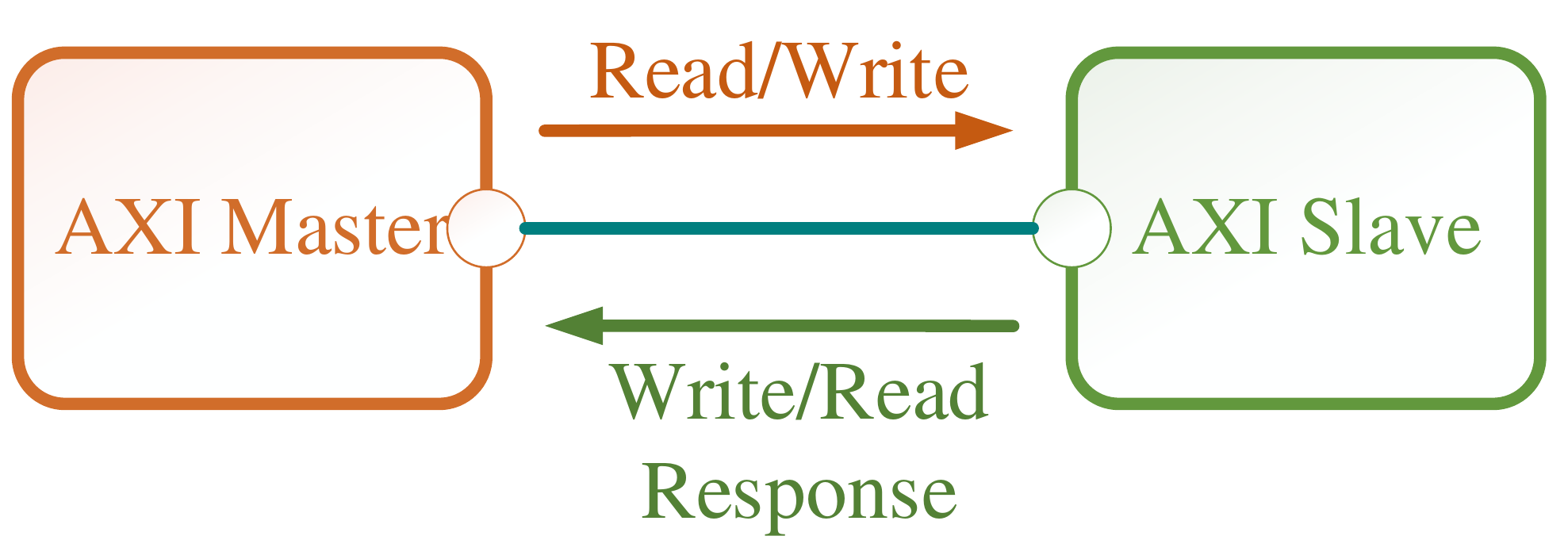}
	\vspace{-0.3cm}
	\caption{AXI master and slave link}
	\label{ax}
		\vspace{-0.3cm}
\end{figure}

In ZSoC, PS and PL are independent of each other and perform the task separately. Thus there is a need of proper communication standard or protocol for efficient implementation. For that Xilinx adopted AXI interface for zynq architecture. This AXI interface is based on Advanced Micro-controller Bus Architecture (AMBA) and synchronizes the data transfer between PS and PL. It has two parts : AXI master and AXI slave. The AXI master always initiates the read/write operation and AXI slave responds to that request as shown in Fig.\ref{ax}. In ZSoC, on 7 out of 9 ports PS acts as master and on the remaining 2 ports PL acts ass master. Three modes are used to configure these ports : AXI lite, AXI stream and AXI memory mapped. The experimental results analysis done in this paper uses AXI lite interface for low throughput requirements between PS and PL on LDACS transceiver.\\

\subsection{Software requirement}
To meet the hardware setup, tools from Mathworks and Xilinx are used to design and simulate the models. In this paper, the analysis is done using MATLAB 2016a and Vivado 2015.2.1. Embedded coder and HDL coder are the specialized tools from Mathworks to target the implementation on PS and PL respectively. In the next section, we will explain each block of the transceiver models. \\

\section{Transceiver Architecture}
In this section, the implementation design of transceiver system using OFDM, F-OFDM and WOLA-OFDM are described. The OFDM implementation of ZSoC has been discussed in \cite{Be1,Be2} and \cite{sashath} and the proposed transceiver is an extension with modifications as per LDACS specifications. The basic transmitter and receiver architecture remains same for all three waveforms. WOLA-OFDM and F-OFDM use some additional blocks to perform the functionality.

\subsection{Stimulus Subsystem}
The stimulus subsystem, reads the input bitstream to be transmitted from the MATLAB workspace. The input bitstream is a biry stream having 864 bits. Out of the 864 bit, 24 bits are transmitted per OFDM frame. Thus overall 36 OFDM frames are transmitted. All the processig is done on the 24 bits and repeates the operations after each new set of 24 bits. Reading of new 24 bits set is done using free running counter which keeps a track of the number of frames. The selector block then selects 24 bits from the incoming stream as the input to the transmitter.

\subsection{Transmitter}
Here, we will discuss about the detailed design architecture of OFDM, F-OFDM, WOLA-OFDM based transmitter.

\subsubsection{orthogonal Frequency Division Multiplexing}
\begin{figure}[h!]
	\centering
	\includegraphics[scale=0.5]{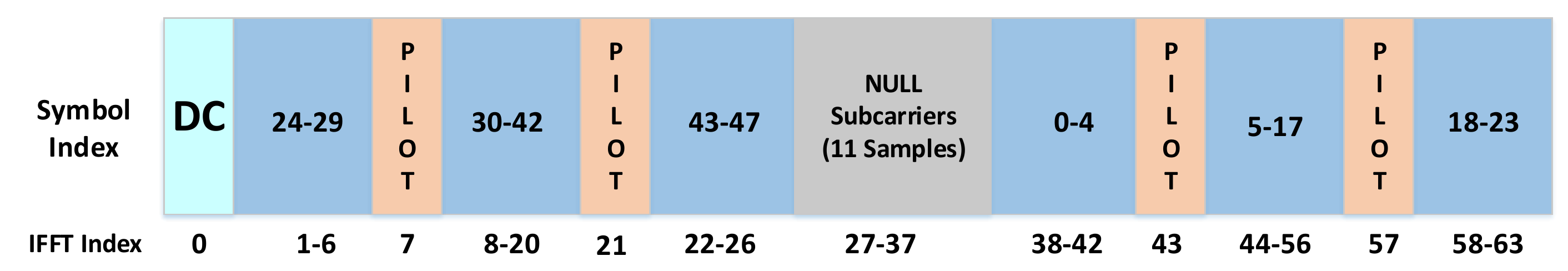}
	\vspace{-0.3cm}
	\caption{Symbols to Subcarrier Mapping}
	\label{SC_map}
		\vspace{-0.3cm}
\end{figure}

The OFDM based transmitter consists of blocks such as scrambler, convolutiol encoder, interleaver, biry phase shift keying (BPSK) modulator, Inverse fast Fourier transform (IFFT) and cyclic prefix adder. The scrambler does the bitwise XOR operation on the incoming input data and a random scrambling sequence generated by linear feedback shift register (LFSR). The same sequence is used to descramble the data at the receiver. This is follwed by convolutiol encoder which uses the generator polynomial of g0 = 133 and g1 = 171. These correspond to a rate 1/2 code with maximum free distance for K = 7. Output of convolution encoder is twice in length of the input. The output of encoder goes to the input to interleaver. It performs two step permutation on coded data and used to handle burst error. The interleaved data is then converted to complex symbols using BPSK modulator. It has -1 and +1 constellation points. After the modulation we get 48 symbols.The OFDM transceiver can also use other modulation schemes such as QPSK, 16 QAM or 64 QAM but here we have used BPSK. These symbols are then mapped to 64 point IFFT points as shown in Fig.~ \ref{SC_map}. Here, 64 subcarriers are used out of which 48 subcarriers are data subcarriers along with the 4 subcarriers containing pilot symbols in each frame. In addition to that, Null subcarriers are added at the remaining locations with single DC subcarriers in the middle.\\

To avoid inter symbol interference, cyclic prefix of length 16 is added to the OFDM symbol. Preambles are added for synchronization purpose. The preamble consists of both short training sequence (STS) and long training sequence(LTS). STS are used for timing acquisition, course frequency acquisition and diversity selection and LTS are used for channel estimation and fine frequency acquisition [7, 8].  For the length of 160 samples, LTS is repeated twice while STS is repeated 10 times. \\

\subsubsection{WOLA-OFDM}
The WOLA-OFDM does Root Raised Cosine (RRC) pulse shaping to soften the edges of rectangular pulse used in OFDM. This pulse shaping is known as windowing. In WOLA-OFDM, some part of the OFDM symbol is copied and applied at the start. The end of one symbol overlaps with the adjacent OFDM symbol. This leads to very less side lobe attenuation and allows wider transmission bandwidth when compared to OFDM. WOLA-OFDM transmitter requires cyclic suffix addition and windowing block in addition to the OFDM architecture.\\

\subsubsection{Filtered OFDM}
The F-OFDM uses linear phase finite impulse response filter instead of time domain windowing for further improvement in out-of-band attenuation. It enables higher transmission bandwidth compared to bandwidth limitation to 498 KHz in OFDM based LDACS system. This also allows transmission in non - contiguous bands and sharing of adjacent frequency band among multiple asynchronous transceivers by using the sub-band filtering approach. In the proposed F-OFDM transceiver, we have used linear phase bandpass filter \cite{SJ, SJ2} of order 150 with a normalized bandwidth of 0.86 and the transition bandwidth of 0.02. The use of filters increases the complexity while compared to the OFDM and WOLA-OFDM.

\subsection{Receiver}
The receiver receives a valid signal from the transmitter which enables the receiver functionality. The preamble detection detects the data frames using auto-correlation. Once the data frame is detected, it is transferred to OFDM demodulation block. For cyclic prefix removal, the starting 16 samples are discarded out of the 80 incoming samples. The remaining 64 samples are given as input to the 64 point FFT block. These 64 subcarriers are then mapped to the 
symbols. Out of 64 symbols, data symbols of lengh 48 are extracted by a selector. The output data symbols goes to the BPSK demodulator to retrieve the interleaved bits. The deinterleaver is used to deinterleave the bits using the pre-defined sequence. After that the bits are decoded using Viterbi decoder with the same generator polynomial as convolutiol encoder in the transmitter. The descrambler uses the corresponding descrambling sequence to retrieve the origil 24 bits. To validate the functiolity, the received data is stored in workspace and compared with the transmitted data. \\

At the receiver end, WOLA-OFDM performs overlap and add operation in which two adjacent OFDM symbols are overlapped with each other and then added to the next symbol to retrieve the data. F-OFDM performs the filtering operation at the receiver. The filter has the same specification as the transmitter.

\section{Implementation transceivers on ZSoC }
In this section, we will discuss Hardware - Software co-design workflow for the implementation of transceiver and its design variants.

\subsection{Hardware - Software co-design workflow}
\begin{figure}[!h]
\centering
	\includegraphics[scale=0.5]{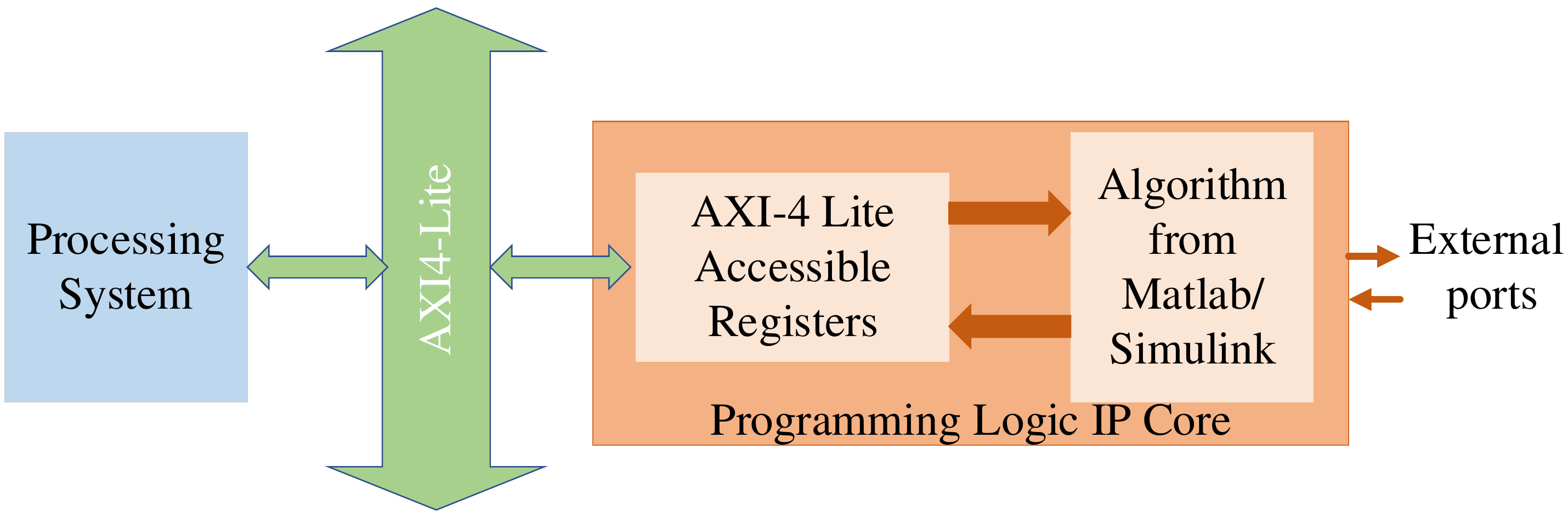}
	\caption{IP core generation Approach}
	\label{ip}
    \end{figure}

To design and simulate the transceiver models Hardware - Software Co-Design approach is being used. This is done using IP core generation in HDL workflow advisor as shown in Fig.~\ref{ip}. It is an important approach to implement any algorithm on ZSoC as it utilizes the heterogeneity of PS and PL. This approach also gives the flexibility to choose which part of the system is best suited to be implemented on PL and which on PS. PS makes easy and faster decision making operations on the other hand PL reduces power consumption and increases speed. The steps for hardware - software Co-Design approach are as follows: \\

\begin{enumerate}
    \item Designing a simulink model for transceivers and set the parameters like number of samples per frame, sampling frequency, total FFT size, Active subcarriers and subcarrier spacing. All the blocks present in the simulink library are not hardware synthesizable. So, while designing the simulink model these blocks need to be avoided.
    \item Differentiate the subsystem of the model which is going to implement on the PL believing that all the other subsystems will target to implement on PS. PL works in sample mode and PS works in frame mode, which requires an appropriate sample to frame and frame to sample conversion at the boundary of PS-PL interface. Fig.~ \ref{flow} shows the design have five functiol blocks 1, 2, 3, 4 and 5. Subsystem consisting of blocks 1, 2 and 3 are implemented on PS and remaining blocks are implemented on PL. Note that, the output to the host computer will come back through the PS.

    \begin{figure}[!h]
\centering
	\includegraphics[scale=0.4]{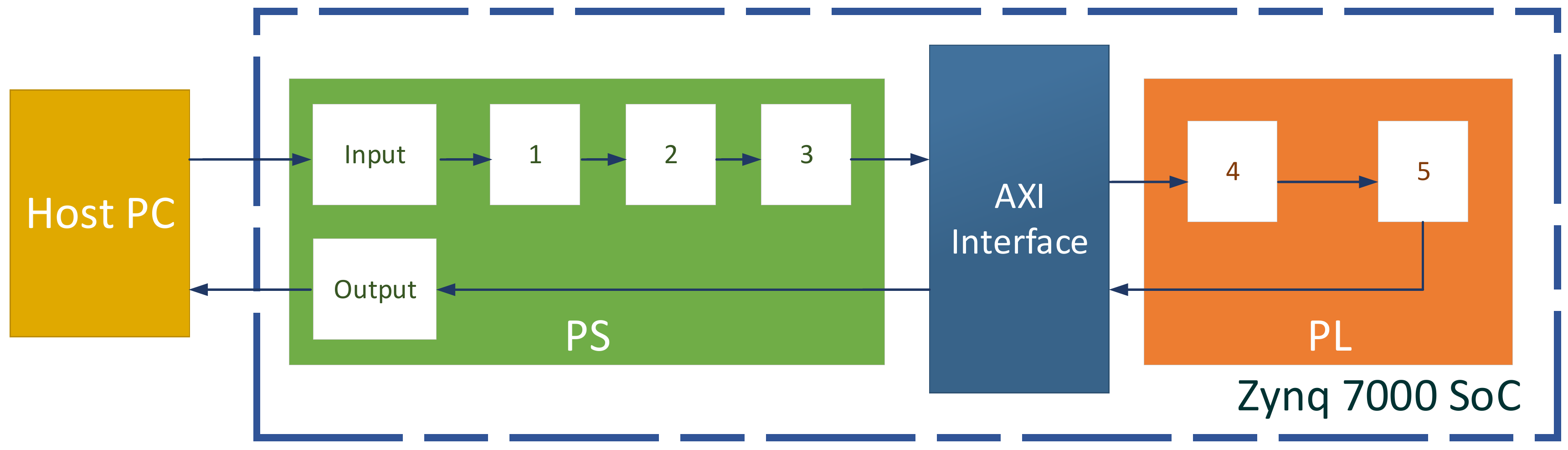}
	\caption{Hardware - Software Co-Design approach for algorithm implementation}
	\label{flow}
    \end{figure}

    \item Then, run the HDL Workflow Advisor to auto-generate an IP Core block for the transceiver design. It automatically generates a Vivado block diagram to combine the DUT with all the AXI interface components and creates an interface model to interact with the PL. It then uses the command line to synthesize, implement and bitstream generation. This bitstream is then used to program the PL.
    \item Filly, by setting the generated interface model to run in exterl mode, simulink uses Embedded coder to generate C code for all the processing blocks. Xilinx Vivado SDK then converts this C code to ARM executable code. When we run the simulation it launches the executable on PS via Ethernet. \\
\end{enumerate}

\subsection{Design Variants}
\begin{figure*}[!h]
  	\includegraphics[width=\linewidth]{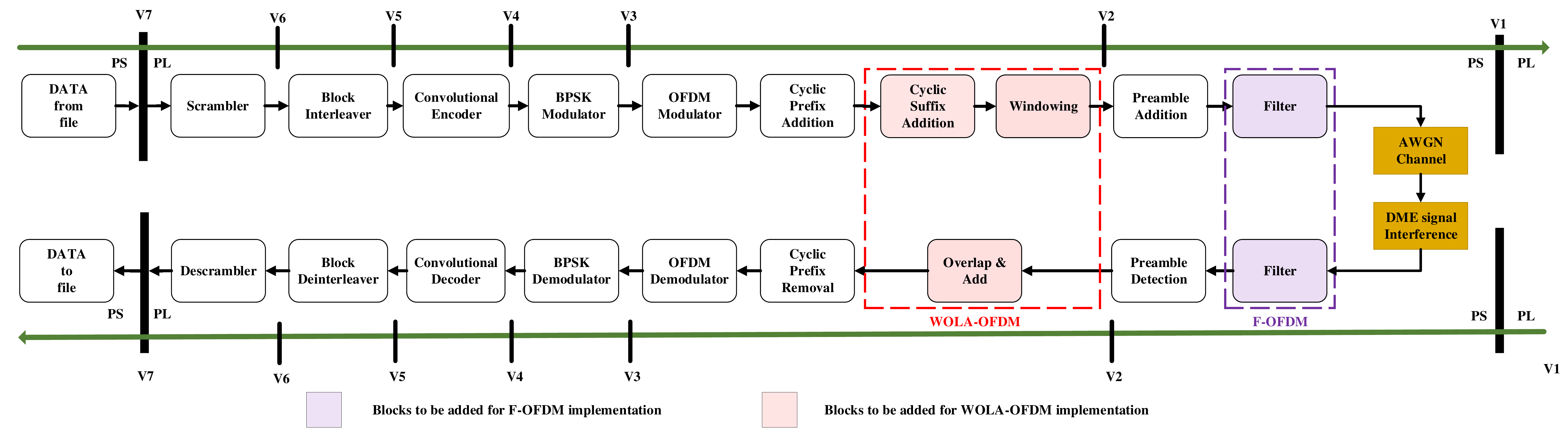}
  	\centering
  	\caption{Block diagram of the LDACS transceiver using various waveforms with seven different versions indicating division between PS and PL}
  	\label{bl_diag}
  \end{figure*}

The transceiver architecture implementation for all three waveforms has seven variant V1-V7 depends upon the subsystem targeted to PS and PL. In the first variant V1, the entire transceiver is implemented on the PS. We then move the components to the PL one-by-one in the subsequent versions i.e V2-V7 as shown in Fig.~ \ref{bl_diag}. Here, we will discuss the variants in detail.\\
  
\begin{figure}[!h]
	\centering
	\subfloat[]{\includegraphics[scale=0.3]{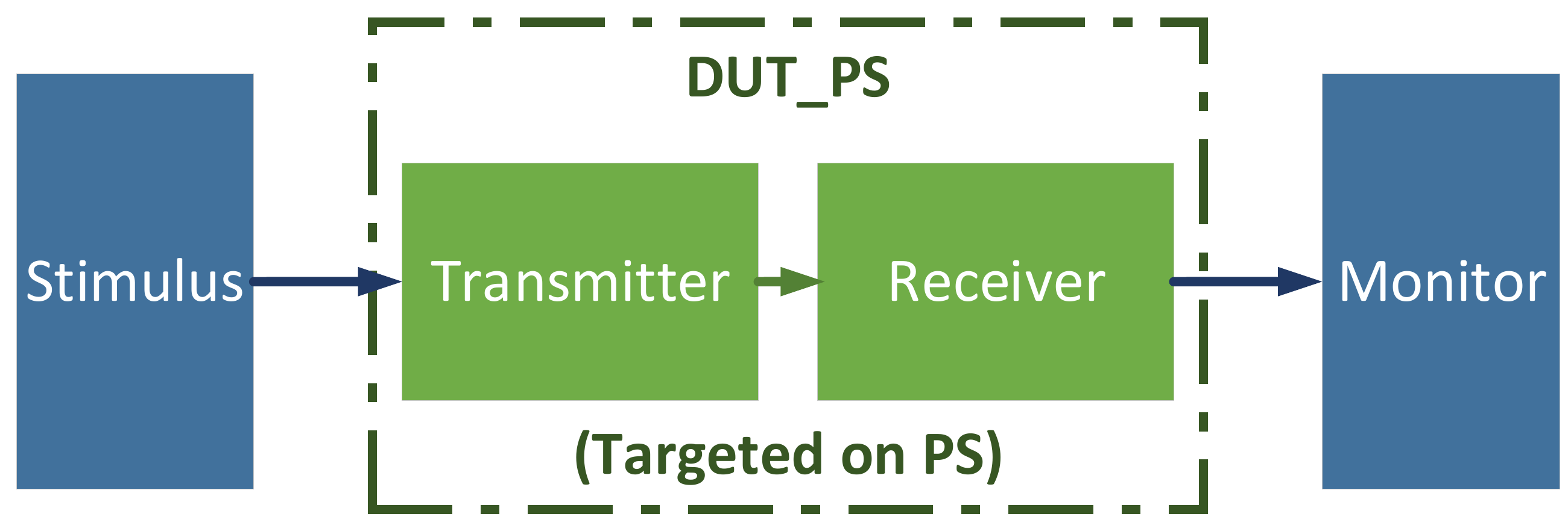}}%
		\label{psd1}
	\hspace{0.6cm}
	\subfloat[]{\includegraphics[scale=0.3]{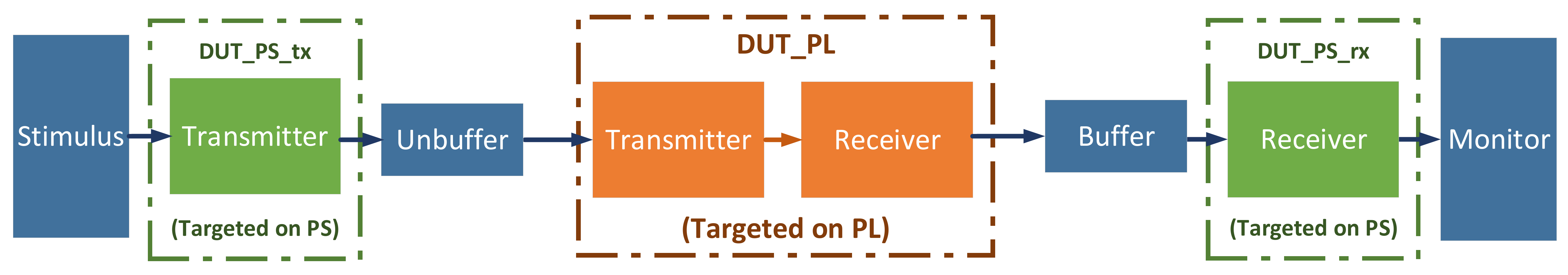}}
		\label{psd2}
   \vskip\baselineskip
	\subfloat[]{\includegraphics[scale=0.3]{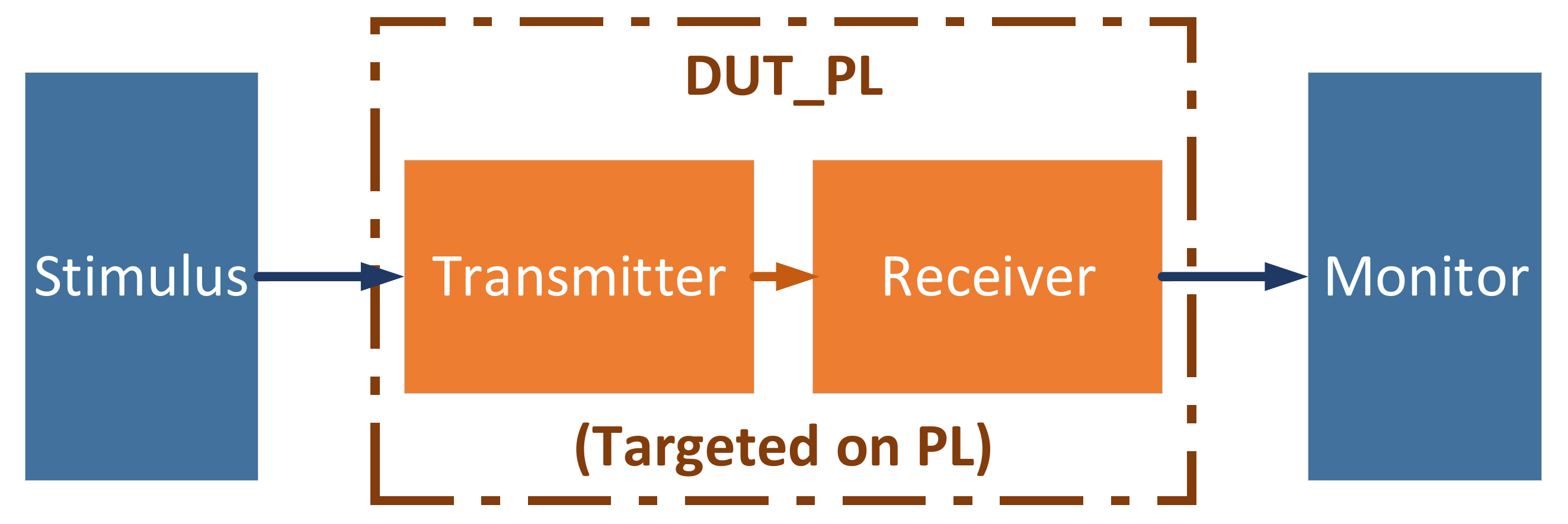}}%
		\label{psd3}
	\label{psd4}\\
	\caption{Top level architecture (a) Variant V1 (b) Variant V2-V6 (c) Variant V7}
	\label{Top}
\end{figure}

The entire transceiver model is implemented on the PS, there is no PL component in variant V1. The simulink model consists of three subsystems, Stimulus, DUT\_PS and Monitor as shown in Fig.~ \ref{Top} (a). The stimulus output is packed into a 32 bit signed integer. Out of these 32 bits, first 24 bits are data bits and remaining are valid bit, a reset bit and zero padding. This integer is then passed to the next DUT\_PS block. The dut-ps has both the transmitter and the receiver operations to be implemented on the PS of the evaluation board. Firstly, the 24 bits are extracted from the input integer then it enables the transceiver functionality depending on the valid bit. Thus, the output of the DUT\_PS block contains 80 samples, each of 16 bit fixed point data type along with the valid and the reset signal being. The sample time for each of the block has been kept equal to the frame time (tpf) of 80ns. The run time has been set to $43 * tpf$ to accommodate the delays encountered. To implement the model on the ZSoC, we follow the Program In Loop (PIL) verification method. In this method, a PIL block is generated for DUT\_PS subsystem and then it is deployed to the hardware. Te generated PIL model is run in the normal mode to load the processor of the target hardware. \\

In the V2 model, the preamble addition and preamble detection operations are implemented in PL while rest of the transceiver functionality are targeted to be implemented on the PS. The model contains five subsystems such as stimulus, DUT\_PS\_tx, DUT\_PL, DUT\_PS\_rx and monitor. The DUT\_PS\_tx and DUT\_PS\_rx subsystems will be implemented on the PS while the DUT\_PL will be targeted to PL. The output of the DUT\_PS\_tx subsystem combines the 16 bit fixed point complex output of the OFDM modulator, valid and reset bits to 32 bit unsigned integer in order to pass the data from PS to PL via AXI interface. 1 PS frame contains 80 such samples, which are then converted to 80 PL sample using unbuffer. The DUT\_PL then extracts the 16 bit complex OFDM sample from the input along with the valid and reset signal. The PL then does the preamble addition and detection as explained previously. The output of the preamble detection is packed into a 32 bit unsigned integer. To generate a complete OFDM frame, a buffer of size 80 is used. This complete frame is passed to DUT\_PS\_rx which performs operations such as OFDM demodulation, BPSK demodulation, deinterleaving, Viterbi decoding and descrambling  on the input data. The blocks in DUT\_PL have sample time as  $1 \mu s$ while the blocks in both PS subsystems have sample time as $80 \mu s$, equivalent to the frame time. \\

similarly, in version V3 - V7 one by one block is transferred to the PL subsystem subsequently. The top model architecture of version V2-V6 is shown in Fig.~\ref{Top} (b). In V7, all the components of the model are implemented on PL subsystem as shown in Fig.~\ref{Top} (c). To verify the model on ZSoC, FPGA In Loop (FIL) approach is used. The implementation is done using HDL workflow advisor which generates a FIL block med after the top-level module and places it in a new model. After this new model generation the HDL Workflow Advisor opens a command window to generate bitstream. Then, the bitstream is loaded to the PL using the FIL block. Then we run the simulation and verify the output of generated model with the output of origil model. The data types of the data transferred between PS an PL in each model version is shown in the  Table.~\ref{res}. \\

\begin{table}[h!]
     \centering
     		\caption{Data Transfer between PS and PL}
    \begin{tabular}{| m{3.5cm} | m{4cm}| m{3.5cm} | m{3.5cm} |}
    \hline
         \textbf{Model Variants} &\textbf{Data Type}& \textbf{Size of 1 element} &\textbf{No. of elements}  \\
         \hline
         V1 & Signed Fixed Point & 16 bits & 80\\
         \hline
        V2 & Signed Fixed Point & 16 bits & 80\\
        \hline
        V3 & Signed Fixed Point & 16 bits & 64\\
        \hline
        V4 & Boolean & 1 bit & 48\\
        \hline
        V5 & Boolean & 1 bit & 48\\
        \hline
        V6 & Boolean & 1 bit & 48\\
        \hline
        V7 & Boolean & 1 bit & 48\\
        \hline
    \end{tabular}
\label{res}
\end{table}

In addition to that, pipelining has been applied in all the seven variants to improve the speed at which it can be clocked. It is done by first checking the max critical path delay locations and then inserting the delays at appropriate location to break the path.\\
\\

\section{Experimental Results}
In this section, we will compare the performance of OFDM, F-OFDM and WOLA-OFDM based LDACS transceiver in terms of PSD, BER, area (resource utilization), power and Critical delay. These results are taken by implementing it on ZC706 evaluation board hardware. The setup requires a host PC, a ZC706 evaluation board and a JTAG and Ethernet cable along with softwares mentioned in the previous sections. As discussed earlier, PL works in sample mode and has fixed sample time of $1\mu s$ and PS works on frame mode consisting of 80 samples per frame having corresponding frame time of $80 \mu s$.\\

\subsection{Power Spectral Density Comparison}
\begin{figure}[!h]
\centering
	\includegraphics[scale=0.7]{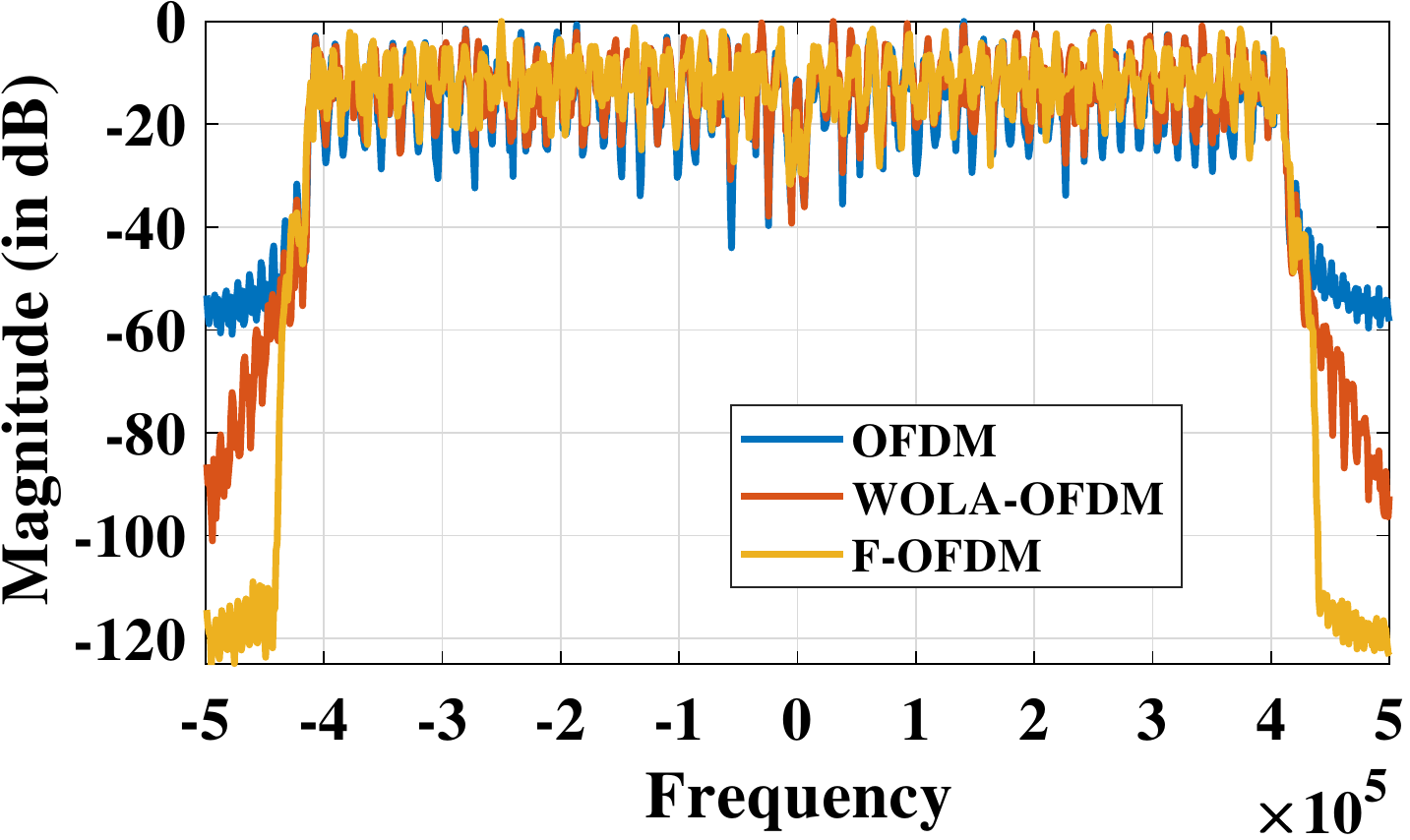}
	\caption{Power spectral density comparison of OFDM, WOLA-OFDM and F-OFDM.}
	\label{PS}
\end{figure}

Here, we will do the comparison of OFDM, WOLA-OFDM amd F-OFDM based LDACS transceiver with respect to their out-of-band emission using the power spectral density (PSD). In LDACS-DME coexistence, DME has total of 1 MHz bandwidth out of which only 498 KHz is used by the existing LDACS system because of the intolerable out of band emission. Our proposed Filtered OFDM waveform has much lower side lobe attenuation and can occupy larger bandwidth which benefits to the air to ground communication. Here, we have performed the demonstration for the transmission bandwidth of 800 KHz as shown in Fig.~ \ref{PS} and side lobe attenuation of F-OFDM comes within the tolerable limit of DME. This concludes that F-OFDM outperforms in LDACS-DME coexistence scenario and can be allocated for larger bandwidth. For validating this waveform ber analysis is also presented in the next subsection.\\

\subsection{Bit Error Rate Comparison}
The ber performance comparison for OFDM, WOLA-OFDM amd F-OFDM transceiver in presence of DME signal is shown in Fig.~ \ref{ber}. It can be observed that WOLA-OFDM has almost similar ber as OFDM. In presence of DME interference, F-OFDM has better ber performance than OFDM because of the less interference to the DME signal. This validates that F-OFDM based LDACS transceiver is a better choice than existing OFDM based LDACS transceiver .\\

\begin{figure}[!h]
\centering
	\includegraphics[scale=0.7]{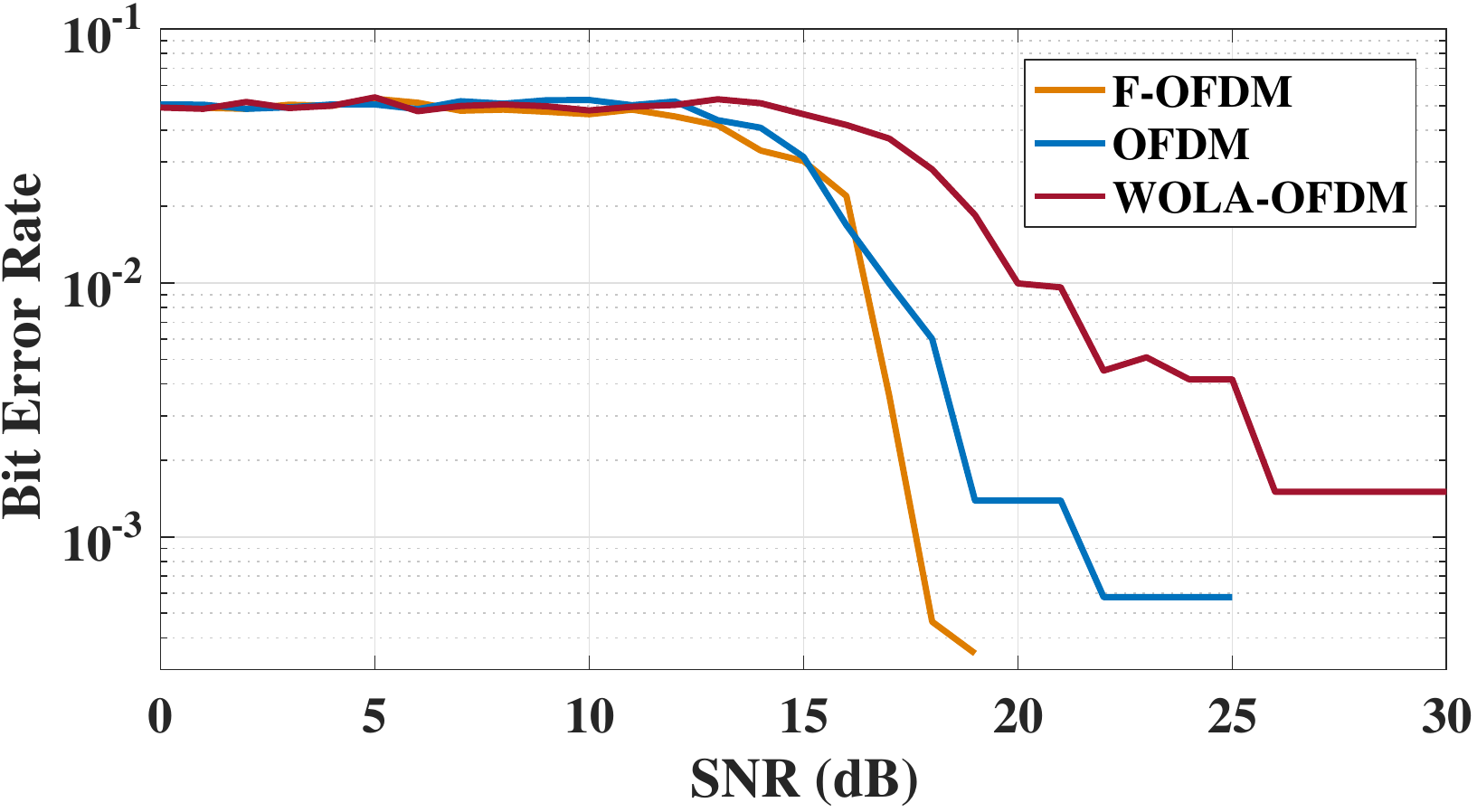}
	\caption{Bit Error Rate comparison of OFDM, WOLA-OFDM and F-OFDM in presence of DME signal.}
	\label{ber}
\end{figure}

\subsection{Resource Utilization and dynamic Power}
In this subsection, we compare the hardware implementation results of the various configurations for all three transceivers. The resource utilization on ZC706 evaluation board is shown in Table ~\ref{result}. Here, we compare the number of flip flops, LUTs, memory LUTs, Registers, DSP48, and Multiplexers. Flip flops and LUTs are the basic building blocks of the FPGA. Registers are mainly used for pipelining. It reduces the critical path delay of the architecture as discussed previously. The critical path delay with our pipeling method is very less as compared to the delay presented in \cite{sashath,Be1}. The critical path delay value for OFDM, WOLA-OFDM and F-OFDM is 9.75 ns, 10.25 ns, 14 ns respectively for all the seven design configurations. As expected, F-OFDM has higher critical path delay because of the filtering operation. \\

DSP-48 are the hardware embedded units specialized to efficiently perform various operations like multiply-accumulator, multiply-adder, counters etc. Usually, FIR filter are realized using these DSP48 elements. In the proposed architectures, we have also used DSP48 for preamble block implementation. As shown in Table.~\ref{result}, DSP48 utilization is around 30\% higher in F-OFDM when compared to OFDM and WOLA-OFDM mainly due to filtering operation. There is slight increase in DSP48 operation in all architectures from V2 to V3 due to FFT/IFFT operations.\\

\begin{table}[h!]
     \centering
     		\caption{Data Transfer between PS and PL}
    \begin{tabular}{| m{2.7cm} | m{2cm}| m{1.7cm} | m{1.7cm}| m{1.7cm} | m{1.7cm}| m{1.7cm} | m{1.7cm} |}
    \hline
         \textbf{Parameter} & \textbf{Waveform}& \textbf{V2} & \textbf{V3} & \textbf{V4} & \textbf{V5}& \textbf{V6} & \textbf{V7}\\
		\hline
        \multirow{3}{*}{\textbf{Flip Flops}} & OFDM & 23806 (5.45\%) & 31617 (7.23\%) & 32628 (7.48\%) & 32628 (7.48\%) & 37738 (8.63\%) & 38193 (8.74\%)\\
		& WOLA-OFDM & 23712 (5.42\%) & 27298 (6.24\%) & 37926 (8.67\%) & 40119 (9.17\%) & 41119 (9.40) & 42791 (9.79)\\
    	& F-OFDM & 33198 (7.59\%) & 40951 (9.37\%) & 41994 (9.61\%) & 42040 (9.62\%) & 44820 (10.25) & 47825 (10.94)\\
		\hline

		\multirow{3}{*}{\textbf{DSP48 units}}& OFDM & 554 (61.55\%) & 570 (63.33\%) & 570 (63.33\%) & 570 (63.33\%) & 570 (63.33\%) & 570 (63.33)\\
		& WOLA-OFDM & 554 (61.55\%) & 570 (63.33\%) & 570 (63.33\%) & 570 (63.33\%) & 570 (63.33\%) & 570 (63.33\%)\\
		& F-OFDM & 830 (92.22\%) & 866 (96.22\%) & 866 (96.22\%) & 866 (96.22\%) & 866 (96.22\%) & 866 (96.22\%) \\
	
		\hline
					
		\multirow{3}{*}{\textbf{Memory LUT}} & OFDM & 396 (0.56\%) & 865 (1.23\%) & 918 (1.30\%) & 922 (1.31\%) & 905 (1.29\%) & 894 (1.27\%)\\
		& WOLA-OFDM & 474 (0.673\%) & 940 (1.34\%) & 972 (1.38\%) & 982 (26.09\%) & 839 (1.19\%) & 828 (1.17 \%)\\
        & F-OFDM & 394 (0.56\%) & 866 (1.23\%) & 869 (1.23) & 871 (1.24\%) & 880 (1.25\%) & 893 (1.27\%)\\
		\hline
					
		\multirow{3}{*}{\textbf{LUTs}} & OFDM & 26141 (11.96\%) & 31687 (14.50\%) & 32555 (14.89\%) & 33509 (15.328\%) & 35509 (16.24)& 36657 (16.77\%)\\
		& WOLA-OFDM & 26221 (11.99\%) & 34816 (15.93\%) & 40439 (18.50\%) & 41114 (18.80\%) & 43432 (19.86 \%) & 44738 (20.46\%)\\
        & F-OFDM & 31983 (14.63\%) & 30703 (14.045\%) & 33321 (15.243\%) & 34782 (15.423\%)  & 41619 (19.04\%) & 44371 (20.30\%)\\
		\hline
					
		\multirow{3}{*}{\textbf{Multiplexers}}& OFDM & 35 & 683 & 1144 & 1217 & 1882 & 1930\\
		& WOLA-OFDM & 35 & 1131 & 1597 & 1673 & 2386 & 2502 \\
		& F-OFDM & 35 & 683 & 1327 & 1401 & 1882 & 1930\\
		\hline
					
		\multirow{3}{*}{\textbf{\pbox{20cm}{\textbf{No. of} \\ \textbf{Registers}}}} & OFDM & 3098 & 3378 & 4479 & 4480 & 7655 & 7780\\
		& WOLA-OFDM & 3101 & 3449 & 4511 & 4512 & 7960 & 8087\\
		& F-OFDM & 3262 & 3982 & 4363 & 4364 & 7933 & 8386\\
		
		\hline
					
		\multirow{3}{*}{\pbox{20cm}{\textbf{Dynamic Pow-} \\ \textbf{er (in Watts)}}} & OFDM & 2.078 & 1.998 & 1.997 & 1.996 & 1.995 & 0.617\\
		& WOLA-OFDM & 2.211 & 2.100 & 2.000 & 2.001 & 1.978 & 0.509\\
		& F-OFDM & 2.361 & 2.357 & 2.353 & 2.349 & 2.286 & 0.898\\
        \hline
    \end{tabular}
\label{result}
\end{table}

So, overall F-OFDM offers better spectral containment without compromising the BER along with less than 50\% of resource utilization. It is also observed that the FPGA resource utilization will increase as we move block to PL from V1-V5, while the PS step time will decrease. However, in all the configurations, the utilization of the FPGA slices and LUTs for the PL section is less than 27\% and 35\% respectively leaving enough resources for higher LDACS layers. \\

\section{Summary}
In this chapter the detailed performance and complexity analysis of various candidate waveforms for LDACS on Zynq System on Chip (ZSoC) platform is presented. The evaluation board ZC706 Zynq-7000 XC7Z045 of programmable logic (PL) such as FPGA and processing system (PS) such as ARM is used for prototyping. The requirement of hardware and tools to implement the models on ZSoc along with the thorough details on the OFDM, F-OFDM and WOLA-OFDM transceiver architecture is discussed in the starting part of chapter. Apart from this, Seven configuration of of the architecture are realized by dividing it into two sections, one for PL and other for PS.\\

The Hardware - Software codesign approach is explained in detail which provides the flexibility to choose which part of the transceiver to implement on programmable logic (PL) such as FPGA and which on processing system (PS) such as ARM to meet the given area, delay and power constraints. Detailed experimental results demonstrate the trade-off between these waveforms with respect to parameters such as, area, delay and power requirements. The PSD and BER results in presence of DME interference shows that F-OFDM perform better than OFDM and WOLA-OFDM.\\

In future, we will deploy the LDACS specifications to all these configuration models. Additionally, we will compare the performance of these models with different word lengths. In addition to that, we will integrate the RF front end AD9361 transceiver to validate the performance over the air transmission.\\

\chapter{Conclusions and Future Works}
In this chapter, a brief synopsis of the contributions as well as conclusions of the work presented in this report is done. Some directions for future work in this research area are also identified.\\

\section{Conclusions}
The challenges in existing air to ground communication due to exponentially increasing air traffic are addressed in this report. The work in this report mainly focuses on the design and prototyping of Ref-OFDM based LDACS transceiver in simulation and hardware. A revised frame structure is proposed for LDACS which is completely compatible with the existing frame structure. The Ref-OFDM waveform uses a bandwidth reconfigurable filter design in which a single prototype filter is used to accommodate all the possible transmission bandwidths. For instance, the Ref-OFDM waveform can switch to OFDM waveform by skipping the filtering operation for narrowband or single user transmission. The dynamic partial reconfiguration can be exploited to switch between the single band and multiband   transmissions. For instance, in case of single band transmission, entire DFT block can be replaced with single adder block. Furthermore, the size of the DFT can be changed on the- fly as per the desired center frequency of the transmission. Such flexibility is not possible in case of OFDM, FBMC and GFDM based LDACS. In addition, Ref-OFDM can be easily extended to a multi-antenna system, unlike FBMC based LDACS. Also, the Ref-OFDM have much better-localized frequency spectrum than OFDM because of the filtering which makes it an attractive alternative to the OFDM based LDACS.\\

The second part of the report focuses on the prototyping of seven design variants (V1-V7) of LDACS transceivers to efficient and reliable implementation on ZSoC. These variants are basically depends on the subsystem implemented on FPGA (PL) and processor (PS). In each subsequent design variant, one block transferred to the FPGA. This is done using hardware - software codesign approach in MATLAB simulink and Xilinx Vivado. It provides the flexibility to choose the part which should be implemented on PL according to the area, power, delay constraints. The transceiver design architecture and its implementation is thoroughly explained in chapter 4. Currently, the basic architecture of OFDM, F-OFDM and WOLA-OFDM are designed, which will be further extended to the LDACS specifications. Experimental results shows that F-OFDM performs better than OFDM and WOLA-OFDM in terms of spectral containment and BER with the trade-off in resource utilization and power consumption.\\

\section{Future Works}
Currently, the proposed revised protocol of Ref-OFDM based LDACS transceivers have the pilot pattern same as the previous frame structure which limits the possible bandwidths with a certain amount of bandwidth gap. This limits the flexibility of the proposed protocol. The work done till now is totally based on the simulation results which are not completely sufficient to claim that the proposed waveform is the superior to all. Some possible future works to make it more feasible for A2GC are discussed below. \\

\subsection{Exploit the revised frame structure along with the theoretical analysis}
To fulfil the requirement of same pilot patterns for having the compatibility with the existing frame structure, the possible bandwidths needs to have 80 KHz gap. The filter design for this uses the CDM and MCDM method but it does not appropriately use the advantages of the MCDM method. When $\omega_{cd} > 0.5\pi$ then we use the MCDM method so that it can accommodate the desired cut-off frequency using a smaller value of decimation factor. But the filter design proposed in this report uses MCDM for the desired cut-off frequency $\omega_{cd} < 0.5\pi$ and $\omega_{cd} > 0.5\pi$ both. For $\omega_{cd} < 0.5\pi$, MCDM uses even higher value of decimation factor which ideally should not happen. Further, we are going to design a new frame structure which can accommodate any bandwidth. The filter will be designed in such a way that it can be used for all the possible transmission bandwidths along with the efficient use of MCDM method. This will lead to a lesser value of decimation factor for narrow bandwidths using CDM and for wider bandwidth using MCDM.\\

We also plan to extend this analysis to complex scenarios where multiple LDACS users share the band between incumbent DME signals. Such deployment is feasible only in the proposed LDACS protocol due to tunable bandwidth and hence, novel approaches need to be explored for the analysis. The work presented in this report support the claims via Monte-Carlo simulation for the proposed candidate for Air to ground communication. In future, we will do the theoretical analysis of the proposed Reconfigurable Filtered OFDM waveform based LDACS to obtain closed-form expressions for its out-of-band emission and bit-error-rate performance in the presence of incumbent DME signals. \\

\subsection{RF front end AD9361 integration with the prototypes on ZSoC}
The designed models till now are implemented on ZSoC and evaluated in terms of resource utilization, PSD. These models can not be used for over the air antenna transmission and reception of the signal until an RF transceiver is not associated with them. The RF transceiver unit combines the RF signal processing, converting, and digitization (and vice-versa) in one single device. \\

We will integrate the RF front end AD9361 transceivers with the designed transceiver models. The AD9361 is a high performance RF agile transceiver designed for broad range of transceiver applications. It combines a RF front end with a flexible mixed-signal baseband section and integrated frequency synthesizer which simplifies the design by providing a configurable digital interface to a processor.\\

\begin{figure}[h!]
	\centering
	\subfloat[]{\includegraphics[scale=0.4]{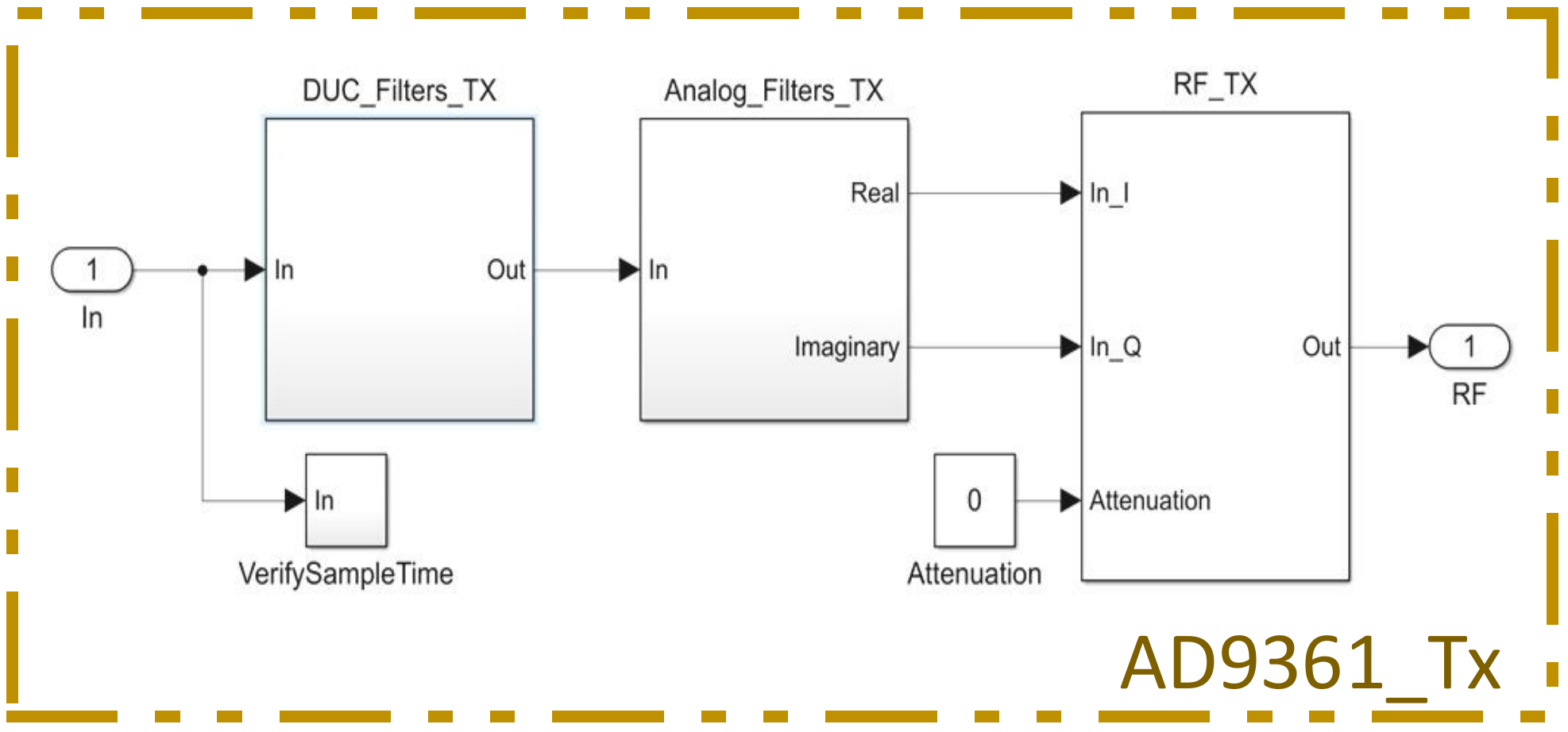}}%
	\hspace{0.6cm}
	\subfloat[]{\includegraphics[scale=0.4]{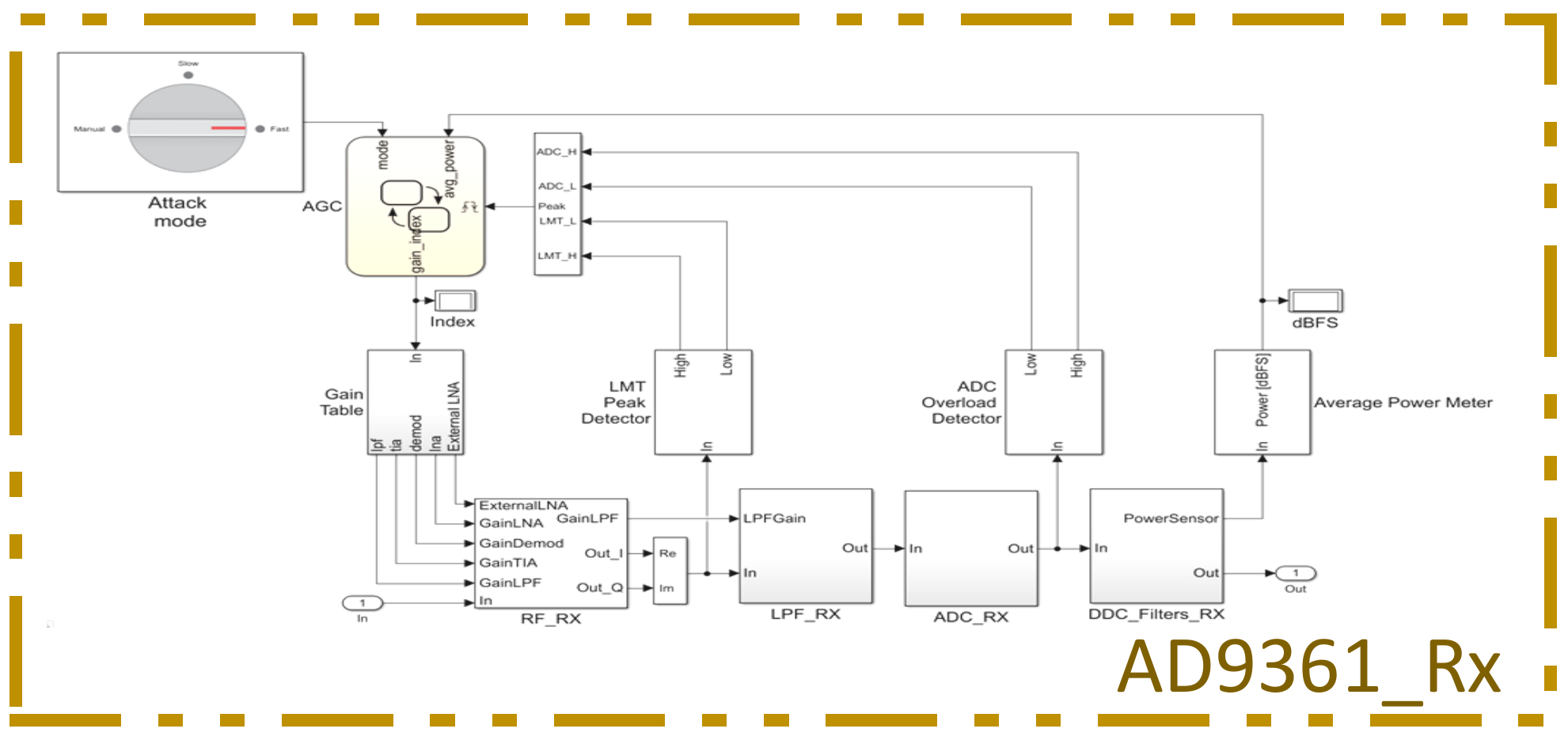}}
	\caption{AD9361 Implementation architecture (a) Transmitter (b) Receiver}
	\label{AD}
\end{figure}

AD9361 transmitter and receiver models will be used from the \lq\lq RF Blockset models for Analog Devices RF Transceivers \rq\rq add on in the simulink library. The AD9361 transmitter and receiver models are shown in Fig.~ \ref{AD} (a) and (b) respectively. The transmitter consists of Digital up-conversion filters (DUC Filters TX)), Analog filters (Analog Filters TX) and RF front end (RF TX). We can modify the settings of the digital up-conversion filters, the analog filters and the digital to analog data converter by using the custom filter configurations. The default configurations for filter are LTE 5 MHz, 10 MHz and 20 MHz. The digital up-conversion filters converts the baseband signal to an intermediate frequency (IF) signal. The sample rate of the input signal should be same as the sample rate of DUC. Digital filter introduces the noise floor. The analog filters are used to shape this noise floor, and provide a continuous time signal processed by the RF front end. The RF transmitter then up converts
the baseband signal around the Local Oscillator (LO) centre frequency using a quadrature modulator. This LO centre frequency is defined in the top mask. Once the signal is transmitted over the air, it is then received by AD9361 receiver. \\

The RF receiver down-converts the signal centred on the same LO frequency to baseband using a quadrature demodulator. The receiver has mainly three components as Low Noise Amplifier (LNA), quadrature demodulator (Mixer) and Trans-impedance amplifier (TIA); the chain is indicated as LMT. The gains of each components are tunable and controlled by the AGC. The analog filters provide a continuous time signal to
the ADC. The ADC models a high-sampling rate third order delta-sigma modulator. The low-pass digital filters convert the highly sampled signal at the output of the ADC to a lower baseband rate.\\

The integration will be same for all the variants. AD9361 has only three sampling rate as mentioned above while our OFDM transceiver has the sampling rate of 1 MHz. To match its sampling rate with the sampling rate of OFDM transceivers a custom filter needs to be designed. The custom filter can be designed with the help of AD9361 Filter Wizard. The passband and stopband frequency for this filter should be according to the input signal. The filter at the receiver side will have the exact same specifications as the transmitter. The output of AD9361 receiver should match the power level of input to the AD9361 transmitter. The synchronization can be handled by identifying the phase of the pilot symbols and shifting the signal accordingly. \\

The FMCOMM-AD9361 board is connected to the FPGA Mezzanine Card (FMC) slot on the Xilinx FPGA board to complete the integration of AD9361 transceiver models with the designed prototypes of OFDM transceivers.\\
In addition to this, there are some possible future works to make it more feasible for A2GC. The existing OFDM transceivers are designed for general 802.11 standard on ZSoC. We aim to modify the designs as per the LDACS specifications for LDACS-DME coexistence scenario. Currently, we have extended the OFDM transceiver designs to F-OFDM. The F-OFDM uses single filter for all the desired bandwidths leads to the high complexity and power consumption. To reduce the complexity and power consumption, we will design our proposed Ref-OFDM by designing a reconfigurable filter model for all the possible bandwidths.\\

For an efficient utilization of vacant spectrum between two adjacent DME signals Ref-OFDM based LDACS was proposed which enables the transmission in multiple narrowband transmissions such as multi-user and multi-band systems. Till now, the designed models for OFDM works only for single bandwidth transmission. We will focus to implement and design the models for multi-user and multi-band transmission using reconfigurable filtered OFDM, which will use one filter to serve all users along with all the bandwidths. \\

\section{Summary}
This section summarizes the conclusions and possible future works in the concerned area of research. A new revised Ref-OFDM based LDACS protocol has been proposed which can adapt to any desired transmission bandwidths. A reconfigurable filter is designed which uses a single filter for all the possible bandwidths. For removing the fixed gap constraint within the possible bandwidths, a new frame structure will be designed in future along with the new advanced filter design. The performance comparison and analysis is done via MATLAB simulation. To claim the better performance of Ref-OFDM based LDACS, we will do the theoretical analysis to obtain closed-form expressions for Ref-OFDM LDACS. For real time hardware validation, the transceiver prototype IEEE 802.11 standards is designed to be implemented on ZSoC. For LDACS-DME coexistence validation, we will modify the designs for LDACS specifications along with the reconfigurable filter design. Also for over the air antenna transmission, AD9361 RF transceiver will be integrated with the design models along with the multi-user and mulit-band transmission for efficient spectrum utilization.\\

Overall, the work planned as part of this thesis involves the design, analysis, validation as well as hardware implementation of the proposed LDACS protocol. I believe that this work will lead to a good comprehensive thesis.

\end{document}